\documentclass[]{aa}
\usepackage{txfonts}
\usepackage{graphicx}
\usepackage{amssymb}
\usepackage{natbib}
\bibpunct{(}{)}{;}{a}{}{,}

\begin{document}

\title{A $\rm 3-5~\mu m$ VLT spectroscopic survey of embedded young low mass stars I}
\subtitle{Structure of the CO ice}
\titlerunning{$\rm 3-5~\mu m$ VLT spectroscopy of low mass YSOs I}

\author{K. M. Pontoppidan\inst{1} \and H. J. Fraser\inst{1}  \and E. Dartois\inst{2}  \and W.-F. Thi\inst{3}  \and E. F. van Dishoeck\inst{1} \and A. C. A. Boogert\inst{4}  \and L. d'Hendecourt\inst{2}  \and A. G. G. M. Tielens\inst{5}  \and S. E. Bisschop\inst{1}}

\institute{Leiden Observatory, P.O.Box 9513, 2300 RA Leiden, The Netherlands \and {Institut d'Astrophysique Spatiale, B{\^a}t. 121, Universit{\'e} Paris XI, 91405 Orsay Cedex, France} \and {Astronomical Institute "Anton Pannekoek", University of Amsterdam, Kruislaan 403, 1098 SJ Amsterdam, The Netherlands} \and {Department of Astronomy 105-24, California Institute of Technology, Pasadena, CA 91125, USA} \and {Kapteyn Astronomical Institute, P.O.Box 800, 9700 AV Groningen, The Netherlands}}

\offprints{K. M. Pontoppidan,
\email{pontoppi@strw.leidenuniv.nl}}
\date{Received / Accepted}

\abstract{Medium resolution ($\lambda/\Delta \lambda =
5\,000-10\,000$) VLT-ISAAC $M$-band spectra are presented of 39 young
stellar objects in nearby low-mass star forming clouds showing the
$\rm 4.67~\mu m$ stretching vibration mode of solid CO. By
taking advantage of the unprecedentedly large sample, high S/N ratio and high spectral resolution, 
similarities in the ice profiles from source to source are identified. It
is found that excellent fits to all the spectra can be obtained
using a phenomenological decomposition of the CO stretching
vibration profile at $\rm 4.67~\mu m$ into 3
components, centered on $\rm 2143.7~cm^{-1}$, $\rm
2139.9~cm^{-1}$ and $\rm 2136.5~cm^{-1}$ with fixed widths of 3.0, 3.5 and $\rm 10.6~cm^{-1}$, respectively. All observed
interstellar CO profiles can thus be uniquely described by a model
depending on only 3 linear fit parameters, indicating that a maximum of 3 specific molecular environments
of solid CO exist under astrophysical conditions. A simple physical model of
the CO ice is presented, which shows that the $\rm 2139.9~cm^{-1}$
component is indistinguishable from pure CO ice. It is concluded, that in the majority of the observed lines of sight, 60-90\% of the CO
is in a nearly pure form. In the same model the
$\rm 2143.7~cm^{-1}$ component can possibly be explained by the longitudinal optical (LO)
component of the vibrational transition in pure crystalline CO ice which appears when the
background source is linearly polarised. The model therefore predicts the polarisation fraction
at $\rm 4.67~\mu m$, which can be confirmed by
imaging polarimetry. The $\rm 2152~cm^{-1}$ feature characteristic
of CO on or in an unprocessed water matrix is not detected toward
any source and stringent upper limits are given. When this is taken into account, the $\rm
2136.5~cm^{-1}$ component is not consistent with the
available water-rich laboratory mixtures and we suggest that the
carrier is not yet fully understood. A shallow absorption band
centered between $\rm 2165~cm^{-1}$ and $\rm 2180~cm^{-1}$ is
detected towards 30 sources. For low-mass stars, this band is correlated with
the CO component at $\rm 2136.5~cm^{-1}$, suggesting the presence of a
carrier different from XCN at $\rm 2175~cm^{-1}$.
Furthermore the absorption band from solid $\rm ^{13}CO$ at $\rm
2092~cm^{-1}$ is detected towards IRS 51 in the $\rho$ Ophiuchi
cloud complex and an isotopic ratio of $\rm ^{12}CO/^{13}CO=68\pm10$ is derived. It is shown that all the observed solid $\rm
^{12}CO$ profiles, along with the solid $\rm ^{13}CO$
profile, are consistent with grains with an irregularly shaped CO ice mantle
simulated by a Continuous Distribution of Ellipsoids (CDE),
but inconsistent with the commonly used models of spherical grains in the Rayleigh limit.

\keywords{Astrochemistry -- Circumstellar matter -- dust, extinction -- ISM:molecules -- Infrared:ISM -- Stars: pre-main sequence}
\thanks{Based on observations obtained at the European Southern Observatory, Paranal, Chile, within the observing programs
164.I-0605 and 69.C-0441. ISO is an ESA project with instruments funded by ESA Member States (especially the PI countries: France, Germany, The Netherlands and the UK) and with the participation of ISAS and NASA.}}
\maketitle

\section{Introduction}

Solid interstellar CO was first reported by \cite{Soifer} who
observed a strong absorption band at $\rm 4.61~\mu m$ in a
spectrum from the Kuiper Airborne Observatory (KAO) along the line
of sight towards the massive young stellar object (YSO) \object{W 33A}.
Although the absorption band later turned out to be dominated by a
CN-stretching carrier, this was the first observational indication
that the ice mantles on dust grains in dense clouds carry
a rich chemistry, as indicated by early chemical models
\citep[e.g.][and references therein]{TielensHagen}. Until this
time the early ideas about ``dirty ice'' were mostly supported by
observations of the $\rm 3.08~\mu m$ water band and the extended
red wing of this band, which was suggested to be due to $\rm NH_3$ - $\rm
H_2O$ complexes and $\rm CH_3OH$ \citep[e.g.][]{Gillett,Willner}.
The presence of both solid CO and XCN towards massive young stars
was confirmed by \cite{Lacy}. The identification and subsequent
study of the ices in general, and of the CO stretching mode at
$\rm 4.67~\mu m$ in particular, was aided by laboratory
spectroscopy of ices intended to simulate interstellar ices, such
as those presented in e.g. \cite{HAG,Sandford,Schmitt89} and more
recently by \cite{Pascale, EAS} and \cite{BP}. The typical
spectral resolution of these studies was
$\lambda/\Delta\lambda\sim 1\,000-2\,000$ ($\rm 1.0-2.0~cm^{-1}$).
However, at this resolving power the $\rm ^{12}CO$ stretching mode
as well as the $\rm ^{13}CO$ stretching mode at $\rm 2092~cm^{-1}$
are in general not fully resolved. This problem has
caused difficulties when interpreting both interstellar and
laboratory spectra. 

The laboratory studies have shown that some band profiles, and in
particular that of the CO stretching vibration mode, are sensitive
to the presence of secondary species in the ice as well as to
thermal and energetic processing of the ices, and attempts have been
made to quantify some of these effects. For instance, the $\rm
4.67~\mu m$ CO stretching vibration profile has been used in
attempts to constrain the concentration of species like $\rm
CO_2$, $\rm O_2$, $\rm N_2$ and the degree to which the CO is
mixed with the water component \citep{tielens, Chiar94, Chiar95, Chiar98}. This information is otherwise
hard or impossible to extract from observations \citep{Bart}. It
has also been suggested that the CO profile may be used as an
indicator of the degree to which the ice has been thermally
processed if different components of the profile correspond to CO
ices in environments of different volatility. For
instance, pure CO may be expected to desorb at lower
temperatures than CO mixed with water \citep{SA88,Collings}.
The laboratory simulations have been used for comparisons with a
large number of spectra from both high and low mass young
stars and quiescent cloud lines of sight probed towards
background late-type giant stars, to constrain
the ice mantle composition for a wide range of different cloud
conditions. Often these studies employ the full range of available
laboratory spectra in order to find the best-fitting combination
of ice mixtures to each source \citep{tielens, Kerr93, Chiar94,
Chiar95, Chiar98, Teixeira}. This approach is known to suffer from
serious degeneracies and is further complicated by the fact that
profiles from ice mixtures with a high concentration of CO are
significantly affected by the shape of the grains \citep{tielens}.
Consequently, the detailed environment of solid CO in interstellar grain
mantles and its role in the solid state chemistry is not strongly constrained by observations. Clearly, a
large, consistent sample of lines of sight is necessary to obtain
further information on the "typical" structure of ices in space.

Recent results from medium to high resolution ($R=5\,000-25\,000$) spectrometers mounted on 8 m class telescopes are beginning
to cast new light on the $\rm 3-5~\mu m$ region of protostellar and interstellar spectra. In particular, new insights have been gained through careful studies of the resolved CO ice band structure and its relation to gas phase CO \citep{AdwinL1489},
using the higher sensitivity to probe new environments \citep{ThiCRBR} and using the $\rm ^{13}CO$ band to constrain both composition and
grain shape corrections \citep{Adwin13CO}.

Here we present medium resolution ($R=5\,000-10\,000$, $\Delta{\tilde{\nu}}=0.4-0.2~\rm cm^{-1}$) $M$-band spectra of 39 low mass young stellar objects obtained with the Infrared Spectrometer And Array Camera (ISAAC)
mounted on the Very Large Telescope (VLT) ANTU telescope of the European Southern Observatory at Cerro Paranal. 
These spectra have allowed us to obtain meaningful statistics on the
general shape of the CO ice profile for the first time. In this work, we show that statistically there is enormous advantage in using a large sample of fully resolved spectra from a well-defined set of young stellar objects. By using the highest possible spectral resolution, we are able to cast new light on the underlying ice structure from a detailed analysis of the CO stretching vibration profile.  
The $L$-band spectra and the analysis of the gas phase CO lines seen in the $M$-band spectra will be presented in later articles.

This article is organised as follows. In Sec. \ref{theSample} the
sample of sources is described including the adopted selection
criteria, observational procedures and main features in the
spectra.  Sec. \ref{AnalFits} presents a phenomenological
decomposition of the CO ice band and the observational results of the decompositions are then
discussed. A simple physical model is described in Sec.
\ref{PhysModel}, which can partly explain the success of the
phenomenological decomposition. The results from the observations
and the modeling are discussed in Sec. \ref{IceDis} and the
astrophysical implications are outlined. Finally, Sec. \ref{Concl}
presents the conclusions and suggests further observations and
laboratory experiments which can be used to test the results
presented in this work. Appendix \ref{lineshapes} reviews the
simplest mechanisms governing solid state line shapes and Appendix
\ref{Comments} gives comments on individual sources in the sample.

\section{The sample}
\label{theSample}
\subsection{Source selection}
The primary objective was to perform a broad survey of the dominant ice components on grains surrounding low-mass embedded objects
in the nearest ($D<500~\rm pc$) star forming clouds in the southern sky.
Emphasis was put on obtaining good S/N ratios at the highest possible spectral resolution in order to study not just the
main $\rm H_2O$ and $\rm CO$ features but also the fainter ice species, probe sub-structure in the ice features, and obtain a census of
gas-phase ro-vibrational emission and absorption towards low mass young stars. The observed sample was primarily extracted from the list
of sources to be observed with the Space Infrared Telescope Facility (SIRTF) as part of the Legacy project {\it From Molecular Cores to Planets}
\citep{Evans03},
but was supplemented with additional sources.
The majority of sources have luminosities of $L_{\rm bol}<10~\rm L_{\odot}$, with only a few brighter sources being included for comparison.
Embedded sources are selected from their published Lada classes if no other information is available indicating youth. The sample includes
class I sources as well as flat spectrum sources defined by having a vanishing spectral index between 2.5 and $\rm 15~\mu m$ \citep{Bontemps}.
In addition, the background star CK 2 in Serpens was observed to probe the quiescent cloud material.

A number of the sources have published $M$-band spectra, albeit at lower spectral resolution, coverage and S/N ratio and were re-observed
in order to get a consistent sample.
The list of observed sources and relevant observational parameters is given in Table \ref{SourceList}.

\subsection{Observations and data reduction}
The ISAAC spectrometer is equipped
with a $1024\times1024$ Aladdin array, which is sensitive in the $\rm 1-5~\mu m$ region. The pixel scale is $\rm 0.148\arcsec~pixel^{-1}$ when using
the spectroscopic objective. The telescope was operated using a standard chop-nod scheme with typical chop-throws of 10-20\arcsec.
Both $L$ and $M$-band spectra of each source were obtained. 
The medium resolution grating was used for the $\rm 4.53-4.90~\mu m$ region, resulting in resolving powers of
$\lambda/\Delta \lambda=5\,000-10\,000$ ($\rm 0.4-0.2~cm^{-1}$). Here a
single setting covering $4.53$ to $\rm 4.77~\mu m$ was always observed and a second setting extending to $\rm 4.85-4.90~\mu m$ was obtained if
necessary, e.g. because of the presence of strong gas phase lines in the first setting. Often, it was possible to place a secondary
source in the slit by rotating the camera. In this way spectra of both components of all binaries with separations
of more than $\sim 1\arcsec$ were obtained.

The data were reduced using our own IDL routines. The individual frames were corrected for the non-linearity of the Aladdin detector,
distortion corrected using a startrace map, and bad pixels and cosmic ray hits were removed before co-adding. Both positive and negative spectral
traces were extracted and added before division by the standard star spectrum. An optimal small shift and exponential airmass correction
between the source and the standard were applied by requiring that the pixel-to-pixel noise on the continuum of the final spectrum is minimized.
Before an exponential airmass correction is applied, the detector and filter response curve has to be removed in order to obtain
a spectrum of the true atmospheric absorption. The combined filter and detector response curve was obtained by fitting an envelope to both
the standard star spectrum and a spectrum of the approximate atmospheric transmission and taking the ratio of the two envelopes.

The spectra were flux calibrated relative to the standard and wavelength calibrated relative to telluric absorption lines in the standard star spectrum.
The uncertainty in the flux calibration is estimated to be better than 30\% for most sources and the wavelength calibration is
accurate to $\rm \sim 5~km~s^{-1}$. The high accuracy in wavelength calibration is due to the large number of well-defined and separated telluric lines
present in a medium resolution $M$-band spectrum. For some of the weaker sources, the wavelength calibration becomes considerably more inaccurate
because of difficulties in determining the small shift always present between the standard star spectrum and the source spectrum.

Finally, optical depth spectra were derived by fitting a blackbody continuum to the regions where no features are expected to appear,
namely around $\rm 4.52 - 4.55~\mu m$ and around $\rm 4.76-4.80~\mu m$. In every case
care was taken not to include gas phase lines and regions affected by strong telluric residual in the fit. In cases of doubt
the fit was compared to the continuum in the $\rm 4.0-4.2~\mu m$ region and adjusted accordingly. However, normally the $L$-band
spectra were not used for the continuum fit, due to uncertainties in the relative flux calibrations between the two spectra.

\subsection{Features in the M band spectra}

The most prominent features in the $\rm 4.5-5.0~\mu m$ region of low mass young stellar objects are the
absorptions from the stretching mode of solid CO around $\rm 4.67~\mu m$ and the multitude of lines from the ro-vibrational fundamental
band of gas-phase CO extending across the entire $M$-band. In addition the region covers the CN stretching band around
$\rm 4.61~\mu m$ as well as the bright Pf$\beta$ hydrogen recombination line
at $\rm 4.65~\mu m$ and the $\rm S(9)$ line from molecular hydrogen at $\rm 4.687~\mu m$.
The VLT spectra in general show the presence of all these features, although the shape and strength of each feature varies greatly
from source to source. For example, the ro-vibrational lines of CO are seen as both broad and narrow absorption lines and as broad and narrow emission lines.
The broad absorption lines are often asymmetric indicating the presence of outflows and infall. The broad emission lines
show a higher degree of symmetry and are clearly caused by hot gas as indicated by the presence of CO molecules excited to the second
and even third vibrational levels. One source showing narrow emission lines (GSS 30 IRS 1) is treated in detail in \cite{Pontoppidan}. All the gas-phase CO lines will be analysed in a later paper (Pontoppidan et al., in prep).
At a resolving power of $\lambda/\Delta\lambda=10\,000$ ($\rm 30~km~s^{-1}=0.2~cm^{-1}$)
all solid state features are completely resolved, including the central parts of the main CO feature and the very narrow $\rm ^{13}CO$ feature.
This is generally not the case for previously obtained astronomical and laboratory spectra at resolutions of $\lambda/\Delta\lambda=1\,000-2\,000$ ($\rm 1-2~cm^{-1}$).

\begin{figure*}
\centering
\includegraphics[width=11cm]{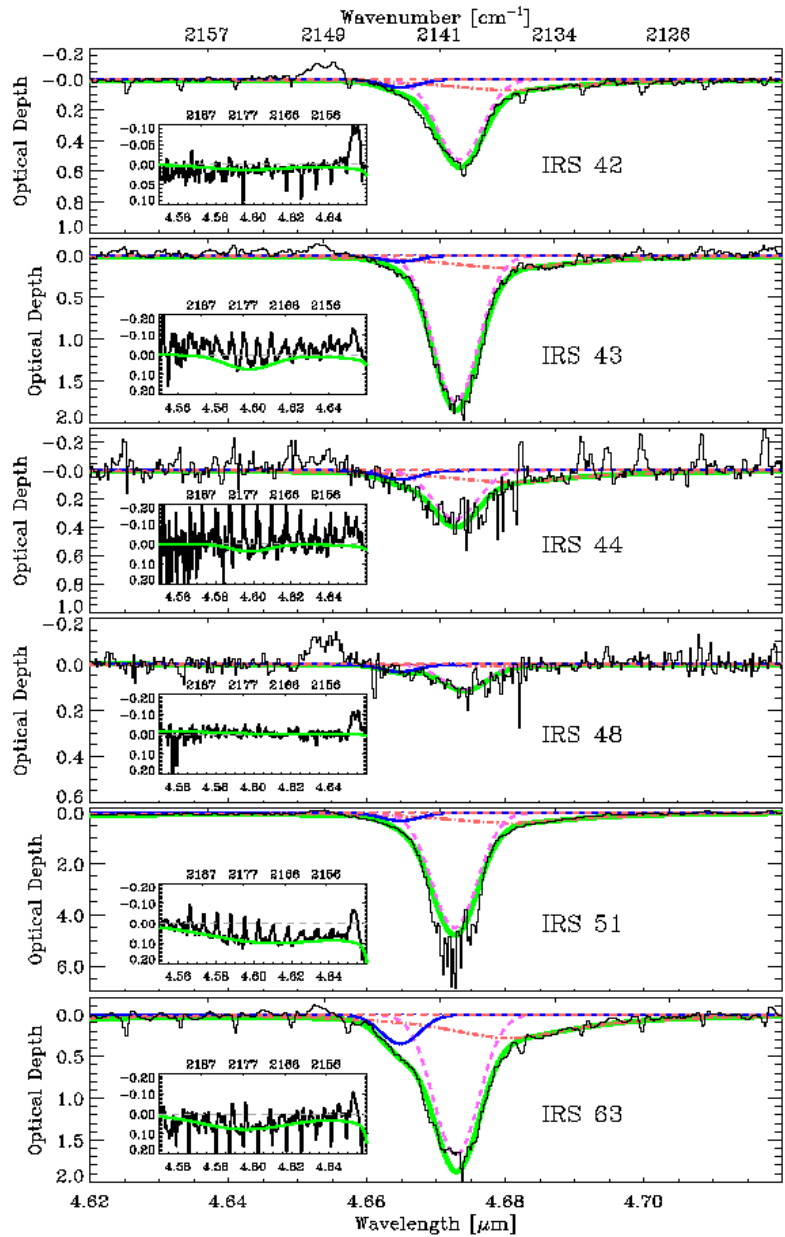}
\caption{VLT-ISAAC $M$-band spectra of sources in the $\rho$ Ophiuchi cloud. The
spectra have been put on an optical depth scale by fitting a blackbody continuum to
parts of the spectrum free of intrinsic features as well as deep telluric lines. In
general, the continuum points are taken in the region 4.52--$\rm 4.55~\mu m$ and
between 4.76--$\rm 4.80~\mu m$. The thick solid curve shows the total phenomenological
model fit to the spectrum and the thin solid, dashed and dot-dashed curves show the individual components (blue, middle
and red, respectively). The inset figures show the fit in the region of the $\rm 2175~cm^{-1}$ feature. }
\label{iceOph}
\end{figure*}

\begin{figure*}
\centering
\includegraphics[width=12cm]{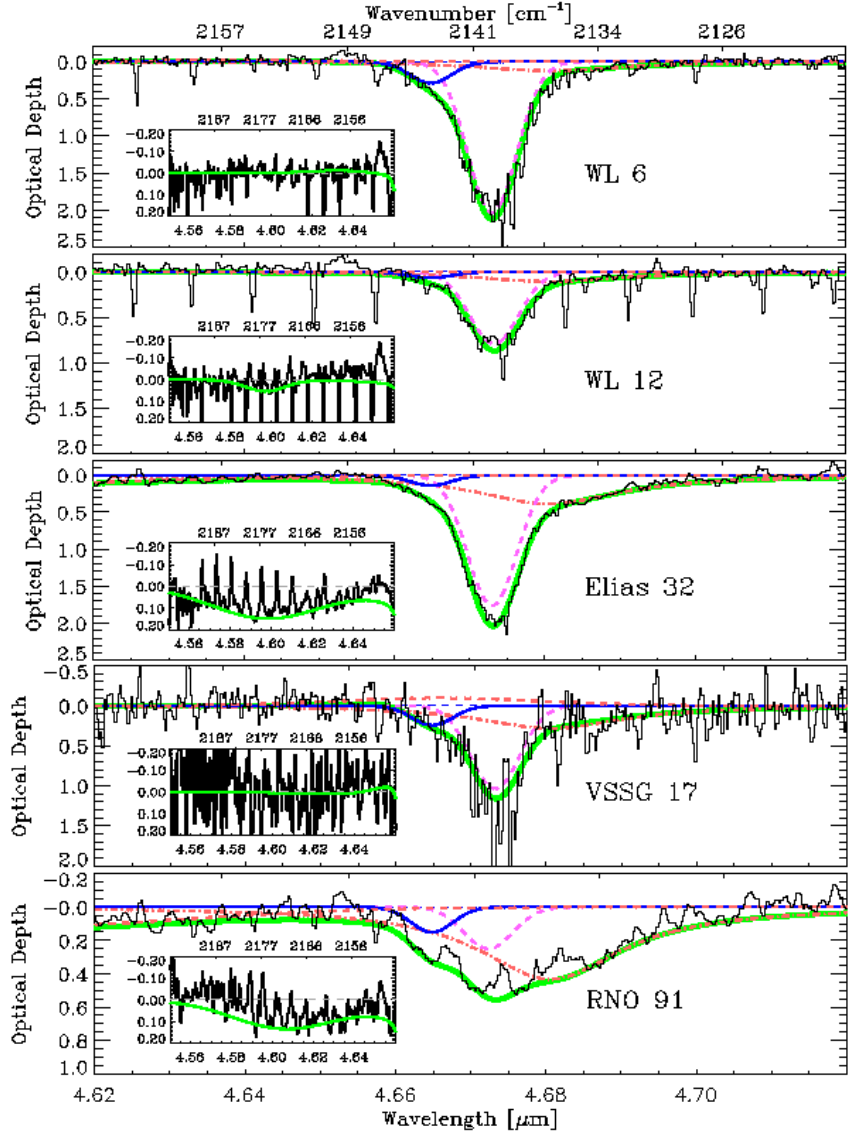}
\caption{$\rho$ Ophiuchi sources, cont. Legend as in Fig. \ref{iceOph}}
\label{iceOph2}
\end{figure*}

\begin{figure*}
\centering
\includegraphics[width=12cm]{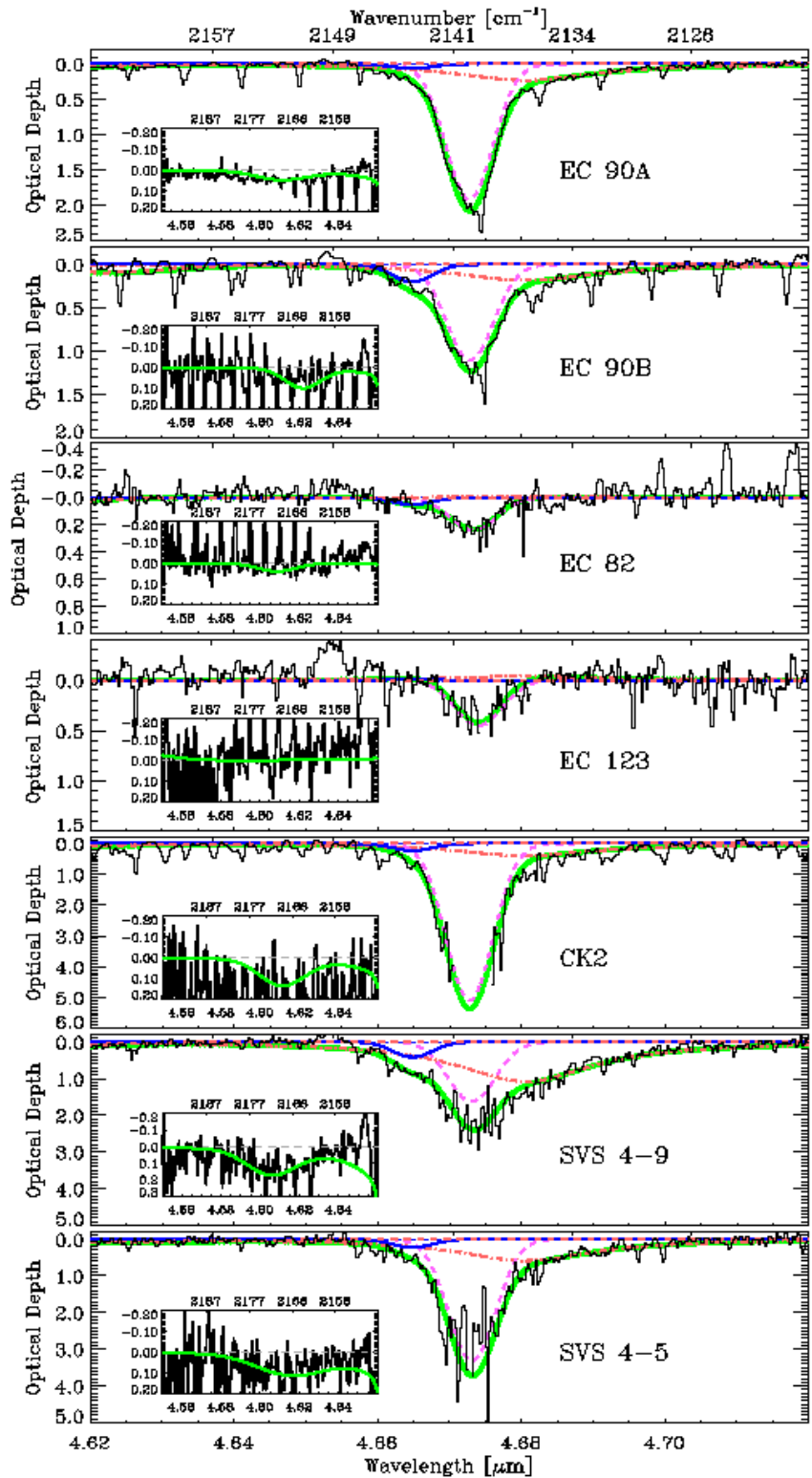}
\caption{Sources in the Serpens cloud. Legend as in Fig. \ref{iceOph}}
\label{iceSerpens}
\end{figure*}

\begin{figure*}
\centering
\includegraphics[width=12cm]{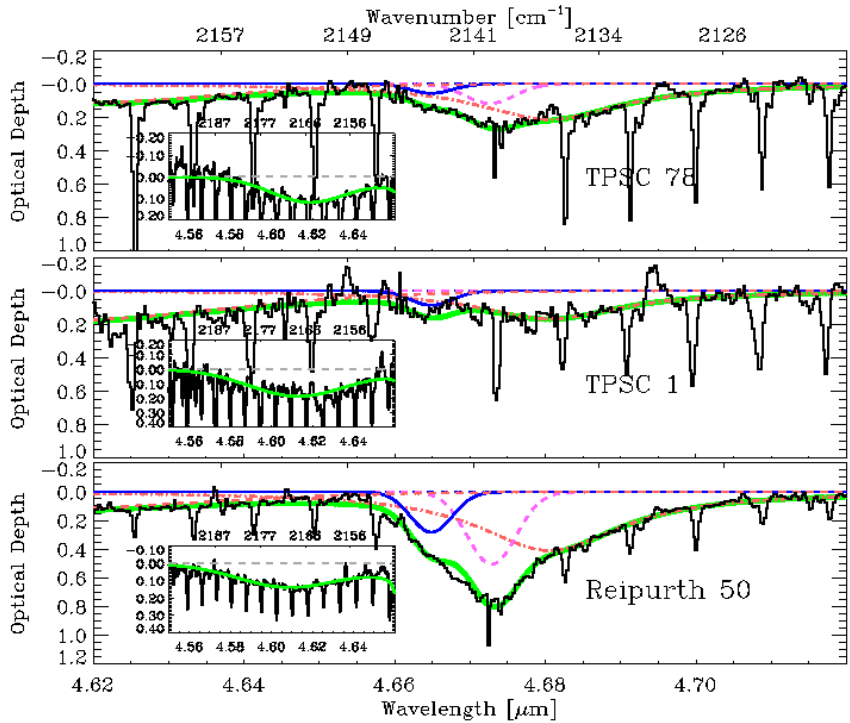}
\caption{Sources in the Orion molecular cloud complex. Legend as in Fig. \ref{iceOph}}
\label{iceOrion}
\end{figure*}

\begin{figure*}
\centering
\includegraphics[width=12cm]{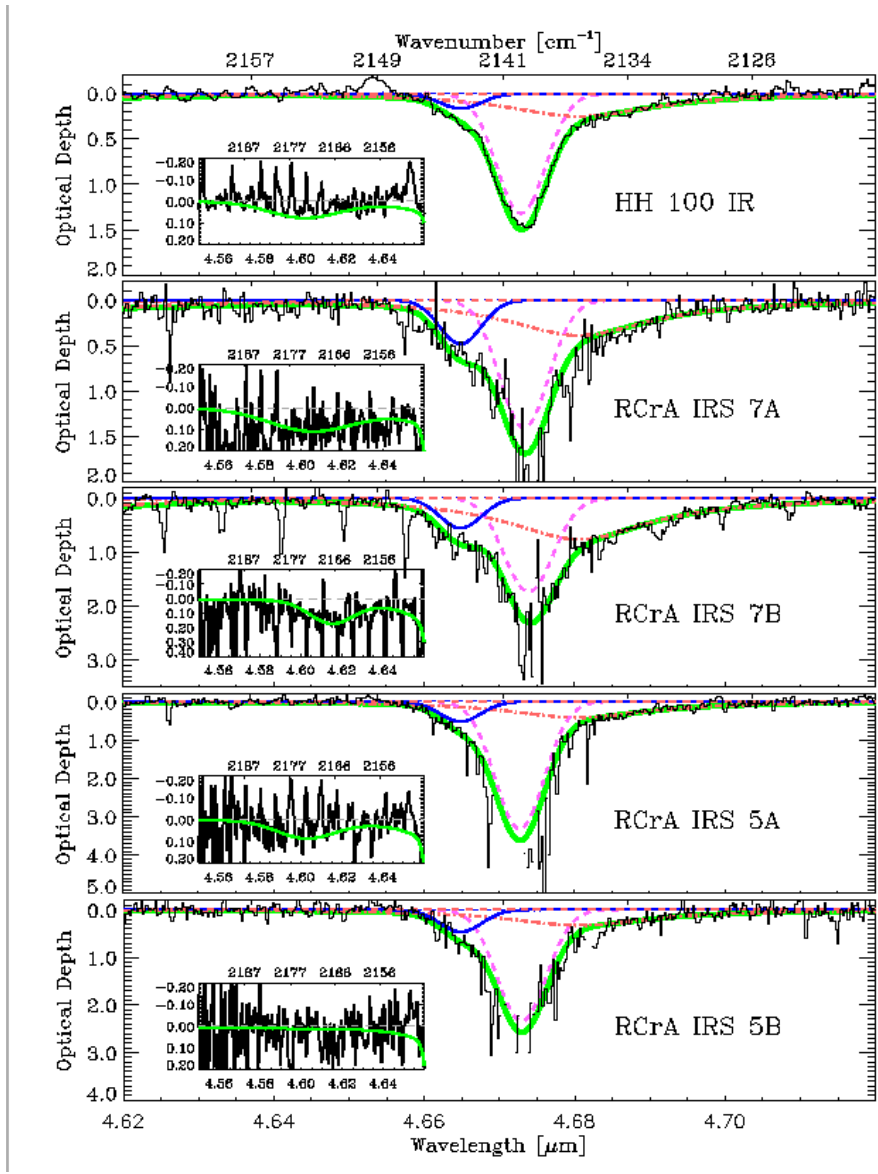}
\caption{Sources in the Corona Australis cloud. Legend as in Fig. \ref{iceOph}}
\label{iceCrA}
\end{figure*}

\begin{figure*}
\centering
\includegraphics[width=12cm]{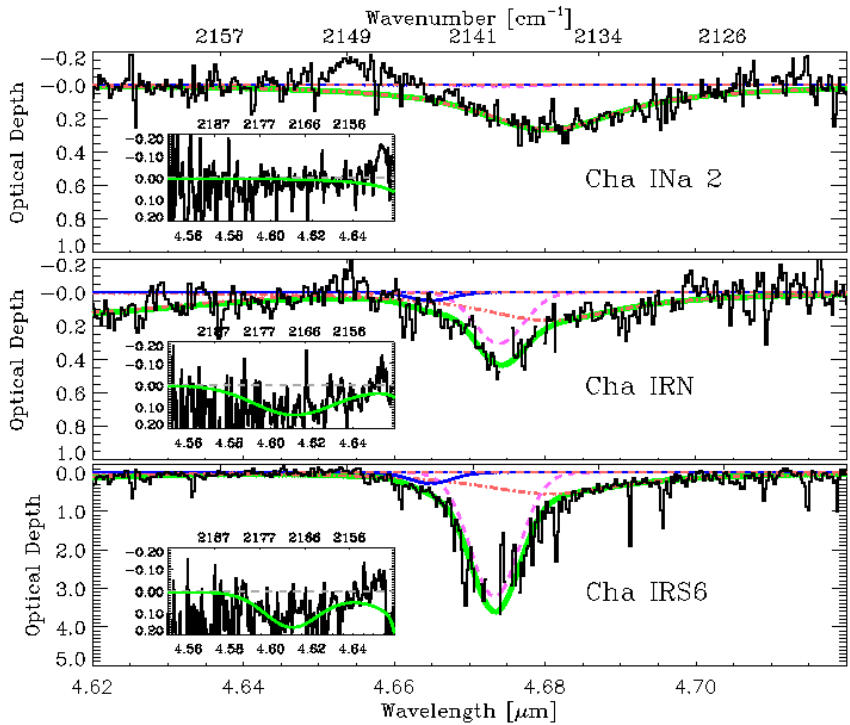}
\caption{Sources in the Chameleon clouds. Legend as in Fig. \ref{iceOph}}
\label{iceCha}
\end{figure*}

\begin{figure*}
\centering
\includegraphics[width=12cm]{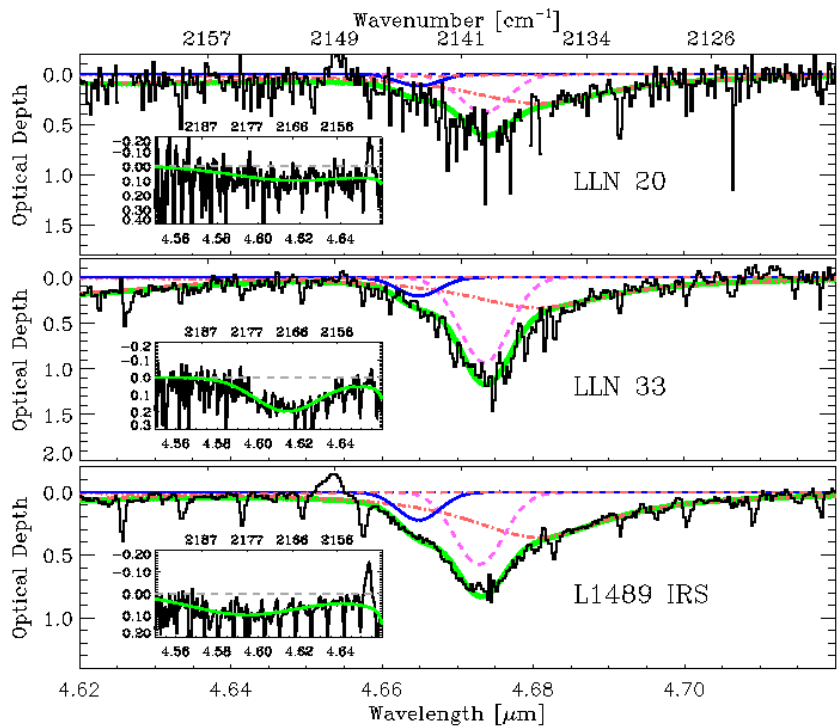}
\caption{Sources in Vela and the Taurus molecular clouds. Legend as in Fig. \ref{iceOph}}
\label{iceVela}
\end{figure*}

The individual $M$-band spectra with detected CO ice are shown in Figs. \ref{iceOph}--\ref{iceVela} on
an optical depth scale. IRS 46, IRS 54, LLN 39 and LLN 47 have no or only marginally detected CO ice bands and are not shown.
The spectrum of GSS 30 IRS 1 is presented in \cite{Pontoppidan} and shows the presence of a very shallow CO ice band. The spectrum of
\object{CRBR 2422.8-3423} is shown in \cite{ThiCRBR} and the spectrum of LLN 17=IRAS 08448-4343 is shown in \cite{ThiThesis}.

\begin{table*}
\centering
\begin{flushleft}
\caption{Parameters of the $M$-band observations}
\begin{tabular}{lllllllll}
\hline
\hline
Source & Lada class & Bolom. & $M$-band  & Date of & Standard star & Int. time & Simbad name\\
&&lum. [$\rm L_{\odot}$] & resolution & observation &  & [min]&\\
\vspace{-0.35cm}\\
\hline
\vspace{-0.35cm}\\
\multicolumn{7}{l}{$\rho$ Ophiuchus}  \\
\hline
IRS 42 &I-II&5.6$\rm^a$&  5\,000&13/8/2001&\object{BS6084}& 20 & \object{BKLT J162721-244142}\\
IRS 43 &I&6.7$\rm^a$&  10\,000&1/5/2002& \object{BS6175} & 25  & \object{BKLT J162726-244051}\\
IRS 44 &I&8.7$\rm^a$&  10\,000 &1/5/2002& \object{BS6084} & 15 & \object{BBRCG 49}\\
IRS 46 &I&0.62$\rm^a$&  10\,000 &1/5/2002& \object{BS6084} & 15 & \object{BBRCG 52}\\
IRS 48 &I&7.4$\rm^a$&  10\,000 &4/5/2002& \object{BS6084} & 20 & \object{BKLT J162737-243035}\\
IRS 51 &I-II&1.1$\rm^a$&  5\,000 &19/8/2001&\object{BS6084}& 20 & \object{BKLT J162739-244316}\\
IRS 54 &I&6.6$\rm^a$&  10\,000&4/5/2002& \object{BS6175} & 10 & \object{BKLT J162751-243145}\\
IRS 63 &I-II&4.2$\rm^b$&  5\,000 &19/8/2001&\object{BS6084}& 20 & \object{GWAYL 4}\\
WL 6 &I&1.7$\rm^a$& 10\,000&4/5/2002& \object{BS6084} & 30 & \object{BBRCG 38} \\
WL 12 &I&2.6$\rm^a$& 10\,000&6/5/2002& \object{BS6084} & 25 & \object{BBRCG 4}\\
GSS30 IRS1 &I&21$\rm^a$& 5\,000&3/9/2001&\object{BS6084}& 20 & \object{BKLT J162621-242306}\\
CRBR 2422.8 &I-II&0.36$\rm^a$& 10\,000&5/5/2002& \object{BS6378} & 35 & \object{CRBR 2422.8-3423} \\
Elias 32 &II&1.1$\rm^a$& 10\,000&1/5/2002& \object{BS6084} & 25 & [\object{B96] Oph B2 5}\\
VSSG 17 &I&3.7$\rm^a$& 10\,000&1/5/2002& \object{BS6084} & 35 & \object{[B96] Oph B2 6}\\
RNO 91 &II&4.7$\rm^b$& 5\,000& 20/8/2001 & \object{BS6147} & 10 & \object{HBC 650}\\
\hline
\vspace{-0.35cm}\\
\multicolumn{7}{l}{Serpens} \\
\hline
EC 90A &I& 35$\rm^d$& 10\,000&1/5/2002& \object{BS6084} & 16 & \object{[B96] Serpens 7}\\
EC 90B &I& 20$\rm^d$& 10\,000&1/5/2002& \object{BS6084} & 16 & \object{[B96] Serpens 7}\\
EC 82 &II&9.7$\rm^d$& 10\,000&1/5/2002& \object{BS6084} & 25 & \object{CK 3}\\
\object{SVS 4-9}  &I&1.5$\rm^d$& 10\,000&6/5/2002& \object{BS7121} & 21 & -- \\
\object{SVS 4-5}  &I&1.2$\rm^d$& 10\,000&6/5/2002& \object{BS7121} & 21 & -- \\
CK 2	      &Background& ? & 10\,000&6/5/2002&\object{BS6378}& 20 & \object{CK 2}\\
EC 123 & II & ~1$\rm^d$& 10\,000&6/5/2002 & \object{BS6378}&20 & \object{[EC92] 123}\\
\hline
\vspace{-0.35cm}\\
\multicolumn{7}{l}{Chameleon} \\
\hline
Cha IRN &I&15$\rm^e$& 10\,000&7/5/2002& \object{BS3485} & 30 & \object{[AWW90]  Ced 111 IRS 5} \\
Cha INa 2 &I&0.6$\rm^f$& 10\,000&6/5/2002& \object{BS4844} & 25 &  \object{[PMK99]  IR Cha INa2}\\
Cha IRS 6A &I&6$\rm^g$& 10\,000&7/5/2002& \object{BS3468} & 30 & \object{[PCW91]  Ced 110 IRS 6A}\\
\hline
\vspace{-0.35cm}\\
\multicolumn{7}{l}{Corona Australis} \\
\hline
HH 100 IRS &I&14$\rm^h$& 10\,000&2/5/2002& \object{BS7121} & 10 & \object{V* V710 CrA}\\
RCrA IRS5A &I&0.6$\rm^e$& 10\,000&2/5/2002& \object{BS7348} & 20 & \object{[B87] 7}\\
RCrA IRS5B &I&0.4$\rm^e$& 10\,000&2/5/2002& \object{BS7348} & 20 & \object{[B87] 7}\\
\object{RCrA IRS7A} &I&3$\rm^i$& 10\,000&2/4/2002&\object{BS6084}& 30 & --\\
\object{RCrA IRS7B} &I&3$\rm^i$& 10\,000&2/4/2002&\object{BS6084}& 30 & -- \\
\hline
\vspace{-0.35cm}\\
\multicolumn{7}{l}{Orion} \\
\hline
Reipurth 50 &I&70$\rm^j$& 10\,000&12/11/2001& \object{BS3468} & 20 & \object{HBC 494} \\
TPSC 78 &I&$\sim 30\rm^{d}$& 10\,000&14/11/2001& \object{BS1790} & 20 & \object{TPSC 78} \\
TPSC 1 &I&$\sim 20\rm^{d}$& 10\,000&14/11/2001& \object{BS1790} & 20 & \object{TPSC 1}\\
\hline
\vspace{-0.35cm}\\
\multicolumn{7}{l}{Taurus} \\
\hline
LDN 1489 IRS &I&3.7$\rm^k$& 5\,000&19/9/2001&\object{BS1543}& 20 & \object{NAME LDN 1489 IR}   \\
\hline
\vspace{-0.35cm}\\
\multicolumn{7}{l}{Vela} \\
\hline
LLN 17 &I&3100$\rm^l$& 800 & 11/1/2001  &\object{BS3185}&10 & \object{IRAS 08448-4343}\\
LLN 20 &I&344$\rm^l$& 10\,000&14/11/2001&\object{BS3468}& 8 & \object{[LLN92] 20}\\
LLN 33 &I&91$\rm^l$& 10\,000&14/11/2001&\object{BS3468}& 12 & \object{[LLN92] 33}\\
LLN 39 &I&807$\rm^l$& 10\,000&14/11/2001&\object{BS1790}& 20 & \object{[LLN92] 39} \\
LLN 47 &I&21$\rm^l$& 10\,000&12/11/2001&\object{BS3468}& 20 & \object{[LLN92] 47}  \\
\hline
\end{tabular}
\label{SourceList}
\end{flushleft}
\begin{list}{}{}
\item[References] $\rm^a$\cite{Bontemps}, $\rm^b$ \cite{Chen95},$\rm^c$ \cite{Chen97}, $\rm^d$ Rough estimate, this work,$\rm^e$ \cite{Chen97},
$\rm^f$ \cite{Persi99}, $\rm^g$ \cite{Persi01}, $\rm^h$ \cite{WLY}, $\rm^i$ \cite{Wilking86}, $\rm^j$ \cite{Manu03},
$\rm^k$ \cite{KCH},$\rm^l$ \cite{LLN}

\end{list}
\end{table*}

\subsection{New detection of solid $\rm ^{13}CO$}
The $\rm 2092~cm^{-1}$ band of solid $\rm ^{13}CO$ has been detected in one source in the sample, namely IRS 51 in the
$\rho$ Ophiuchi cloud. The source is a transitional Class I-II object and shows a very deep ($\tau\sim 5$) $\rm ^{12}CO$
ice band. The $\rm ^{13}CO$ band has an optical depth of 0.12 and a width of $\rm 1.2~cm^{-1}$. The detection is analysed
in Sec. \ref{13COsec}. This is only the second $\rm ^{13}CO$ feature reported in the literature after \object{NGC 7538 IRS 9} \citep{Adwin13CO}
and the first towards a low mass source.

\section{Analytical fits}
\label{AnalFits}

\subsection{A phenomenological approach}
\label{pheno}

One approach to the analysis of interstellar ice bands,
especially in the case of the CO stretching vibration, is via a
mix-and-match scheme, whereby laboratory spectra of differing
chemical composition, relative concentrations, temperature and
degree of processing are compared to astronomical spectra to find
the best fits \citep[e.g.][]{Chiar94,Chiar95,Chiar98,Teixeira}. 
While the presence of solid CO can be unambiguously identified using the laboratory data,  
the mix-and-match approach is seriously affected by the degeneracy introduced by the dependancy of the CO vibration profile on many experimental
parameters. In Sec. \ref{LabStrategy} we will return to this point and discuss the best strategies for comparing laboratory data to interstellar spectra.

Due to the degeneracy encountered when attempting to match CO ice bands in space with a database of interstellar ice simulations, a
more phenomenological approach is attempted here. Results from laboratory work can then be applied subsequently to explain the
general trends identified using a phenomenological method of analysis. Similar approaches to the analysis of ice bands
were used by \cite{Keane01} for the $\rm 6.85~\mu m$ band, \cite{Adwin13CO} for the CO band and \cite{ThiThesis} for the 
$\rm 3.08~\mu m$ water band, although for smaller samples or for individual sources.

The motivation is
to drastically reduce the number of free model parameters and at the same time break the degeneracy by requiring the ice profiles of the entire sample to be described by a single model. The literature suggests that 3 distinct components are fundamental to the CO stretching vibration profile.  It is well known that the CO band can be decomposed into a narrow component centered around $\rm 2140~cm^{-1}$ and a
broader red-shifted component situated somewhere in the region $\rm 2130-2138~cm^{-1}$ \citep{tielens}. Furthermore, \cite{AdwinL1489}
detected the presence of a third narrow blue component at $\rm 2143~cm^{-1}$ in the spectrum of the low mass young star L1489 in Taurus.
We will in the remainder of the text refer to the $\rm 2134~cm^{-1}$ feature as the red component (rc), the $\rm 2140~cm^{-1}$ feature
as the middle component (mc) and the $\rm 2143~cm^{-1}$ feature as the blue component (bc).

The qualitative identification proposed in the literature of the three components is as follows: The red component is traditionally
identified as CO ice in a mixture with hydrogenated molecules such as $\rm H_2O$ or $\rm NH_3$,
while the middle component is thought to be connected with pure CO or CO in a non-hydrogenated mixture containing species
such as $\rm O_2$, $\rm N_2$ or $\rm CO_2$. The physical difference between the two components is that hydrogen-bearing
species form pseudo-chemical bonds between the hydrogen and the CO molecules, while the non-hydrogenated mixture only interacts via weak van der Waals
forces. Traditionally, in the astronomical literature the two components are referred to as ``polar'' and ``apolar''. However, 
while it is true that the traditional mixtures used to reproduce the red component are dominated by polar molecules, the 
chemical mechanism governing the profile of this component are bonds with the hydrogen atoms of the surrounding $\rm H_2O$, $\rm CH_3OH$
and $\rm NH_3$ molecules. Therefore, in the astronomically relevant case, the presence of hydrogenated species is what physically distinguishes the red component from the other two components and not their polarity. Incidentally, the presence
of hydrogen bindings in the red component also determines a number of known characteristics such as lower volatility, ability to trap CO molecules in
micropores etc.
Henceforth in this article we will refer to the two components as hydrogen-bonding and van der Waals-bonding components, 
which more accurately
describes the chemical and physical structures in the ice mantles.

A simple way of treating different components of the CO ice band, is to postulate that the entire band is fundamentally a
superposition of contributions from CO stretching vibrations where
the CO is situated on or in a discrete number of
environments. The term environment is defined in the broadest sense
and refers to an average configuration of nearest molecular
neighbours to a CO molecule for a given ice. The environment
depends on the abundance of other species as well as the structure
of the ice, such as phase, density, porosity etc. Each
environment should give rise to a certain spectral line with a
center frequency and a width determined by the average interaction
between a CO molecule and its nearest neighbours. A detailed discussion
and justification is given in Sec. \ref{IceDis} along with implications for the
interpretation of the spectra presented in this paper. 

\subsection{The analytical model}
To test the simple postulate on the large sample of YSO's
observed here, every CO ice band is assumed to be decomposed into
the three components mentioned in the literature. Since the
presence of discrete components suggests that only a small number
of environments are fundamental to ice on interstellar grains, it
is also required that {\it each of the three components has a
constant central position and width such that only the relative
intensities are allowed to vary from source to source}. In
practice the position (but not the width) of the middle component
is allowed to vary freely because it is the most sensitive to
grain shape effects. The derived center position may then serve to
test the degeneracy of the problem. It is essential to note, that
if the fitted center positions of the middle component for the
entire sample vary by more than is allowed by the statistical
uncertainty of the data, then either the fit is degenerate or our
assumptions are incorrect.

Since there are sources in our sample where only one component in the CO ice dominates, it is straightforward
to determine the best centers and widths for each component. In this way TPSC 78, SVS 4-9, Cha INa 2 and RCrA IRS 7A are used to determine the
position and width of the red component, while L1489, Reipurth 50, IRS 63 and Elias 32 fix the parameters of the blue component.

The fundamental analytical shapes which can be chosen for each component
are Gaussian and Lorentzian profiles.
If three Lorentz curves are used in the fit, it quickly becomes
clear that the blue side of the CO band fits very poorly
in all cases. This is due to the steepness of the blue wing in
the astronomical spectra. Since a Lorentz profile has a very broad
wing, it is not appropriate for the middle and the blue component
and consequently a Gaussian is adopted for these two components.
At the same time, use of a Lorentz curve for the red component
gives significantly better results than using a Gaussian for all
three components. Consequently, two Gaussians and a Lorentzian were
used for the blue, middle and red component, respectively. 
Physical reasons for this choice are given in detail in
Sec. \ref{LorentzSec}.

In addition to the three components used for the CO ice band, a Gaussian is placed at about $\rm 4.61~\mu m$ to account for possible
absorption by $\rm OCN^-$. In this case our choice of a Gaussian is not as significant as for the blue and the middle component. A Gaussian
gives in most cases a better fit than a Lorentzian, but due to the weakness of the feature other types of profile cannot be ruled out.
This last component  does not overlap with the CO band, and has rarely any influence on the three component fit and thus both
center and width of this component are allowed to vary.

Before attempting to fit the decomposition to the data a correction for the systemic
velocity was applied to the model. The motion of the Earth in addition to the LSR velocity of the sources can shift the spectrum
by up to $\rm \pm 40~km~s^{-1}$ corresponding to $\rm \pm 0.29~cm^{-1}$. The fits improve significantly when
a proper velocity correction is applied. Finally, the model spectra are convolved with a Gaussian to match the spectral resolution of the ISAAC spectra. Even
at $\rm R=5\,000$ ($\rm \sim 0.4~cm^{-1}$) this correction must be done to obtain consistent results.

\begin{figure}
\centering
\includegraphics[width=8.5cm]{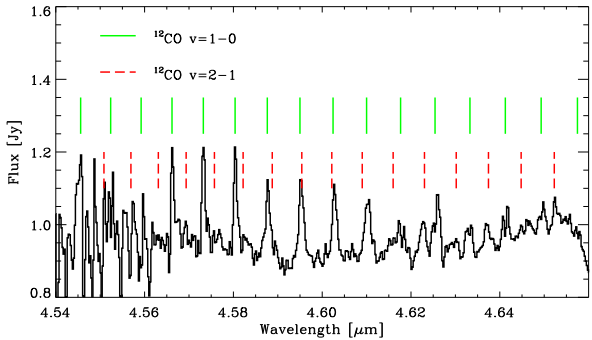}
\caption{Example of the identification of gas phase CO emission lines. Shown is the spectrum of Elias 32. In this
case all the lines are resolved. A weak and broad underlying $\rm 4.61~\mu m$ feature is seen.}
\label{gasExample}
\end{figure}

The common presence of CO gas phase lines requires that care is taken when fitting a Gaussian to the $\rm 4.61~\mu m$ band. Especially
when broad emission lines are present,
a careful selection of unaffected parts of the spectrum for the ice fit must be made.
This is evident in e.g. IRS 43, Elias 32 and HH 100 IR, and indeed in most of the sources to some extent. In general the
CO gas-phase lines were identified and small areas between the lines were picked by eye for the fitting. In Fig. \ref{gasExample}
an example is shown of the line identification. This also prompts a concern in the case of spectra of low signal-to-noise, which
may contain a significant filling-in of the $\rm 4.61~\mu m$ region from gas phase emission lines, which individually may remain undetected.
This may be the case in SVS 4-5, Cha IRN, Cha IRS 6, RCrA IRS 7B and RCrA IRS 5B.

Only the points on the spectrum which are not affected by CO
gas-phase lines are used in the fit of the phenomenological ice
band. In cases where broad emission lines with a complex
structure are present, it is especially difficult to determine
which points to use for the fit. This is in a few cases reflected
in a reduced $\chi^2$ value significantly greater than unity.
However, since most of the broad gas lines are very weak compared
to the main ice features, the final fitting parameters are not
expected to be severely affected. Note that the broad lines are
fully resolved and the selection of points will not improve with
higher spectral resolution.

\subsection{The CO ice band structure}

The common parameters adopted for the decomposition are
shown in Table \ref{COpars}. The best fit parameters are given in
Tables \ref{FitPars} and \ref{FitOCN} and the decompositions are
shown overlaid on the spectra in Figs.
\ref{iceOph}--\ref{iceVela}. The uncertainties on optical depths are given
as statistical 3$\sigma$ errors. Due to the systematic errors introduced by the continuum fit, the presence of gas-phase CO lines
and the phenomenological decomposition we estimate that the statistical uncertainties on the optical depths of the red, middle and blue
components are underestimates of the total error when the features are strong. However, since the fitting parameters for the $\rm 4.61~\mu m$ feature are left free and the
feature is not blended with other components, the statistical 1$\sigma$ error on $\tau({\rm 2175~cm^{-1}})$ is probably a good estimate of the total error. For consistency reasons the errors on this quantity are also given as 3$\sigma$ errors in Table \ref{FitOCN}.

The empirical fits are in general excellent, as is seen from the reduced $\chi^2$ values in Table \ref{FitPars} and by visual inspection.
The distribution of fitted positions of the middle component is shown in Fig. \ref{centerpos}. It is found to vary between $\rm 2139.7~cm^{-1}$ and $\rm 2140.1~cm^{-1}$ which corresponds to a standard deviation of $\rm 0.16~cm^{-1}$.
Since the mean uncertainty of
the positions is $\rm 0.12~cm^{-1}$, the scatter in the fitted center positions is to a large extent caused by statistical uncertainties
in the fitting and not necessarily
by a real effect. An additional scatter may be introduced from uncertainties in the velocity correction of the model. The accuracy
of the velocity correction is illustrated by the differences between the velocity shift of the CO gas-phase lines and the
calculated velocity shift. If no velocity correction is applied, the spread in center
positions of the middle component increases significantly to a standard deviation of $\rm 0.20~cm^{-1}$.
The parameters for the blue and the red component are slightly more uncertain, since they are both shallower and the red one significantly broader than
the middle component.

In view of the large number of lines of sight probed, the large
range of source parameters and the great variety in CO ice
profiles, this result is highly surprising. In simple terms, it is
seen that {\it all CO bands towards both YSOs and background stars can be well fitted with an empirical
model depending on only 3 linear parameters.} Therefore our
original postulate from Sec. \ref{pheno} that the CO stretching
vibration mode can be decomposed into a small number of unique
components holds. This strongly suggests that there is more
physics contained in the decomposition than previously assumed.
The approach of fitting the same simple structure to every
astronomical spectrum has thus provided three
fundamental, identical line components that are common to all the
lines-of-sight observed, and which must be taken into account in
subsequent experimental and theoretical modeling of interstellar ice mantles.

\begin{table}
\centering
\begin{flushleft}
\caption{Best fitting CO ice profile decomposition}
\begin{tabular}{lll}
\hline
\hline
Component & Center & FWHM$^a$\\
& [$\rm cm^{-1}$] & [$\rm cm^{-1}$]\\
\hline
red &$2136.5 \pm 0.3$& 10.6\\
middle &$2139.9 \pm 0.16$ & 3.5 \\
blue &$2143.7 \pm 0.3$& 3.0 \\
\hline
\end{tabular}
\begin{list}{}{}
\item[$^a$] FWHM of Lorentzian profile for the red component and of Gaussian profile for the middle and blue components.
\end{list}
\label{COpars}
\end{flushleft}
\end{table}

\begin{table*}[ht]
\centering
\begin{flushleft}
\caption{Best fits to the CO ice band}
\begin{tabular}{lllllllll}
\hline
\hline
Cloud & Source & center(mc) & $\tau$(rc)$^{a,c}$  & $\tau$(mc)$^{b,c}$ & $\tau$(bc)$^c$  & Vel. corr.$^d$ & CO gas vel.$^e$ & $\chi^2$\\
&&$\rm cm^{-1}$&&&&$\rm km~s^{-1}$&$\rm km~s^{-1}$&\\
\vspace{-0.35cm}\\
\hline
$\rho$ Oph & IRS 42           &$2139.78 \pm 0.02  $   &$0.072 \pm 0.006 $ & $0.53 \pm 0.02 $ &$ 0.05 \pm 0.01   $ & +22   & +20      & $ 0.82 $\\
           & IRS 43           &$2140.01 \pm 0.01  $   &$0.15 \pm 0.02 $   & $1.75 \pm 0.02 $ &$ 0.07 \pm 0.02   $ & -20   & --       & $ 1.76 $\\	
           & IRS 44           &$2140.08 \pm 0.08  $   &$0.08 \pm 0.03 $   & $0.34 \pm 0.05 $ &$ 0.07 \pm 0.03   $ & -20   & -12      & $ 0.85 $\\	
           & IRS 46	      &$2140.24 \pm 0.43  $   &$0.09 \pm 0.03 $   & $0.11 \pm 0.06 $ &$ 0.08 \pm 0.06	$ & -20   & --       & $ 1.81 $\\    
           & IRS 48	      &$2139.47 \pm 0.07  $   &$0.01 \pm 0.01 $   & $0.11 \pm 0.02 $ &$ 0.03 \pm 0.01	$ & -20   & --       & $ 2.36 $\\
           & IRS 51	      &$2140.12 \pm 0.01  $   &$0.42 \pm 0.02 $   & $4.54 \pm 0.15 $ &$ 0.33 \pm 0.03	$ & +23   & --       & $ 2.30 $\\
           & IRS 54	      &--		      &$<0.06	      $   & $<0.12	   $ &$<0.06		$ & -20   & --       & $ 1.89 $\\
           & IRS 63	      &$2140.01 \pm 0.01  $   &$0.28 \pm 0.02 $   & $1.69 \pm 0.03 $ &$ 0.35 \pm 0.02	$ & +23   & +18      & $ 2.84 $\\
           & WL 6	      &$2139.94 \pm 0.03  $   &$0.12 \pm 0.01 $   & $2.04 \pm 0.12 $ &$ 0.29 \pm 0.05	$ & -20   & --       & $ 1.45 $\\
           & WL 12	      &$2139.82 \pm 0.03  $   &$0.11 \pm 0.02 $   & $0.79 \pm 0.04 $ &$ 0.07 \pm 0.04	$ & -20   & -9       & $ 2.10 $\\
           & GSS30 IRS1       &$2139.22 \pm 0.32  $   &$0.06 \pm 0.02 $   & $0.07 \pm 0.06 $ &$<0.04		$ & +23   & +20      & $ 0.57 $\\
           & CRBR 2422.8      &$2139.89 \pm 0.04  $   &$0.47 \pm 0.03 $   & $4.38\pm 0.60^h$ &$ 0.13 \pm 0.07	$ & -17   & -9       & $ 2.35 $\\
           & Elias 32	      &$2139.95 \pm 0.01  $   &$0.39 \pm 0.02 $   & $1.76 \pm 0.05 $ &$ 0.14 \pm 0.03	$ & -24   & --       & $ 1.04 $\\
           & VSSG 17	      &$2139.86 \pm 0.13  $   &$0.28 \pm 0.30 $   & $1.04 \pm 0.36 $ &$ 0.25 \pm 0.32	$ & +22   & +20      & $ 0.30 $\\
           & RNO 91	      &$2140.25 \pm 0.27  $   &$0.43 \pm 0.02 $   & $0.25 \pm 0.02 $ &$ 0.15 \pm 0.04	$ & -19   & --       & $ 2.85 $\\
\hline
Serpens    & EC 90A	      &$2140.07 \pm 0.01  $   &$0.24 \pm 0.01 $   & $2.09 \pm 0.03 $ &$ 0.06 \pm 0.02   $ & -23   & -35      & $ 9.84 $\\
           & EC 90B	      &$2140.00 \pm 0.01  $   &$0.19 \pm 0.02 $   & $1.11 \pm 0.02 $ &$ 0.21 \pm 0.03   $ & -23   & -35/-120 & $ 5.12^*$\\
           & EC 82	      &$2139.82 \pm 0.12  $   &$<0.03         $   & $0.24 \pm 0.05 $ &$ 0.05 \pm 0.05   $ & -22   & -18      & $ 0.72 $\\
           & EC 123	      &$2139.50 \pm 0.50  $   &$<0.12         $   & $0.45 \pm 0.15 $ &$ <0.09    	$ & --    & --       & $ 0.47 $\\
           & CK 2	      &$2140.06 \pm 0.05  $   &$0.40 \pm 0.07 $   & $5.11 \pm 0.60 $ &$ 0.23 \pm 0.12   $ & --    & --       & $ 0.73 $\\
           & SVS 4-9	      &$2139.93 \pm 0.16  $   &$1.07 \pm 0.10 $   & $1.63 \pm 0.30 $ &$ 0.42 \pm 0.14   $ & -22   & --       & $ 0.52 $\\
           & SVS 4-5	      &$2139.89 \pm 0.07  $   &$0.59 \pm 0.09 $   & $3.32 \pm 0.71 $ &$ 0.23 \pm 0.14   $ & -22   & --       & $ 0.60 $\\
\hline
Orion      & TPSC 78          &$2140.13 \pm 0.04  $   &$0.21 \pm 0.01 $   & $0.12 \pm 0.01 $ &$ 0.06 \pm 0.01   $ & +8    & +8       & $ 5.98 $\\
           & TPSC 1           &$2140.60 \pm 0.88  $   &$0.17 \pm 0.03 $   & $0.06 \pm 0.01 $ &$ 0.09 \pm 0.06   $ & +8    & +8       & $ 0.74 $\\
           & Reipurth 50      &$2140.02 \pm 0.01  $   &$0.412 \pm 0.003 $ & $0.51 \pm 0.01 $ &$ 0.28 \pm 0.003  $ & +7    & +10      & $ 12.1 $\\
\hline
CrA        & HH 100 IRS	      &$2140.01 \pm 0.01  $   &$0.25 \pm 0.01 $   & $1.32 \pm 0.01 $ &$ 0.17 \pm 0.01   $ & -23   & --       & $ 82.5^*$\\
           & RCrA IRS7A	      &$2139.82 \pm 0.06  $   &$0.39 \pm 0.04 $   & $1.41 \pm 0.17 $ &$ 0.48 \pm 0.11   $ & -24   & -24      & $ 1.14 $\\
           & RCrA IRS7B	      &$2139.55 \pm 0.07  $   &$0.77 \pm 0.06 $   & $1.74 \pm 0.27 $ &$ 0.56 \pm 0.10   $ & -24   & -24      & $ 0.75 $\\
           & RCrA IRS5A	      &$2140.08 \pm 0.02  $   &$0.42 \pm 0.02 $   & $3.33 \pm 0.18 $ &$ 0.53 \pm 0.04   $ & -22   & --       & $ 3.40 $\\
           & RCrA IRS5B	      &$2139.98 \pm 0.10  $   &$0.33 \pm 0.06 $   & $2.34 \pm 0.57 $ &$ 0.47 \pm 0.17   $ & -22   & --       & $ 0.77 $\\
\hline
Cha        & Cha INa 2	      &--		      &$0.27 \pm 0.02 $   & $<0.03         $ &$<0.06            $ & +3    & 0        & $ 0.43 $\\
           & Cha IRN	      &$2139.54 \pm 0.16  $   &$0.17 \pm 0.05 $   & $0.31 \pm 0.10 $ &$ 0.05 \pm 0.07   $ & +3    & --       & $ 0.42 $\\
           & Cha IRS 6A	      &$2139.82 \pm 0.05  $   &$0.55 \pm 0.06 $   & $3.22 \pm 0.47 $ &$ 0.28 \pm 0.11   $ & +3    & --       & $ 0.36 $\\
\hline
Vela       & LLN 17           &$2140.30 \pm 0.13  $   &$0.32 \pm 0.02 $   & $0.19 \pm 0.03 $ &$ 0.14 \pm 0.04   $ & -15   & --       & $ 1.00 $\\
           & LLN 20	      &$2139.74 \pm 0.13  $   &$0.29 \pm 0.04 $   & $0.38 \pm 0.08 $ &$ 0.12 \pm 0.07   $ & -10   & --       & $ 2.05 $\\
           & LLN 33	      &$2139.70 \pm 0.06  $   &$0.33 \pm 0.03 $   & $0.94 \pm 0.09 $ &$ 0.21 \pm 0.05   $ & -10   & --       & $ 1.04 $\\
           & LLN 39	      &$2139.97 \pm 0.10  $   &$0.03 \pm 0.01 $   & $0.02 \pm 0.01 $ &$ 0.004 \pm 0.003 $ & -10   & --       & $ 7.72 $\\
           & LLN 47	      &$2138.78 \pm 0.43  $   &$<0.01         $   & $<0.02         $ &$<0.06            $ & -10   & -10      & $ 3.24 $\\
\hline
Taurus     & LDN 1489 IRS     &$2140.04 \pm 0.04  $   &$0.36 \pm 0.02 $   & $0.58 \pm 0.03 $ &$ 0.23 \pm 0.04   $ & -11   & -10      & $ 0.99 $\\
\hline
Add.       & GL 2136$^f$      &$2139.50 \pm 0.06  $   &$0.17 \pm 0.01 $ & $0.15 \pm 0.02 $ &$ 0.02 \pm 0.02 $ & --    & --       & $ 0.96 $\\
sources    & NGC 7538 IRS1$^f$&$2140.04 \pm 0.03  $   &$0.10 \pm 0.01 $ & $0.19 \pm 0.01 $ &$ 0.10 \pm 0.01 $ & --    & --       & $ 2.09 $\\
           & RAFGL 7009S$^f$  &$2140.19 \pm 0.04  $   &$0.89 \pm 0.02 $ & $0.66 \pm 0.06 $ &$ 0.19 \pm 0.03 $ & --    & --       & $ 3.82 $\\
           & W33A$^f$	      &$2140.90 \pm 0.08  $   &$0.75 \pm 0.05 $ & $0.59 \pm 0.10 $ &$ <0.05         $ & --    & --       & $ 1.01 $\\
           & RAFGL 989$^f$    &$2139.90 \pm 0.30  $   &$0.18 \pm 0.03 $ & $0.44 \pm 0.10 $ &$ 0.14 \pm 0.03$  & --    & --       & $ 0.98 $\\
\hline
\end{tabular}
\begin{list}{}{}

\item[$^a$] Solid CO column densities of the red component can be calculated using eq. \ref{Nred}. 
\item[$^b$] Solid CO column densities of the middle component can be calculated using eq. \ref{Nmiddle}. 
\item[$^c$] Uncertainties are $3\sigma$.
\item[$^d$] Calculated absolute velocity shift of the spectra, including Earth motion and LSR velocity of the parent molecular cloud.
\item[$^e$] Measured absolute velocity shift of CO ro-vibrational gas phase lines, where available.
\item[$^f$] ISO-SWS archive data.
\item[$^h$] The optical depth given for the (saturated) CO band of CRBR 2422 is different from that of \cite{ThiCRBR} since a 
laboratory profile was used rather than the emperical profile presented here.
\item[$^*$] The goodness-of-fit estimate for this source is affected by broad CO
 gas phase emission lines.
\end{list}
\label{FitPars}
\end{flushleft}
\end{table*}

\begin{table}
\centering
\begin{flushleft}
\caption{Best fits to the $\rm 2175~cm^{-1}$ band}
\begin{tabular}{llll}
\hline
\hline
Source & center & FWHM  & $\rm \tau(2175~cm^{-1})^a$  \\
\vspace{-0.35cm}\\
\hline
\vspace{-0.35cm}\\
\multicolumn{4}{l}{$\rho$ Ophiuchus}  \\
\hline
IRS 42         &$2175.7 \pm 2.5 $ &$ 23 \pm 6  $ &$0.012 \pm  0.006 $ \\
IRS 43         &$2175.0 \pm 1.2 $ &$ 14 \pm 3  $ &$0.07  \pm  0.03 $ \\
IRS 44         &$2174.8 \pm 2.5 $ &$ 11 \pm 6  $ &$0.04  \pm  0.05 $ \\
IRS 46         &$2165.0 \pm 3.2 $ &$ 24 \pm 8  $ &$0.06  \pm  0.03 $ \\
IRS 48         &$2194.3 \pm 8.6 $ &$ 29 \pm 11 $ &$0.013 \pm  0.015 $ \\
IRS 51         &$2170.2 \pm 1.2 $ &$ 36 \pm 4  $ &$0.089 \pm  0.006 $ \\
IRS 54	       &--                &--		 &$<0.1 	    $ \\
IRS 63	       &$2176.2 \pm 0.8 $ &$ 26 \pm 2  $ &$0.073 \pm  0.009 $ \\
WL 6	       &--                &--		 &$<0.06	    $ \\
WL 12	       &$2175.2 \pm 1.3 $ &$ 11 \pm 3  $ &$0.05  \pm  0.04 $ \\
GSS30 IRS1     &--                &--		 &$<0.075	    $ \\
CRBR 2422.8    &$2175.2 \pm 1.1 $ &$ 9  \pm 2  $ &$0.09 \pm  0.05 $ \\
Elias 32       &$2175.1 \pm 0.8 $ &$ 29 \pm 2  $ &$0.14 \pm  0.03 $ \\
VSSG 17	       &--                &--		 &$<0.2 	    $ \\
RNO 91	       &$2171.2 \pm 0.4 $ &$ 28 \pm 1  $ &$0.125 \pm  0.009 $ \\
\hline
\vspace{-0.35cm}\\
\multicolumn{4}{l}{Serpens} \\
\hline
EC 90A	       &$2170.5 \pm 0.7 $ &$ 17 \pm 2  $ &$0.073 \pm  0.009 $ \\
EC 90B	       &$2163.6 \pm 0.4 $ &$ 10 \pm 1  $ &$0.10 \pm  0.02 $ \\
EC 82	       &--                &--		 &$<0.06	    $ \\
SVS 4-9	       &$2171.3 \pm 1.9 $ &$ 14 \pm 4  $ &$0.15 \pm  0.10 $ \\
SVS 4-5	       &$2166.6 \pm 3.4 $ &$ 19 \pm 8  $ &$0.10 \pm  0.10 $ \\
CK 2           &--                &--		 &$<0.15	    $ \\
EC 123         &--     	          &--		 &$<0.3 	    $ \\
\hline
\vspace{-0.35cm}\\
\multicolumn{4}{l}{Chameleon} \\
\hline
Cha IRN	       &$2168.6 \pm 3.0 $ &$ 22 \pm 6  $ &$0.13 \pm  0.07 $ \\
Cha INa 2      &--  	          &--		 &$<0.09	    $ \\
Cha IRS 6A     &$2169.1 \pm 2.1 $ &$ 14 \pm 4  $ &$0.15 \pm  0.14 $ \\
\hline
\vspace{-0.35cm}\\
\multicolumn{4}{l}{Corona Australis} \\
\hline
HH100 IRS      &$2173.2 \pm 0.2 $ &$ 19 \pm 1  $ &$0.072 \pm  0.006 $ \\
RCrA IRS5A     &$2172.7 \pm 0.9 $ &$ 16 \pm 2  $ &$0.08 \pm  0.02 $ \\
RCrA IRS5B     &--                &--		 &$<0.09	    $ \\
RCrA IRS7A     &$2171.5 \pm 2.2 $ &$ 23 \pm 5  $ &$0.10 \pm  0.06 $ \\
RCrA IRS7B     &$2167.0 \pm 1.2 $ &$ 12 \pm 2  $ &$0.15 \pm  0.08 $ \\
\hline
\vspace{-0.35cm}\\
\multicolumn{4}{l}{Orion} \\
\hline
Reipurth 50    &$2170.2 \pm 0.1 $ &$ 27 \pm 1  $ &$0.129 \pm  0.003 $ \\
TPSC 78	       &$2165.3 \pm 0.2 $ &$ 20 \pm 1  $ &$0.111 \pm  0.006 $ \\
TPSC 1	       &$2167.9 \pm 0.7 $ &$ 26 \pm 2  $ &$0.18 \pm  0.03 $ \\
\hline
\vspace{-0.35cm}\\
\multicolumn{4}{l}{Taurus} \\
\hline
LDN 1489 IRS  &$2176.9 \pm 1.2 $ &$ 29 \pm 3  $ &$0.09 \pm  0.02 $ \\
\hline
\vspace{-0.35cm}\\
\multicolumn{4}{l}{Vela} \\
\hline
LLN 17         &$2169.2 \pm 0.3 $ &$ 27 \pm 1  $ &$0.13 \pm  0.01   $ \\
LLN 20	       &$2166.2 \pm 2.6 $ &$ 31 \pm 9  $ &$0.09 \pm  0.04   $ \\
LLN 33	       &$2167.4 \pm 0.7 $ &$ 17 \pm 2  $ &$0.19 \pm  0.04   $ \\
LLN 39	       &$2173.4 \pm 1.3 $ &$ 24 \pm 3  $ &$0.010 \pm  0.003 $ \\
LLN 47	       &--                &--            &$<0.03            $ \\
\hline
\vspace{-0.35cm}\\
\multicolumn{4}{l}{Additional sources} \\
\hline
GL 2136$^b$       &$2164.4 \pm 0.2 $ &$ 24 \pm 2  $ &$0.13 \pm  0.01 $ \\
NGC 7538 IRS1$^b$ &--	             &--	    &$<0.03	     $ \\
RAFGL 7009S$^b$   &$2168.2 \pm 0.2 $ &$ 25 \pm 1  $ &$0.47 \pm  0.02 $ \\
W33A$^b$          &$2165.7 \pm 0.1 $ &$ 26 \pm 1  $ &$1.19 \pm  0.03 $ \\
RAFGL 989$^b$     &$2169.7 \pm 1.2 $ &$ 23 \pm 3  $ &$0.04 \pm  0.01 $ \\
\hline
\end{tabular}
\label{FitOCN}

\begin{itemize}
\item[$^a$] 3$\sigma$ errors are given.
\item[$^b$] ISO-SWS archive data.
\end{itemize}

\end{flushleft}
\end{table}

\subsection{The $\rm 4.61~\mu m$ band}

A weak absorption band between $\rm 2200~cm^{-1}$ ($\rm 4.55~\mu m$) and $\rm 2150~cm^{-1}$ ($\rm 4.650~\mu m$)
is detected towards 30 of the 39 sources. Absorption in this spectral region can be attributed to the stretching mode of
CN bonds. The most likely species to contain the CN bond around young stars is the $\rm OCN^-$ ion
\citep{SG,Pendleton, Novozamsky}. The derived band parameters along with upper limits are given in Table \ref{FitOCN}.
The widths and center positions of the band seem to vary significantly. In particular, many of the bands are centered closer to
$\rm 2175~cm^{-1}$ than to the $\rm \sim 2165~cm^{-1}$ found for the bands discovered previously, mostly towards high-mass sources. Also laboratory spectra give
center positions of the $\rm OCN^-$ band and for the CN-stretch, which in general are located red-ward of many of the observed center
positions. For example \cite{Hudson} and \cite{Fleur} find that the centers of the CN-stretch in interstellar ice analogs vary between 2158 and $\rm 2170~cm^{-1}$.
Most of the fitted centers in our sample are placed between 2170 and $\rm 2180~cm^{-1}$. There is a clear tendency that
the higher mass stars have redder $\rm 4.61~\mu m$ band centers than lower mass stars. The average position for the intermediate
mass stars Reipurth 50, LLN 20, LLN 33 together with the low mass stars TPSC 1 and TPSC 78, which are located only $30\arcsec$ from the
trapezium OB association in Orion, is $\rm 2167\pm 2~cm^{-1}$. The rest of the detected bands have an average position of
$\rm 2173\pm 4~cm^{-1}$.
For simplicity the band will hereafter be referred to as the $\rm 2175~cm^{-1}$ band.

\subsection{Correlations}
\label{correlations}

\begin{figure}
\centering
\includegraphics[width=8.5cm]{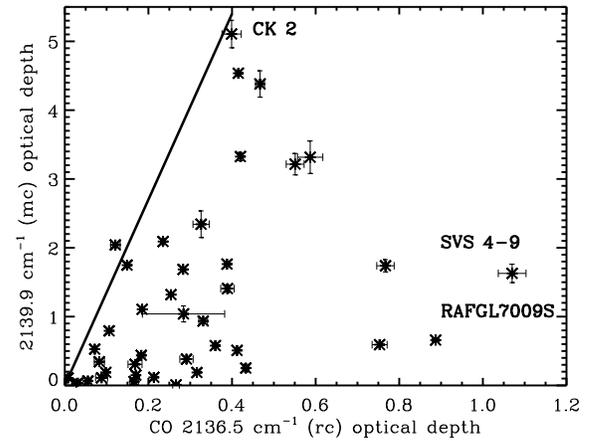}
\caption{Red component optical depth against middle component optical depth. The solid line corresponds to
16\% of the total amount of CO present in a hydrogen-bonding environment under the assumption that the red component
is carried by a water-rich mixture. The percentage can be calculated using Eqs. \ref{Nmiddle} and \ref{Nred}. }
\label{RedMiddle}
\end{figure}

\begin{figure}
\centering
\includegraphics[width=8.5cm]{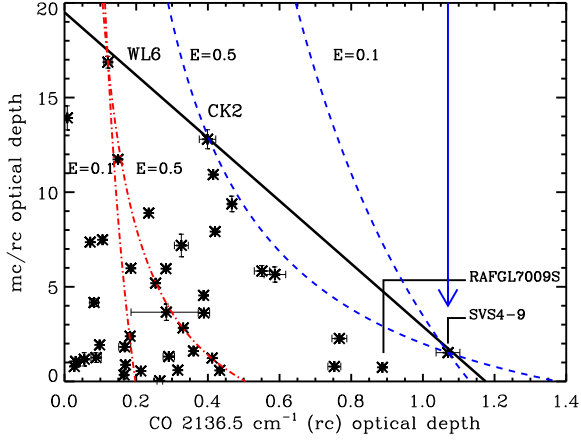}
\caption{Red component optical depth against the ratio of the middle component optical depth and the red component
optical depth. The solid arrow indicates the possible evolutionary track of a source if the pure CO does not migrate into a porous water
ice during warm-up. The solid line indicates the excluded region. The dashed curves indicate evolutionary tracks for SVS 4-9 assuming 
that migration does occur with efficiencies of 0.1 and 0.5. The dot-dashed curves are evolutionary tracks for WL 6 with migration (see Sec. \ref{Interpretations} ).}
\label{RedMiddleRat}
\end{figure}

\begin{figure}
\centering
\includegraphics[width=8.5cm]{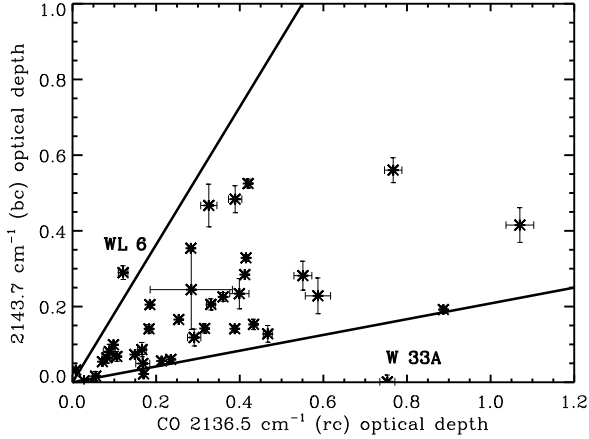}
\caption{Red component optical depth against blue component optical depth. The solid lines indicate excluded regions.}
\label{RedBlue}
\end{figure}

\begin{figure}
\centering
\includegraphics[width=8.5cm]{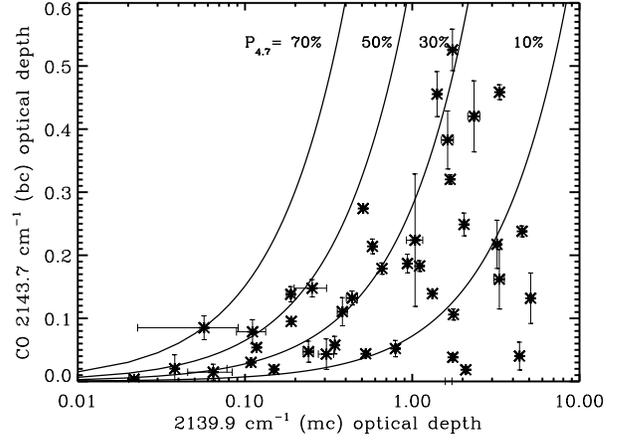}
\caption{Middle component optical depth against blue component optical depth. The solid lines are curves of constant polarisation fraction calculated using Eq. \ref{Poleq}. The blue component optical depths have been slightly corrected by using a Lorentz oscillator model for the middle component rather than a Gaussian. See Secs. \ref{LorentzSec} and \ref{LOTO}.}
\label{BlueMiddle}
\end{figure}

\begin{figure}
\centering
\includegraphics[width=8.5cm]{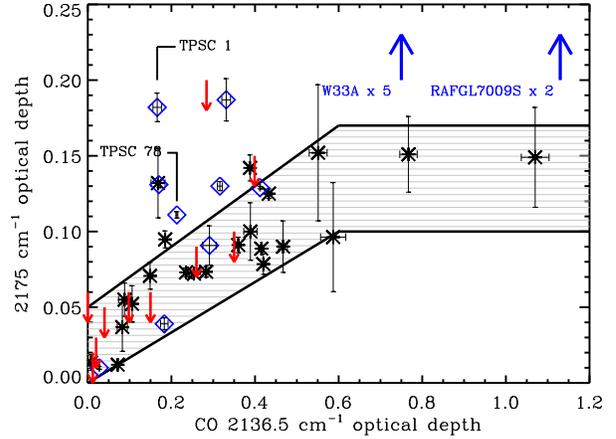}
\caption{Optical depth at $\rm 2175~cm^{-1}$ against optical depth of the red component. Stars indicate low-mass young stars defined
by having $L_{bol}<50~L_{\odot}$. Diamonds indicate intermediate to high-mass stars with $L_{bol}>50~L_{\odot}$. Small arrows indicate
upper limits on the $\rm 2175~cm^{-1}$ optical depth. The points of the high-mass stars W 33A and RAFGL 7009 S lie outside the plotting range and are indicated by large arrows. Although TPSC 1 and TPSC 78 are low-mass stars, they are marked as high-mass stars due to their location in the high-mass star forming Orion core. The marked region indicates the possible range of the correlation for low mass stars. }
\label{COOCN}
\end{figure}

Correlation plots for the derived optical depths of the components of the decomposition are shown in
Figs. \ref{RedMiddle}--\ref{COOCN}. Possible interpretations of the correlations, non-correlations and exclusion regions are given in Sec. \ref{Interpretations}. All correlations except Fig. \ref{COOCN} only show points with firm detections of the relevant CO components. 
A decomposition of the CO ice band taken from ISO-SWS ($\lambda/\Delta \lambda\sim 2000$) archival spectra of the high-mass stars \object{W 33A}, \object{GL 989}, \object{GL 2136}, \object{NGC7538 IRS 1} and \object{RAFGL 7009S} have been added to the sample for comparison. The parameters derived for these sources are included in Table \ref{FitPars} and \ref{FitOCN}. The error bars in the correlation plots are $\pm 1\sigma$.  

No correlation is found between the red and the middle component (see Fig. \ref{RedMiddle}). Sources exist where virtually only the
red component is visible, such as TPSC 78, TPSC 1 and
Cha INa 2, while other sources are dominated by the middle component. However, no source with a deep middle component
has been found where the red component is altogether absent. 
This effect is seen in the figure as an exclusion region in the upper left corner where no points are located. The region is indicated by a solid line.  

In Fig. \ref{RedMiddleRat} the red component optical depth is plotted against the ratio of the middle and red component optical depths.
Evidently no direct correlation is seen. However, the region in the upper right corner of the figure is excluded, which means that
a deep red component results in a low strength of the middle component relative to the red component. 
This plot has been included to investigate if the ratio between the two components can be used as a processing indicator.

The plot of the red and the blue component shown in Fig. \ref{RedBlue} also shows that certain regions in parameter space are 
excluded as indicated by solid lines.
However, it should be noted that the blue component is the most susceptible
to systematic effects created by the artificial decomposition due
to the dominating presence of the nearby middle component. Thus caution should be exercised when interpreting this
correlation plot in areas where the blue component is weak. In essence, a strong blue component is excluded if the red component is weak as seen by the absence of sources in the upper left corner of the figure. The absence of sources in the lower right corner indicates that a deep red component excludes a weak blue component

Fig. \ref{BlueMiddle} shows the depth of the middle component plotted against the blue component. 
The curves indicate constant polarisation fractions of the background source as required if the blue component is assumed to be due to
the LO-TO splitting of crystalline CO as discussed in Sec. \ref{LOTO}.

Finally, a trend between the red component and the $\rm 2175~cm^{-1}$ component is evident in Fig. \ref{COOCN}.
In this correlation plot the sources have
been divided into low-mass and high-mass stars to search for evidence for enhanced $\rm OCN^-$ abundances around high-mass stars. High-mass
stars have been defined as stars more luminous than $\rm 50~L_{\odot}$ (bolometric). This definition is flawed by the fact that
the luminosity of a young star is dependent on age as well as inclination, but due to the inherent difficulties in
determining spectral classes of embedded
stars no better definition was found. Roughly, the definition will divide bona fide embedded sources into young stars lighter and heavier than
$\rm \sim 2~M_{\odot}$, respectively. The low-mass stars
exhibit a nearly linear relation between the red component optical depth and the $\rm 2175~cm^{-1}$ feature for $\tau(\rm rc)$ up to 0.6
after which the relation apparently saturates and flattens out. However, the flattening in the relation only depends on a few points and
should thus be interpreted with caution.
Some of the high-mass stars follow the same relation, but many show a dramatic excess absorption around $\rm 2175~cm^{-1}$. When including the
massive young stars W 33A and RAFGL 7009S a rough correlation can also be seen for the high-mass stars, although much steeper than for the
low-mass stars.
No sources are found significantly under the best-fitting line and no $\rm 2175~cm^{-1}$ absorption bands are detected for sources without
CO ice bands. Within our detection limits, there seems to be an almost one-to-one relationship between the presence of a $\rm 2175~cm^{-1}$ feature
and the red component of the CO ice. 

\subsection{Possible interpretation of correlations}
\label{Interpretations}

In Sec. \ref{correlations} the optical depths of the four separate solid state components identified in the spectra have been 
plotted against each other. The regions of the parameter space covered by the observed sources and especially the regions
where no sources fall may provide clues to the nature and evolution of the components.  

If the middle component is ascribed to pure CO and the red component to CO in a water-rich environment, one relevant question is how the volatile, pure CO interacts with the water-rich CO when the ice is thermally processed. 
If processing causes the evaporation of some of the carrier of the
middle component while the rest migrates into a porous water ice
believed to be located below, such as has been suggested by
\cite{Collings}, an increase in red component optical depth would
be expected when the ratio of middle to red component decreases.
Due to different absolute column densities along different lines of sight, this effect may be hard to detect without
comparing to an independent ice temperature indicator, such as the $\rm 3.08~\mu m$ water band. However,
the exclusion regions seen in Figs. \ref{RedMiddle} and \ref{RedMiddleRat} still provide useful
constraints. The fact that the sources with the deepest red components seem to have the smallest middle components
compared to the red as shown with a solid line in Fig. \ref{RedMiddleRat} indicates that this redistribution of CO may take place.
If this holds then Fig. \ref{RedMiddle} would indicate that some redistribution of CO has taken place for all sources, since 
the red component is always present with a minimum depth relative to the middle component. This is indicated in Fig. \ref{RedMiddle} with a solid line,
corresponding to 16\% of the total amount of CO embedded in water. The interpretation is then that the least processed CO
profiles are those closest to the solid line, and for some reason the minimum fraction of CO in water along any line of sight is 16\%.

Sketches of possible evolutionary tracks with or without migration of the CO ice into the water ice during warm-up have been drawn in Fig. \ref{RedMiddleRat}. The solid track shows the middle component evaporating independent of the red component at temperatures below 90 K for SVS 4-9. This scenario of no CO migration raises the question of why no "progenitors" have been found of the sources in the lower right corner of Fig \ref{RedMiddleRat}, such as SVS 4-9, RCrA IRS 7B and the massive YSOs W 33A and RAFGL 7009S. According to
the simple evolutionary track without migration, these progenitors should be found in the upper right corner of the plot.
If migration does occur, i.e. if the CO of the middle component is deposited on a layer of porous water ice, the first part of the evolutionary 
track changes due to a growth of the red component simultaneously with the evaporation of the middle component. For a constant efficiency, $E$, of migration, such that ${\rm d}N_{\rm rc}=-E {\rm d}N_{\rm mc}$, the evolutionary track is:

\begin{equation}
\frac{N_{\rm mc}}{N_{\rm rc}} = \frac{1}{E}\frac{(N_{\rm rc,0}-N_{\rm rc})+N_{\rm mc,0}}{N_{\rm rc}},
\end{equation}

where $N_{\rm mc}$ and $N_{\rm rc}$ are the column densities of the middle and red components, respectively, while  $N_{\rm mc,0}$ and $N_{\rm rc,0}$ are the observed column densities. In Fig. \ref{RedMiddleRat}, evolutionary tracks for migration efficiencies of 0.1 and 0.5 have been
drawn through the extreme sources WL 6 and SVS 4-9. It is seen that the CO profile of SVS 4-9 can be explained by evolution from a quiescent line of sight, such as that probed by the background source CK 2, if the migration efficiency is large. This is consistent with the lack of sources above the solid line in the diagram. The tracks through WL 6 show how this source can evolve into sources in the lower left part of the diagram. The distribution of sources in Fig. \ref{RedMiddleRat} can thus be taken as evidence that migration of CO molecules into a porous water ice does occur upon warm-up
of the circumstellar grains (but see Sec. \ref{red}).   

Amorphous water ice only becomes porous when it is deposited at low temperatures (10--20 K).  The migration efficiency drops with increasing deposition temperature and if the ice is deposited at temperatures higher than 70 K, the pores do not form at all \citep{Collings}. 
If the distribution of sources in Fig. \ref{RedMiddleRat} is interpreted as a result of migration, an ice structure created at a low deposition temperature is favoured. Since water is thought to be formed on interstellar grain mantles rather than being deposited, this may put constraints on the formation of $\rm H_2O$, since a similar porous structure must then be formed as the water is chemically assembled on the grain surfaces.

The trend between the red component and the $\rm 2175~cm^{-1}$ band has not previously been
observed, probably because earlier samples contained a large fraction of intermediate and high mass stars.
The observation that high-mass stars exhibit a radically different correlation suggests that two different
components contribute to the absorption between 2155 and $\rm 2185~cm^{-1}$.
Since the $\rm 2175~cm^{-1}$ profiles towards the low mass stars also
tend to be centered redder than laboratory experiments allow for the CN-stretch, we propose that a second weak absorber is present
with a center at  $\rm 2170-2180~cm^{-1}$ and a FWHM of $\rm 15-25~cm^{-1}$ and that, based on the correlation with the red component, a likely carrier is CO in a presently unidentified
binding site. One significant implication of the proposed correlation is that no absorption
at $\rm 2136.5~cm^{-1}$ is expected for any source if
the $\rm 2175~cm^{-1}$ band is not present. This is supported by the sensitive upper limits given in Fig. \ref{COOCN} for a number of sources which contain abundant water ice.

The excess seen in some high mass stars is then
an indication of the formation of $\rm OCN^-$ by thermal, UV or other forms of processing not present in low mass stars. The $\rm OCN^-$ band in these sources floods the weaker underlying band and the correlation disappears.

A number of alternative explanations may be explored. For example, it is known that a libration band of water ice creates a broad band centered on
$\rm 2200~cm^{-1}$. This libration mode could create an excess absorption, which when requiring the continuum to fit the blue edge of the
spectra, may mimic the observed $\rm 2175~cm^{-1}$ band. This is a valid objection, since the depth of the
feature depends on the assumption that the blue edge of the spectrum can be used as continuum. However, the water libration band
has a FWHM of about $\rm 300~cm^{-1}$ and the total depth is only $\sim2.5\%$ of the $\rm 3.08~\mu m$ main water band. This means
that the expected excess optical depth from the water combination mode when using a continuum fixed at 2100 and $\rm 2200~cm^{-1}$ is at most
$\tau\sim 0.01$ for a $\tau\sim 5$ main water band.

\begin{figure}
\centering
\includegraphics[width=8.5cm]{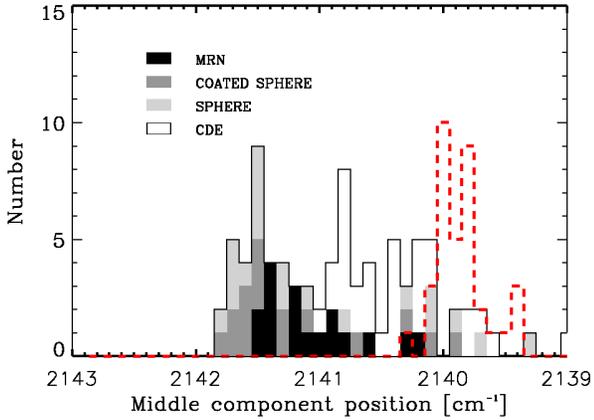}
\caption{Distribution of fitted middle component positions. The thick, dashed curve shows the distribution derived from the astronomical spectra and corrected for the systemic velocities of the sources. Only middle component positions with $1\sigma$ errors of less than $\rm 0.15~cm^{-1}$ have been included.
The thin curve shows the distribution of the center positions of the non-hydrogenated ices from the ice analog database by E97.
Only ice analogs with FWHM between 3 and $\rm 4~cm^{-1}$ have been included in the distribution.
All the laboratory ices are corrected for grain shapes as indicated in the figure.}
\label{centerpos}
\end{figure}

\section{Physical modeling of the solid CO band}
\label{PhysModel}
\subsection{The middle component}

Having identified a simple and well-defined decomposition of the astronomical CO ice band, a unique basis is provided, which can be tested against
models of interstellar CO ice. Since the shape of the middle component is the best
constrained, it is instructive to compare it to laboratory spectra of van-der-Waals interacting ice mixtures corrected for different grain shapes. 
Such a comparison is not intended to provide unique constraints on the composition of the CO ice responsible for the middle component, but
illustrates well how narrow the observed range of profiles is, compared to the possible range of shapes depending on ice composition and grain shape corrections. For this purpose the optical constants from the laboratory database of non-hydrogenated mixtures by \cite{Pascale} (hereafter E97) are used. The laboratory spectra are corrected for
grain shape effects using the standard four particle shapes almost exclusively used in the literature, namely spheres,
coated spheres, a continuous distribution of
ellipsoids (CDE) and an MRN \citep{MRN} size distribution of coated spheres. The parameters for the grain shape correction from E97 have been adopted. For details the reader is referred to \cite{tielens, Pascale, BH}. 
In general, the grain shape corrections are independent of grain size as long as
the grains are small, i.e. $2\pi a/\lambda\ll 1$, and the Rayleigh approximation is valid. Clearly, the requirement
is only marginally satisfied at $\rm 5~\mu m$ and breaks down for grains larger than $\rm 0.1~\mu m$. This may be
a serious concern in the vicinity of protostars, where grain growth is expected to occur. Strictly, grain shape corrections for
wavelengths shortwards of $\rm 10~\mu m$
should be applied using a full numerical solution of the field interactions with the particle.
Nonetheless, we assume in the following that the Rayleigh condition holds given the limited constraints on the detailed grain size distributions in the observed lines of sight. In essence
it is assumed that the extinction is dominated by small grains. This also implies that the effects of scattering out of the line
of sight are ignored. For small particles, the ratio of average absorption to scattering cross sections can be approximated with
$\sim (18 \pi/3) \times \hat{\nu}^{-3} V^{-1}$, where $\hat{\nu}$ is the wave number and $V$ is the particle volume, which only becomes less than unity
for $a\gtrsim \rm 10~\mu m$ at a frequency of $\rm 2140~cm^{-1}$. For the CO band it is a good approximation to ignore scattering
effects, while they become important for bands at higher frequencies, such as for the water band at $\rm 3~\mu m$. Scattering
effects on the CO band may become observable in regions with grain growth, such as in circumstellar disks. 

The coated sphere model was calculated using identical volumes of core and mantle, while the MRN model was calculated using
a constant mantle thickness of $\rm 0.01~\mu m$; models with thicker mantles approximate the results for solid ice spheres. For these models, a constant baseline was subtracted from the absorption cross sections to remove the absorption from the silicate core.

A Gaussian was fitted to the CO band to determine the peak position and FWHM of the laboratory ice absorption. Shape corrected laboratory ices
with FWHM larger than $\rm 4~cm^{-1}$ and smaller than $\rm 3~cm^{-1}$ were discarded since they are inconsistent with the astronomical spectra.
The distribution of the fitted center
positions is compared to the distribution of center positions of the laboratory spectra in Fig. \ref{centerpos}.
Only five different laboratory spectra of non-hydrogenated ice mixtures are in the narrow parameter range defined by the observed middle component using the standard grain shapes out of a possible 228.

The variety of laboratory mixtures that are consistent
with the observational peaks range from pure CO at 10 K with a CDE
grain shape, a $\rm H_2O$:CO:$\rm O_2$ = 1:20:60 at 30 K for coated spheres, to
$\rm H_2O$:CO:$\rm O_2$:$\rm N_2$:$\rm CO_2$=1:50:35:15:3 at 10 K for all the grain shapes. 
The latter mixture gives a reasonable center and width for all grain shapes since the dielectric function is weak. The shift and the broadening
are therefore caused by the presence in the mixture of $\rm CO_2$ and $\rm O_2$, respectively, rather than by grain shape effects. 
It will be shown in Sec. \ref{13COsec} that the shape of the $^{13}$CO band rules out the multi-component mixtures.
This range in possible mixtures illustrates
the degeneracy inherent in fits to the middle component. However, it also shows that the range of ice mixtures in which the carrier
of the middle component is present must be very limited since most of the non-hydrogenated ice analogs are excluded, i.e. the laboratory spectra
coupled with commonly used grain shape corrections show a much larger range in centers and band widths than those
commensurate with the observations. This is encouraging, since it points towards a simple solution to the questions regarding the
composition and structure of CO-rich ice on interstellar grain mantles.

We will proceed to show that the simplest physical model of pure CO ice is sufficient to explain all the middle components
observed. Furthermore, it will be shown that in nearly all cases perfect fits are achieved using the CDE grain shape.

\subsection{Lorentz oscillators as models for binding sites}
\label{LorentzSec}

The simplest possible physical model of the absorption of a molecule in the solid phase is that
of a Lorentz oscillator. This is fundamentally a classical model but yields expressions similar to
those of simple quantum mechanical models \citep[see e.g.][]{Gadzuk, Ziman}, albeit with significant conceptual differences.
Interactions with the surrounding medium may of course alter the
potential energy surface of the oscillator, affecting the line shape. 

In the classical picture the CO molecules are seen as a set
of identical springs with mass $m$, corresponding to the reduced mass of the CO molecule, charge $e$,
spring constant $K$ and damping constant $b$. The last two parameters are affected by the
bond of a CO molecule to the surrounding molecules, and for each configuration of a CO molecule and its nearest neighbours
a unique set of $K$ and $b$ exists. In an amorphous ice there is a continuum of different configurations for a classical CO molecule,
which when averaged over a large number of CO molecules results in a characteristic set of $K$ and $b$. In the simplest model
the configurations of the CO molecules are random, resulting in a ``pure de-phasing'' broadening.  See Appendix \ref{lineshapes} for the technical argument for modeling solid state dielectric functions with Lorentz Oscillators.

The complex dielectric function for a Lorentz oscillator is:

\begin{equation}
\epsilon = \epsilon_0 + \frac{\omega_p^2}{\omega_0^2 - \omega^2 - i\gamma\omega},
\end{equation}
where $\omega_p^2 = e^2\mathcal{N}/m\varepsilon_0$ is the plasma frequency, $\omega_0^2=K/m$ and $\gamma = b/m$.
$\mathcal{N}$ is the number density of oscillators. Finally $\epsilon_0$ determines the dielectric function at frequencies
which are low compared to the electronic excitation frequencies. The parameter is basically the low-frequency wing of the
combined dielectric functions of all the electronic transitions.

A thick slab, i.e. much thicker than a monolayer, of pure
amorphous CO ice is expected to show just a single dominant
environment, namely CO on CO, and the simple model should fit well
to the measured optical constants. In the case of CO this
is well known: the vibrational spectrum of adsorbed CO is commonly
used as a probe of surfaces in the chemical literature \cite[see e.g.][and references therein]{Somorjai94}.
Sub-monolayer coverages of CO can be used to probe
the underlying surface structure, physical and chemical
behaviour of CO on the surface, interactions between CO and other
adsorbates at the surface, and to identify the range of binding
sites favorable to CO on any particular surface.

Note that dielectric functions superpose such that additional binding sites and environments
can be included in the model simply by adding single Lorentz oscillators in $\epsilon$-space. In this picture,
the optical constants of more complicated ice mixtures can also be reproduced by fitting a sufficient number of Lorentz oscillators
accounting for specific environments including other ice species,
although the problem may quickly become too degenerate to yield useful physical information.

Naturally, this is a very simplified picture and many other effects may play a role. For instance, some hydrogenated molecules, such as $\rm H_2O$, complicate
the picture significantly by forming hydrogen bonds with the CO molecules, thus breaking the assumption
of pure de-phasing. However, for weakly interacting environments such as that of CO interacting with itself or with $\rm N_2$ or $\rm O_2$, such
a model may yield useful information. This is a break with the traditional view that solid state features from amorphous ices have a continuous range of shapes depending on the abundance of other species, but should be easy to
verify experimentally using high resolution laboratory spectroscopy in the limit of a low concentration of contaminating molecules,
which interact with CO molecules only through van der Waals
bonds, such as $\rm O_2$, $\rm N_2$ and $\rm CO_2$. In the following the term ``environment'' means a sum of unique sets of nearest neighbours giving rise to a sum of Lorentz oscillators. We will show that in the case of pure CO a simplified physical model can be used to fit the data. 

\begin{figure*}
\centering
\includegraphics[width=17cm]{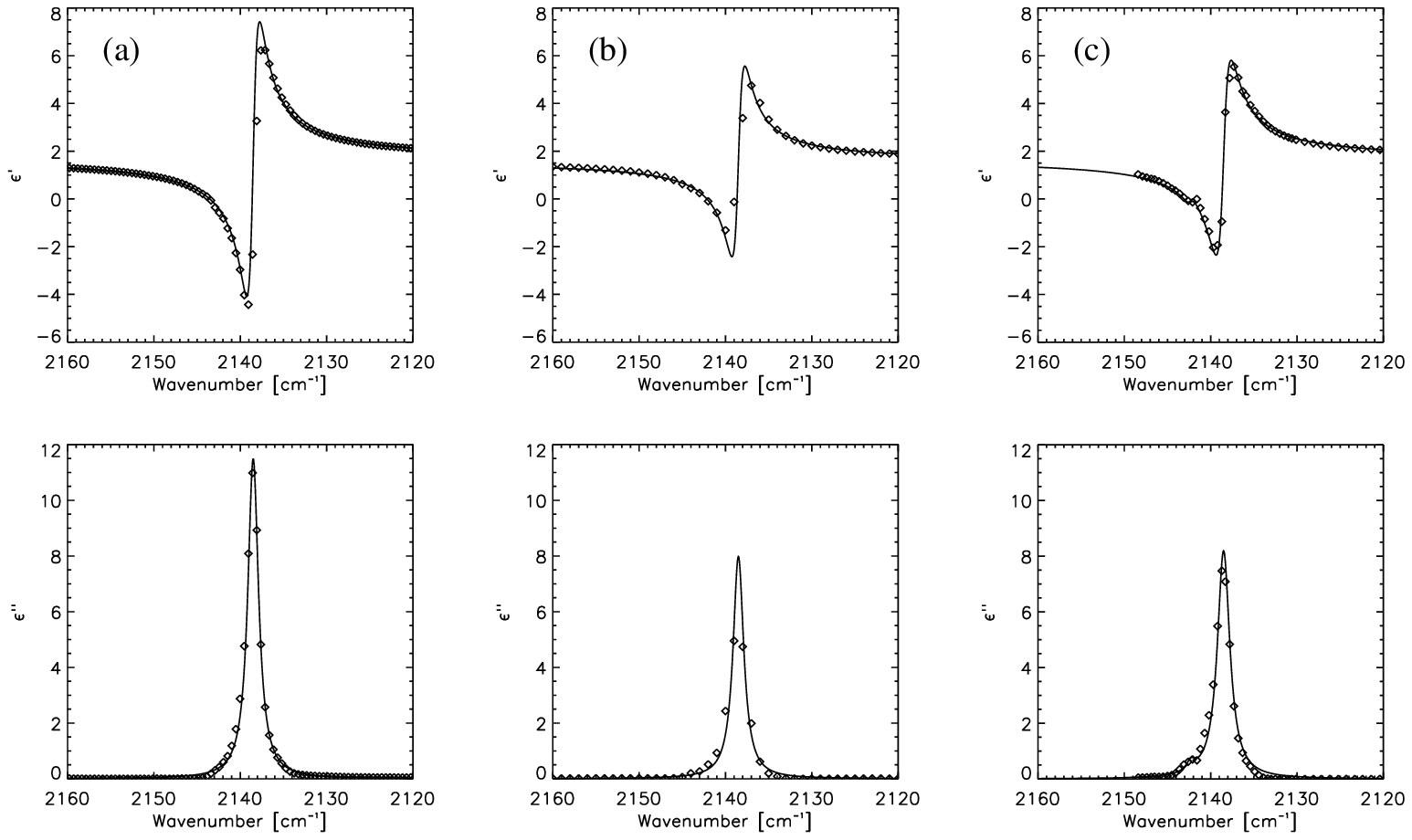}
\caption{Fit of a single Lorentz oscillator to the dielectric function of a pure CO ice at 10 K from E97 (a), \cite{BP} (b) and \cite{EAS} (c).
The best fitting parameters are: (a): $\rm \gamma = 1.5~cm^{-1}$, $\rm \omega_p = 195~cm^{-1}$ and $\epsilon_0=1.67$.
(b): $\rm \gamma = 1.5~cm^{-1}$, $\rm \omega_p = 160~cm^{-1}$ and $\epsilon_0=1.57$. (c):
$\rm \gamma = 1.75~cm^{-1}$, $\rm \omega_p = 175~cm^{-1}$ and $\epsilon_0=1.67$. All have $\rm \omega_0=2138.5~cm^{-1}$ Additionally
a second component has been fitted in panel (c) with parameters: $\rm \gamma = 1.2~cm^{-1}$, $\rm \omega_p = 30~cm^{-1}$ and $\omega_0=2142.3$.}
\label{PureCOFit}
\end{figure*}

\subsection{Oscillator density as fitting parameter}
There is a standing controversy in the literature regarding the accuracy of the determination of optical
constants from laboratory measurements. Since the grain shape correction is sensitive to small differences in
optical constants, the derived band profiles of mixtures with a high concentration of CO can vary significantly when
using different optical constants from the literature \citep[e.g. E97 and][]{Adwin13CO}. This is particularly true for pure
CO. The resulting differences in the dielectric functions are illustrated in Fig. \ref{PureCOFit}, where the optical constants
for pure CO from  \cite{BP} (hereafter BP) and \cite{EAS} (hereafter EAS) have been used to calculate the dielectric functions
in panels (b) and (c), respectively.
The main difference between the functions is in the plasma frequency, resulting in strengths of $\epsilon$ varying by more than 20\%.
Also, the fits to a single Lorentz oscillator are slightly worse for the dielectric functions from BP and EAS.
There is a second component clearly present in the spectrum from EAS with $\omega_0 = \rm 2142.3~cm^{-1}$, which has been included in
the fit. This may be a contaminant. Also, recent results and work by \cite{Collings} clearly shows the presence of this second peak, attributed to LO-TO splitting. Further discussion of this effect is found in Sec. \ref{LOTO} of this article.
The width ($\gamma$) and center positions ($\omega_0$) of the dielectric
functions are very similar in all the experiments.
Within the model, the only parameter which can change the plasma frequency, and thus the optical constants, is the number of
oscillators per unit volume, $\mathcal{N}$. In other words, the difference between the laboratory spectra of identical mixtures can
to a large extent be explained by differing ice density, differing ice porosity, or, less likely,
by dilution by a relatively inactive molecule such as $\rm N_2$. Uncertainties in the determination
of ice thickness in lab experiments may also play a role in the differences of the optical constants obtained by different groups.

\begin{figure*}
\centering
\includegraphics[width=10.5cm]{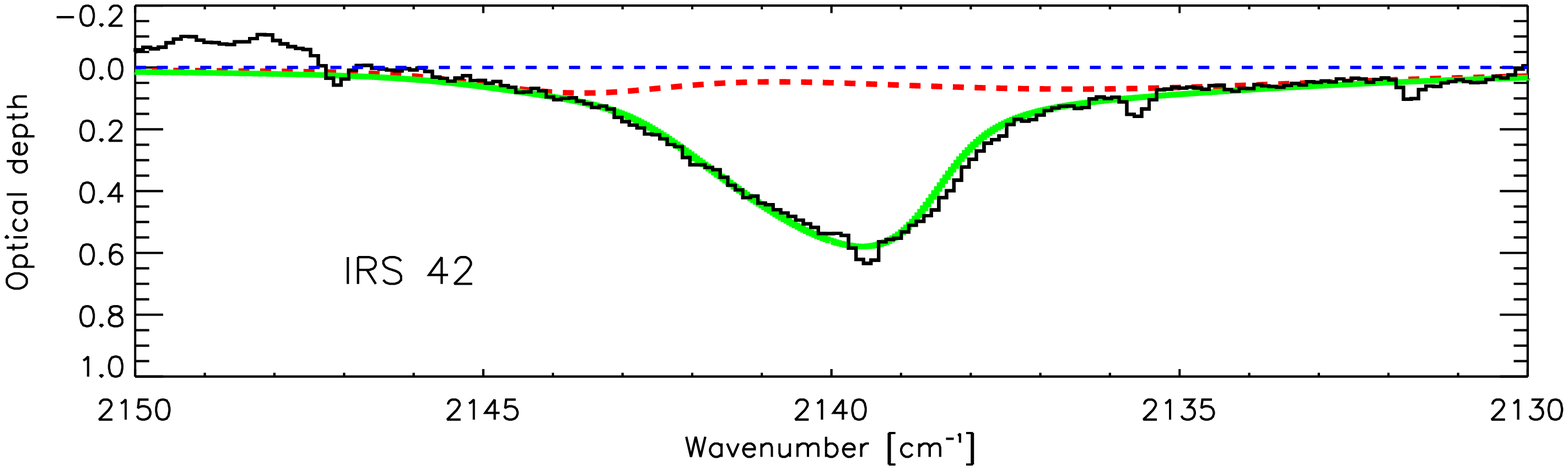}
\includegraphics[width=10.5cm]{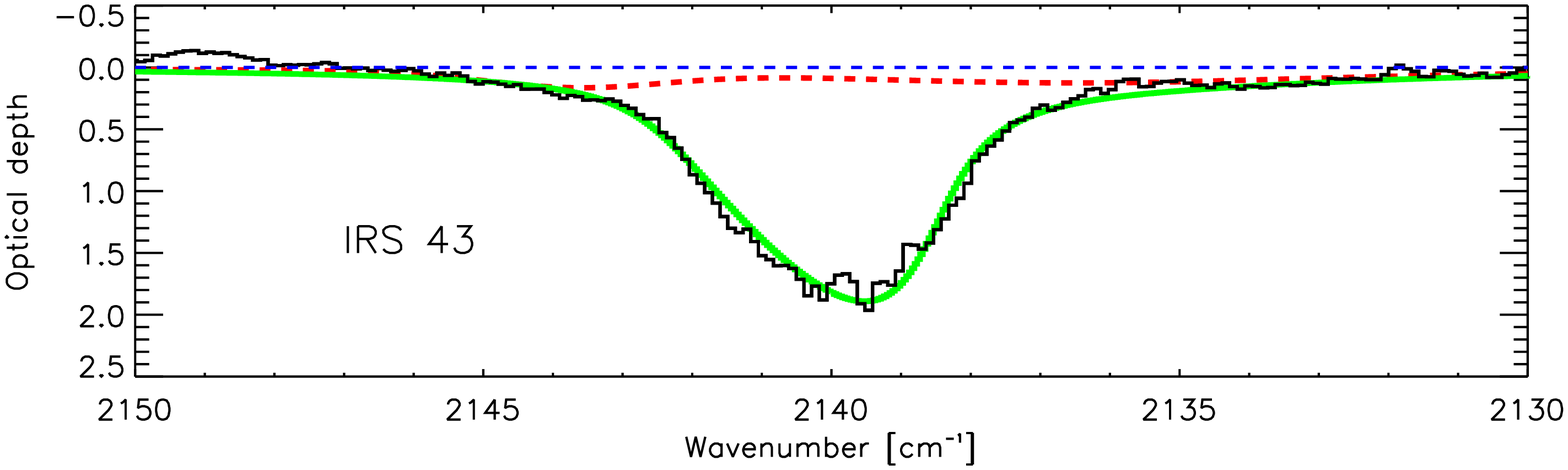}
\includegraphics[width=10.5cm]{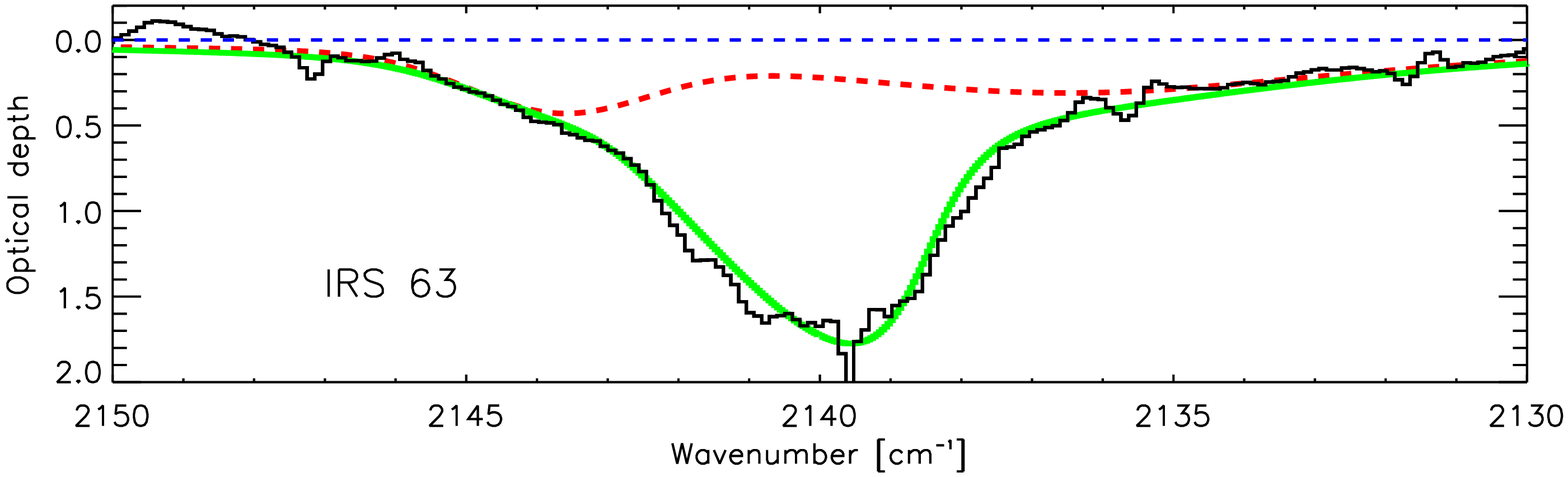}
\includegraphics[width=10.5cm]{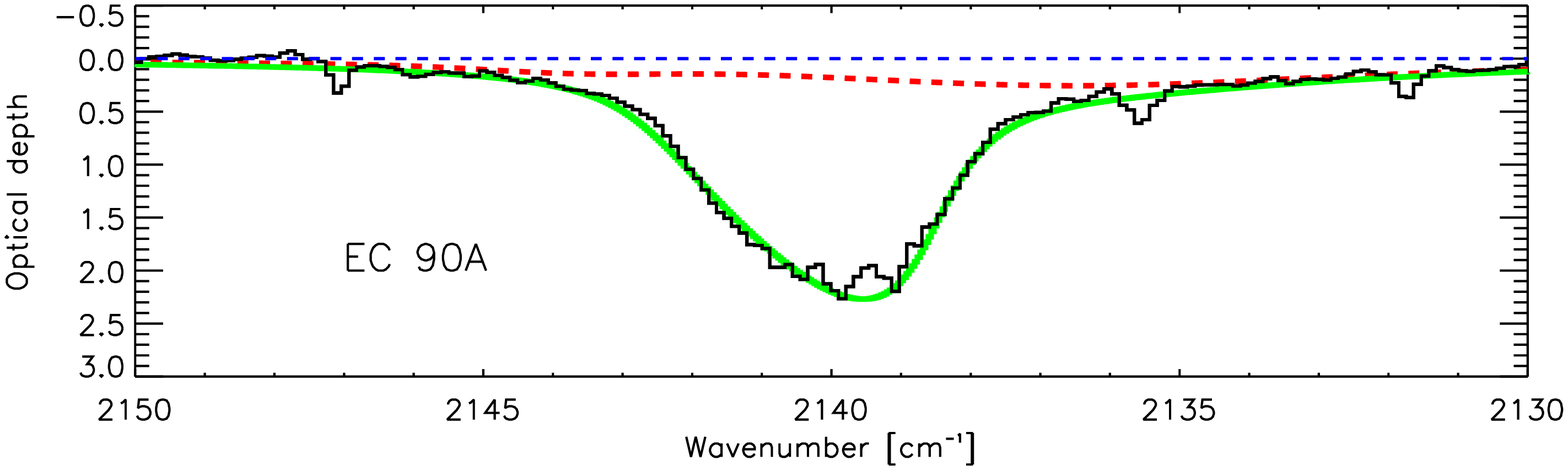}
\includegraphics[width=10.5cm]{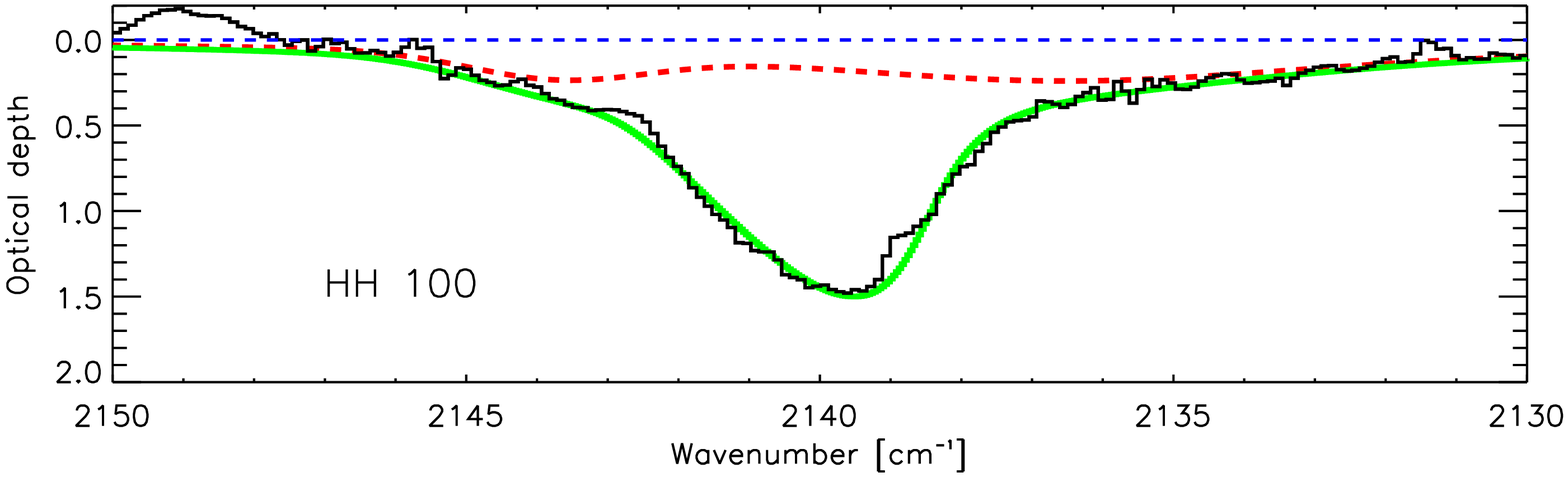}
\includegraphics[width=10.5cm]{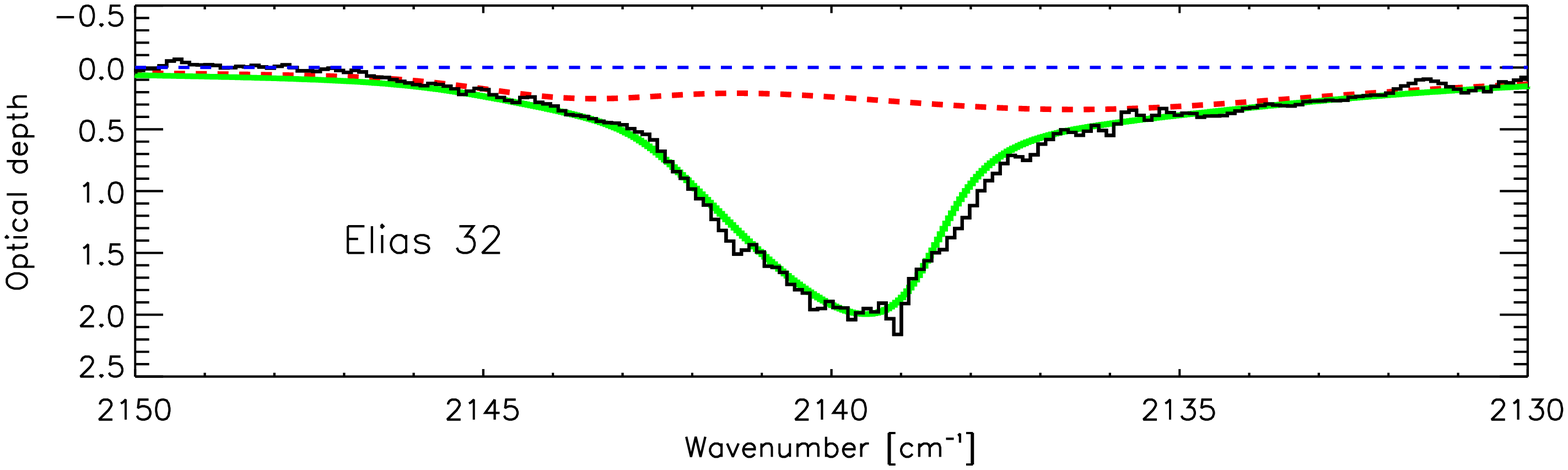}
\caption{Fit of a pure CO Lorentz oscillator model corrected for CDE grain-shape effects to the best observed spectra
which are dominated by the middle component.
The red and blue components have been added to the shape-corrected Lorentz oscillator as in the
purely phenomenological model (cf. Table \ref{FitPars}). 
Often, the depth of the blue component had to be
lowered by up to 30\% for sources with a deep middle component when using a Lorentz oscillator due to the
presence of a blue wing in the CDE corrected profile. For the parameters of the adopted Lorentz oscillator the fit to the optical
constants from E97 are used, except for the plasma frequency which is allowed to vary from $\rm 170-180~cm^{-1}$. The dashed curve
shows the sum of the red and blue components. A dashed straight line indicates the continuum level at zero optical depth.}
\label{LOfitTOspectra}
\end{figure*}

\begin{figure*}
\centering
\includegraphics[width=11cm]{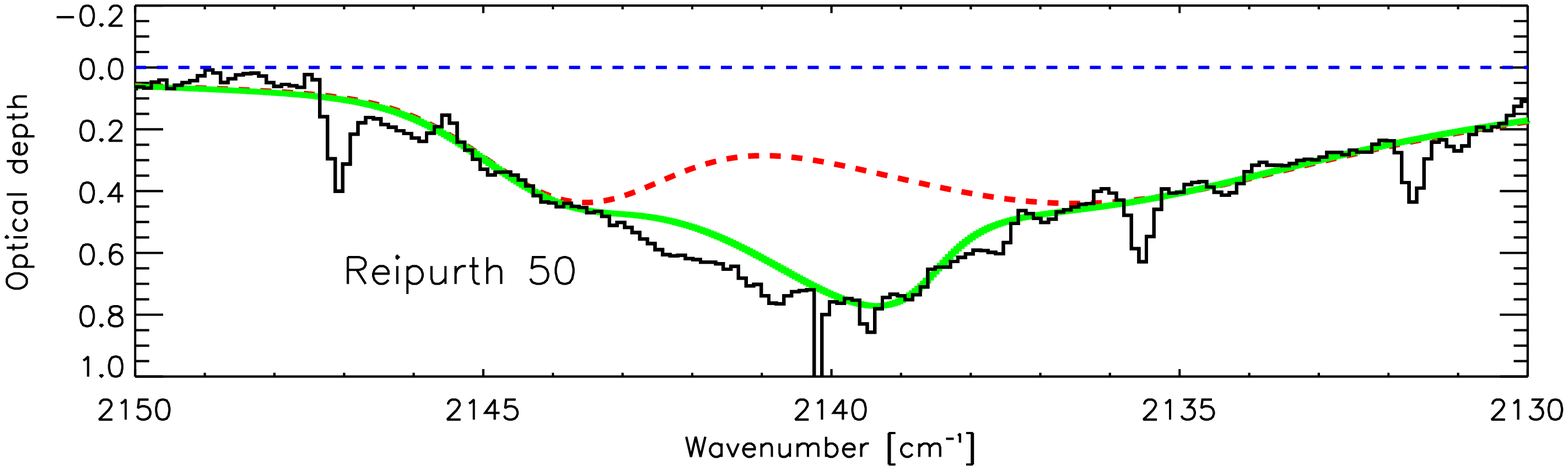}
\includegraphics[width=11cm]{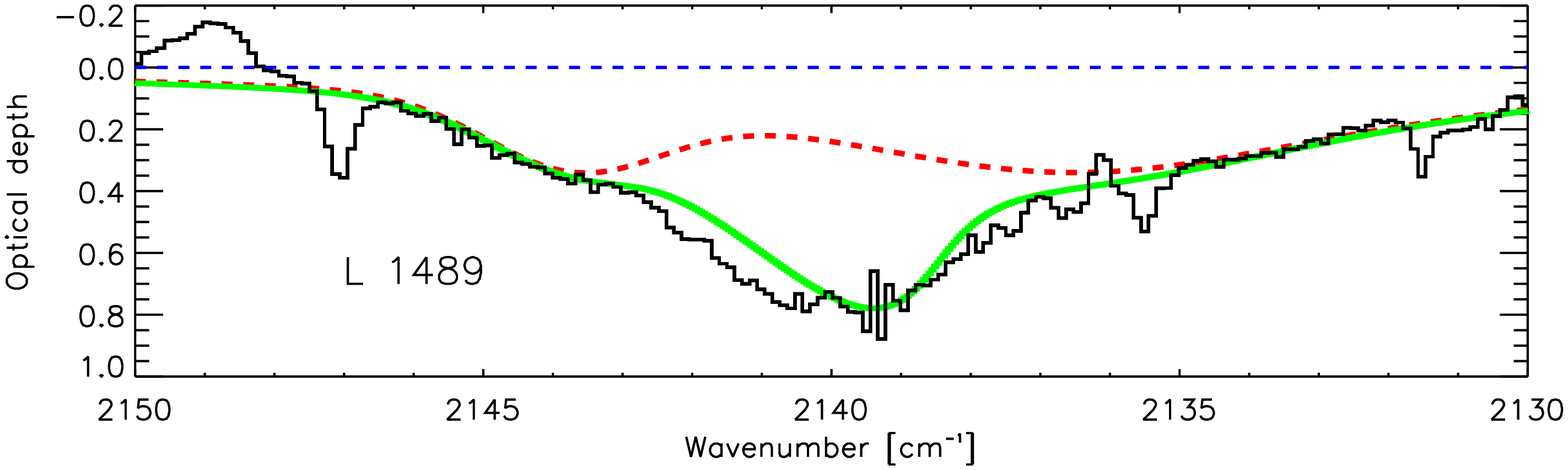}
\caption{Same as Fig. \ref{LOfitTOspectra} for sources dominated by the red and blue components. For these sources a clear deviation from the 
pure CO model is seen between $\rm 2140~cm^{-1}$ and $\rm 2143~cm^{-1}$.}
\label{LOfitTOspectra2}
\end{figure*}

Fits using a single Lorentz oscillator as a model of the middle component to the interstellar spectra with the highest S/N ratios are shown
in Figs. \ref{LOfitTOspectra} and \ref{LOfitTOspectra2}. The model uses the same phenomenological components for the
red and blue components, but now adopts the Lorentz oscillator fit to the pure CO spectrum by E97.
The middle component is grain-shape corrected using
CDE particles. The CDE particles are used simply because they fit the data extremely well and because this model
is believed to simulate irregular particles well by including a maximal range of possible shapes. 
The CDE particles have a specific distribution and have thus no free parameters.
The actual shapes of interstellar grains are expected to be entirely different. The plasma frequency is allowed to vary to account for different ice densities or for small differences in dilution.
$\omega_{p}$ was found to vary slightly between values of 170 to $\rm 180~cm^{-1}$ corresponding to densities 8 to 13\% smaller than the
E97 laboratory ice but very similar to the densities of the BP and EAS ices. Overall the fits are excellent and often clearly an improvement
over the Gaussian model. Note that the fact that the grain-shape corrected Lorentz oscillators so closely emulates Gaussian profiles
indicates that the blue and
middle components are ices with a high concentration of CO, thus making grain shape effects important, while the red component is likely
much more dilute, since it is close to the Lorentzian profile.

\begin{figure}
\resizebox{\hsize}{!}{
\includegraphics{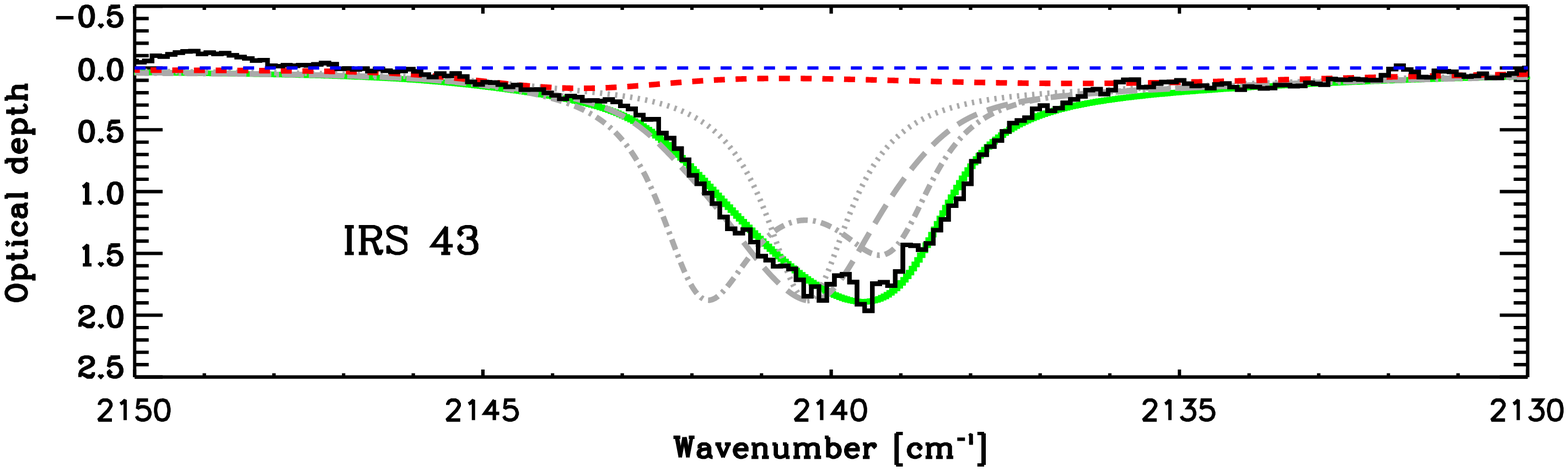}}
\caption{Comparison of different grain shape corrections for a Lorentz Oscillator pure CO. Solid line: CDE;
Dotted line: Identical spheres; Dash-dotted line: identical coated spheres with equal volume core and mantle;
Dashed line: MRN distribution of coated spheres with a $\rm 0.01~\mu m$ mantle thickness.}
\label{OtherShapes}
\end{figure}

\begin{figure}
\resizebox{\hsize}{!}{
\includegraphics{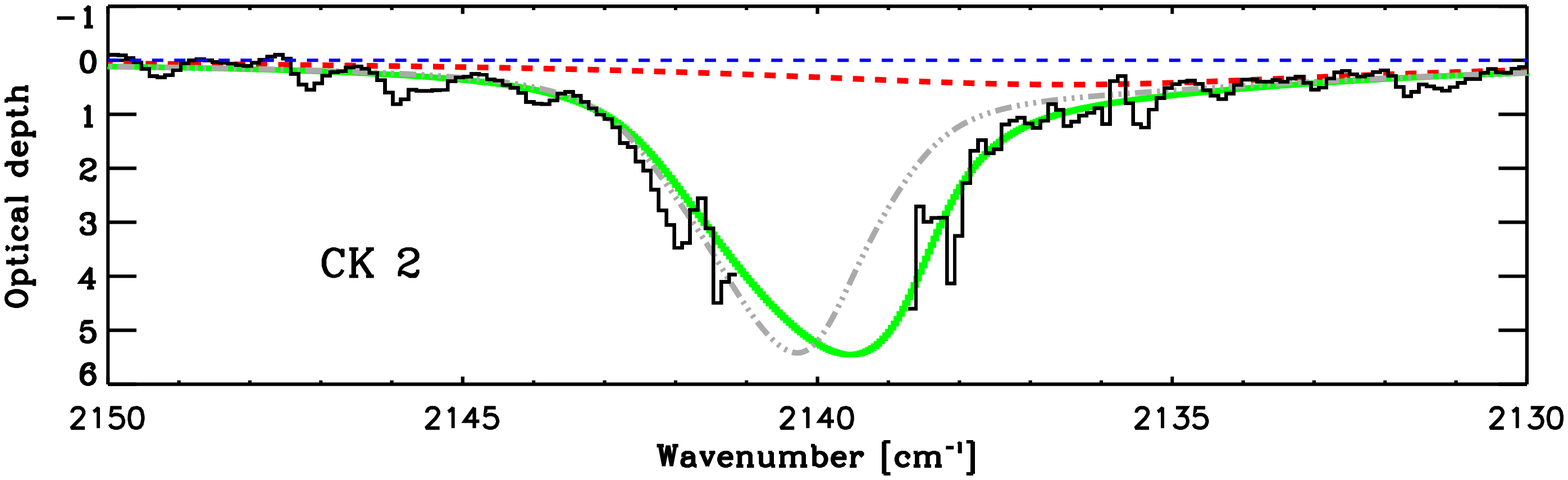}}
\caption{Model profiles using a Lorentz oscillator for pure CO at 10 K corrected for CDE grains (solid curve)
and MRN-distributed coated spheres (dot-dashed curve) compared to the spectrum of CK 2. }
\label{CK2profile}
\end{figure}

It is instructive to repeat the exercise for the other three grain shapes used in the literature. Fig. \ref{OtherShapes} shows
the CO profile for IRS 43 compared with the pure CO profile corrected for spherical grains, coated spheres and coated spheres
with an MRN size distribution. Clearly none of the other Rayleigh limit grain shapes fit the observed profile. If any of these grain shapes
is to be made to fit, some ice mixture is needed to broaden the profile. Alternatively, the MRN size distribution can be modified.
Since the MRN distribution is a power law with exponent $-3.5$, it is dominated by small grains. This is reflected in the
CO profile by the fact that it peaks at $\rm 2140.2~cm^{-1}$; i.e. a large mantle to core volume ratio shifts the profile to the
blue. If a power law grain size distribution is to fit the red wing of the observed profile, the distribution must be much shallower in
order to include a larger fraction of large grains. A broader shape-corrected profile can be obtained if the exponent is
changed to a value closer to $-1$. This may be interpreted as evidence for grain growth near the young stars.
However, this scenario requires that a change in profile is seen in YSO's as compared to quiescent
dense clouds. Fig. \ref{CK2profile} shows the observed CO profile of the well-studied background star CK 2.
This is most likely a K0/G8 supergiant seen through a dense part of the Serpens molecular cloud core \citep{CE96}. While the CO profile towards CK2
is severely affected by absorption lines intrinsic to the star as well as being saturated, it is clear that the wings
of the CO profile are consistent with the shape given by CDE grains and inconsistent with MRN grains. 
In conclusion, the data presented here do not
provide evidence for a strong difference in grain shapes in the quiescent medium compared to lines of sight towards YSOs. 
The similarity between the CO profile towards CK 2 and the other observed profiles also show that the separation of a possible contribution
from solid CO in quiescent foreground clouds may be very difficult using only the solid state band. Gas-phase observations of rotational
lines are thus needed  to estimate the foreground contribution to a CO band towards a YSO in a rigorous manner \citep{AdwinElias29}.

The excellent fit to the middle component as a single CO environment is strong evidence that the non-hydrogenated
or van-der-Waals interacting component of interstellar
ices is pure CO with at most a 5-10\% contaminants. This percentage corresponds to the range of oscillator strengths of the pure CO 
dielectric function (from 170 to $\rm 180~cm^{-1}$) used to fit the data. With the current lack of knowledge about the true
grain shapes in the interstellar medium, it can be concluded that within the quality of the presented data,
the middle component is indistinguishable from the profile of pure CO
along all observed lines of sight where the middle component dominates the CO profile. Deviations are seen in sources
dominated by the red component. This is most clearly evident in Reipurth 50 and L 1489 as shown in Fig. \ref{LOfitTOspectra2}.
While the pure CO component is still present, one or perhaps two new features appear around $\rm 2141~cm^{-1}$ and $\rm 2142~cm^{-1}$.
The features may also be present in IRS 63. These features have optical depths of only 0.1 -- 0.2 and may therefore be
present in all spectra, but are simply swamped by the deep middle component when this dominates. The centers of these new components 
coincide well with the bulk of the laboratory spectra in Fig. \ref{centerpos}, and may be evidence for a small contribution from mixed CO ices. 

For the pure CO combined with CDE grains the column densities can be found from:

\begin{equation}
N_{\rm Pure, CDE}=6.03~{\rm cm^{-1}}\times \tau_{\rm max}\times A_{\rm bulk}^{-1},
\label{Nmiddle}
\end{equation}
where $A_{\rm bulk}$ is the band strength of the bulk material. The numerical factor takes into account the $\tau=1$ band equivalent width and
the fact that the grain shape effect changes
the effective band strength with respect to the band strength measured in the laboratory with a factor of $0.71=\int C_{\rm CDE}d\hat{\nu}/\int C_{\rm slab}d\hat{\nu}$ in this specific case. $C_{\rm CDE}$ and $C_{\rm slab}$ are the absorption coefficients for pure CO CDE grains and for a pure CO slab, respectively.  In the following we use a laboratory band strength of $\rm 1.1\times 10^{-17}~cm~molec^{-1}$ \citep{Gerakines95}.

\subsection{$\rm ^{13}CO$ in IRS 51}
\label{13COsec}
\begin{figure}
\resizebox{\hsize}{!}{
\includegraphics{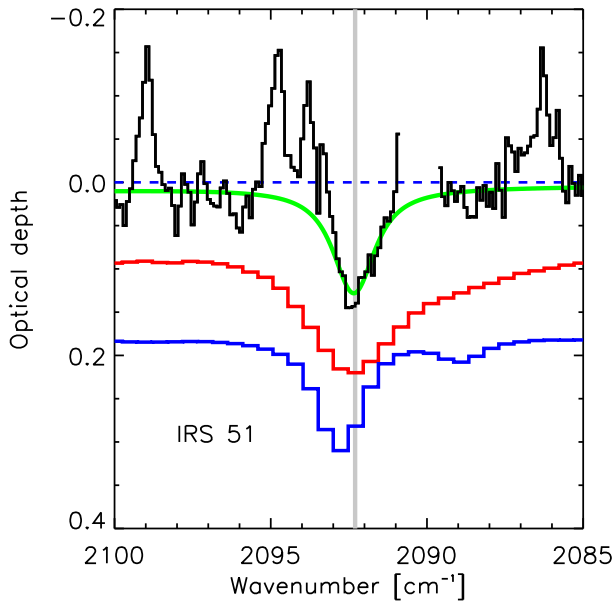}}
\caption{The solid $\rm ^{13}CO$ band of IRS 51. The upper solid line is a Lorentz oscillator fit for a pure CO ice.
The blue wing is somewhat affected by a broad CO gas phase emission line. The middle and lower curves are laboratory spectra of
binary mixtures at 10 K rich (1:1) in $\rm O_2$ and $\rm N_2$, respectively. The vertical line indicates the center of the pure CO profile.}
\label{13CO}
\end{figure}

Solid $\rm ^{13}CO$ is detected towards one of the sources, namely IRS 51 (Fig. \ref{13CO}).
Since the isotopic ratio of $\rm ^{12}CO$ to $\rm ^{13}CO$ is expected to be between 65 and 75 in the gas phase for typical molecular cloud
conditions \citep{Langer93} and between 55 and 85 in the solid phase \citep{Adwin13CO2}, $\rm ^{13}CO$ is highly dilute in the ice. An important consequence is that grain shape effects are unimportant
and the solid $\rm ^{13}CO$ band therefore offers the chance to disentangle, with respect to the band profile, the effects of the ice matrix from those of the grain shape \citep{Adwin13CO}.

For the $\rm ^{13}CO$ band towards IRS 51 the same procedure as for the middle component of the $\rm ^{12}CO$ band is repeated.
A Lorentz oscillator was fitted to the dielectric function by E97. The best fitting parameters are
$\rm \omega_p=22~cm^{-1}$, $\rm \omega_0=2092.3~cm^{-1}$
and $\rm \gamma = 1.5~cm^{-1}$. Again the plasma frequency is adjusted to fit the observed band. A good fit to the
interstellar profile is found with $\rm \omega_p=21~cm^{-1}$, although the blue wing contains flux from the $v=2-1$ P(6) ($\rm 2193.40~cm^{-1}$)
and the $v=1-0$ P(12) ($\rm 2194.86~cm^{-1}$) emission lines
from gas phase $\rm ^{12}CO$. The red wing of the observed band is fully consistent with a pure CO ice. The ratio of the integrated
imaginary part of the Lorentz oscillator dielectric functions gives the isotopic ratio of solid CO along the line of sight,
$N({\rm ^{12}CO})/N({\rm ^{13}CO})=68\pm 10$
for IRS 51. The uncertainty reflects the fitting uncertainty in the saturated $\rm ^{12}CO$ band assuming that the shape of the
band is identical to the other sources. Comparisons with laboratory spectra of binary mixtures with $\rm O_2$ and $\rm N_2$ at 10 K from E97 are also shown in Fig. \ref{13CO}. Both laboratory mixtures give significantly worse fits to the IRS 51 spectrum, the $\rm O_2$-rich mixture being too broad, and the $\rm N_2$-rich mixture being shifted to the blue. Heating the $\rm O_2$-rich mixture to 30 K narrows the profile, but not enough to equal the
quality of the fit of that of pure CO. It may also be considered unlikely to find most of the volatile solid CO component in dense clouds at a temperatures
higher than 20 K. 

The results for the solid $\rm ^{13}CO$ band agree well with the conclusions reached by \cite{Adwin13CO} for the $\rm ^{13}CO$ band
towards the high mass source NGC 7538 IRS9, who also found both isotopic bands of CO to be fitted well with pure CO. Furthermore, the isotopic ratio for NGC 7538 IRS9 of $71\pm 15$ is remarkably similar to that found for IRS 51. Indeed, the shapes of the two $\rm ^{13}CO$ bands match closely, in agreement with the similarities observed among the $\rm ^{12}CO$ bands.

\subsection{The red component}
\label{red}
The red component in the interstellar CO ice profile was first assigned by \cite{Sandford} and \cite{tielens}  to CO mixed in a
hydrogen-bonding ``polar'' mixture containing species
such as $\rm H_2O$ or $\rm CH_3OH$. The previous $M$-band surveys towards young embedded stars all concluded that this assignment is roughly consistent with the data. The ice mixtures found to provide the best fits in the literature \citep{Kerr93,Chiar94,Chiar95,Chiar98,AdwinL1489}
are almost exclusively pure $\rm H_2O$-CO mixtures containing 25 or 5\% CO, but with temperatures ranging from 10 to 100 K. Occasionally
good fits are found with irradiated $\rm H_2O$-$\rm CH_3OH$ mixtures. The identification of a phase of CO mixed in water ice
is also indirectly supported by observations, since large columns of solid, generally amorphous, water ice are known
to exist along the same lines of sight. Nonetheless, we find that the assignment of the red component to an amorphous CO-$\rm H_2O$ mixture presents some inconsistencies between the observations and both old (high vacuum) and new (ultra-high vacuum) laboratory results.

Laboratory spectra of CO co-deposited with water at low temperature ($T<20~\rm K$) show two distinct absorption peaks. One well defined peak
near $\rm 2152~cm^{-1}$ and one generally stronger peak near $\rm 2138~cm^{-1}$, although different experiments disagree somewhat
on the peak position of the lower frequency peak, some placing it at $\rm 2136-2137~cm^{-1}$ \citep{Sandford, Schmitt89}. This shift
may be related to the degree to which the CO and the water are mixed, since CO deposited on a water surface shows absorption
at $\rm 2138~cm^{-1}$. In recent work by \cite[e.g.][]{Devlin, Palumbo, Manca01,Collings} strong experimental evidence is presented that the
$\rm 2152~cm^{-1}$ peak is due to the CO bonding with the hydrogen in OH dangling groups as was suggested by e.g. \cite{Schmitt89}. In the same picture the $\rm 2137~cm^{-1}$ peak is the normal CO bond site also found in pure CO but perturbed by a water ice surface. This holds in particular for
CO trapped in micropores inside the amorphous water ice. Therefore the $\rm 2152~cm^{-1}$ feature represents the principal binding site
for CO on water, the presence of which would be strong evidence for CO interacting with an OH containing species. 
The  $\rm 2152~cm^{-1}$ bond is suppressed by warm-up and is known to completely
disappear for $T \rm >80~K$. Also irradiation by UV
photons or bombardment by energetic particles destroys the bonds responsible for the $\rm 2152~cm^{-1}$ feature. In general, warm-up
of a $\rm H_2O-CO$ mixture also narrows the $\rm 2136~cm^{-1}$ feature and shifts it towards the red. Irradiation also shifts the
feature to the red, but in contrast to pure thermal processing broadens the band slightly. All of these processing effects
are irreversible. Therefore the search for the
$\rm 2152~cm^{-1}$ feature is essential because it has the potential to not only uniquely confirm the association of the red component
with OH-bearing species, but also to provide a sensitive temperature and processing indicator.

\begin{figure}
\resizebox{\hsize}{!}{
\includegraphics{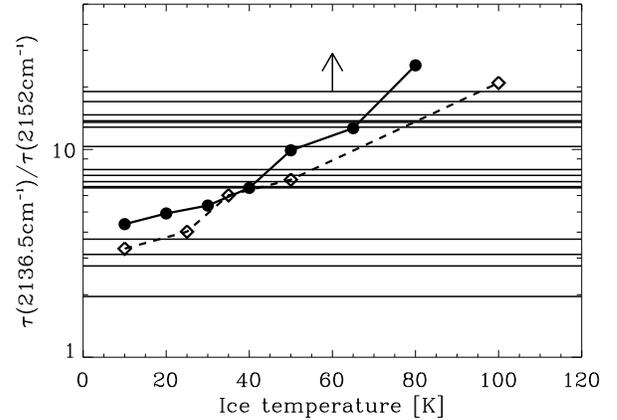}}
\caption{Lower limits on the ratio between the red component at $\rm 2136.5~cm^{-1}$ and the $\rm 2152~cm^{-1}$ OH-CO bond for
the highest quality interstellar spectra. The horizontal lines indicate the limits while the curves show the expected
ratios from co-deposited mixtures of $\rm H_2O$ and CO. Solid line: $\rm H_2O:CO = 20:1$ \citep{Sandford}. Dashed line: $\rm H_2O:CO = 4:1$ \citep{Schmitt89}}
\label{waterbindings}
\end{figure}

None of our interstellar spectra show any evidence of the $\rm 2152~cm^{-1}$ OH-CO bond. The lower limits on the ratio between the optical depth of
the red component and the $\rm 2152~cm^{-1}$ feature for the best quality spectra are shown in Fig. \ref{waterbindings}. They are compared to the
temperature-dependent ratios from the laboratory spectra by \cite{Sandford} and \cite{Schmitt89}. Assuming that the
laboratory spectra used are suitable as interstellar ice analogs, the limits on 8 sources are good enough
to exclude ice temperatures below 60 K while an additional 6 sources can exclude ice temperatures below 40 K. However, the profiles
for ice mixtures at temperatures higher than 60 K have FWHM of only $\rm \sim 7~cm^{-1}$, making them inconsistent with
the broader profiles observed along all lines of sight. It can be concluded
that the available laboratory spectra in general exclude the low temperature non-processed hydrogenated ices, otherwise traditionally found to be consistent with interstellar spectra, due to the absence of CO in a OH bonding site (the $\rm 2152~cm^{-1}$ feature). Alternatively, the interstellar ices may in general be strongly irradiated, but this requires an explanation why no other signatures of irradiation are seen along many lines of sight, such as $\rm OCN^-$, aliphatics, etc. Strong irradiation may also be inconsistent with the strength of the typical UV-field in a dark cloud, since UV photons can only
be formed as a secondary effect to cosmic ray hits due to extinction except at cloud surfaces.

A significant problem in the interpretation of the red component is thus revealed. On the one hand, there is clearly evidence in the
data for the presence of a CO component embedded in the water ice and for migration of a pure CO component into a porous water ice (Sec. \ref{Interpretations}). The shape of the red component, a high efficiency of CO migration and the general shape of water ice bands towards low-mass YSOs all suggest that most of the water ice column has a fairly low temperature \citep{AdwinElias29,ThiThesis}. On the other hand the absence
of the $\rm 2152~cm^{-1}$ feature seems to exclude this. It is a challenge for laboratory studies to explore how the CO-OH bonds can be avoided
or efficiently destroyed under interstellar conditions.  Possible explanations may include the destruction of CO-OH bonds on interstellar time scales ($>10^3$ years). Also, the formation of water molecules on grain surfaces may produce an ice structure different from structures obtained by deposition. Finally, the possibility that the red component is not due to CO interacting with water should not be ruled out, although it seems unlikely given the ubiquitous presence of water ice in grain mantles. 

For the red component, assuming grain shape effects are negligible, the column densities can be found from:

\begin{equation}
N_{\rm rc}=16.0~{\rm cm^{-1}}\times \tau_{\rm max}\times A_{\rm rc}^{-1},
\label{Nred}
\end{equation}
where $A_{\rm rc}$ is the band strength of the bulk material. It assumed that the CO concentration small so grain shape effects can be ignored. The numerical factor is then the equivalent width of a $\tau=1$ band. In the following a band strength for a water-rich mixture of $A=1.1\times 10^{-17}~\rm cm~molec^{-1}$ is used \citep{Gerakines95}.

\subsection{The blue component}
Since the blue component was only identified recently in the high resolution
Keck-NIRSPEC spectrum of L1489 IR by \cite{AdwinL1489}, the only available statistics are from our work.
In the present sample, it is found that the component is particularly prominent as a distinct shoulder in the profile
in L1489 IR and in Reipurth 50.
The feature is detected as a well-defined shoulder in many of the other sources such as HH 100, Elias 32, SVS 4-9 and IRS 63,
but has a smaller ratio with the middle component. 

\cite{AdwinL1489} suggest that the blue component can be identified with CO mixed in a
$\rm CO_2$-rich ice ($\rm CO_2/CO>1$) or a mixture with less $\rm CO_2$ but with significant amounts of $\rm O_2$
and $\rm N_2$ present. While reasonable fits can be obtained with a suitable $\rm CO_2$-containing laboratory mixture,
we will argue in the following sections that another candidate explanation exists, namely the longitudinal optical (LO) component from pure
crystalline CO, which appears when the background infrared source is polarised. The presence of this component
can be independently tested by measuring the linear polarisation fraction at $\rm 4.7~\mu m$.

\subsection{LO-TO splitting in crystalline $\alpha$-$\rm CO$?}
\label{LOTO}

\begin{figure}
\resizebox{\hsize}{!}{
\includegraphics{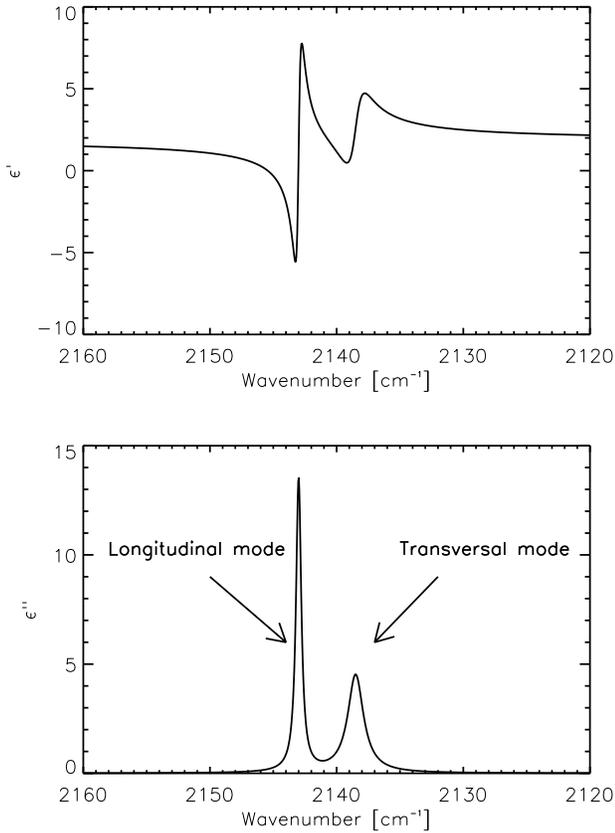}}
\caption{The adopted model of the complex dielectric function of $\alpha$-CO for p-polarised light.}
\label{LOTOdielec}
\end{figure}

Laboratory spectroscopy of multi-layered crystalline CO ($\alpha$-CO) using a p-polarised (light polarised parallel to the plane of incidence)
infrared source shows a splitting of the CO-stretching vibration mode \citep{Chang,Collings}. The splitting is
due to vibrations perpendicular and parallel to the ice surface which are generally referred to as the longitudinal optical (LO)
and the transversal optical (TO) modes, respectively. In this article model parameters for LO-TO split CO are
derived using the laboratory work of \cite{Chang}, unless otherwise stated. For thin ice layers ($\lesssim 30$ mono-layers), the two modes
show very narrow profiles.  According to \cite{Chang}, the LO and TO modes have widths of $\rm 0.25~cm^{-1}$ and
$\rm 0.85~cm^{-1}$, respectively. The LO mode disappears for a single monolayer. For thicker layers ($\gtrsim 100$ monolayers),
the TO mode exhibits a profile very similar to the
profile seen in unpolarised light, i.e. $\rm \omega_0=2138.5~cm^{-1}$ and $\rm FWHM=2\gamma=1.5~cm^{-1}$. The LO mode
gives for thick layers a strong very narrow and blue-shifted profile with  $\rm \omega_0=2143~cm^{-1}$ and $\rm 2\gamma=0.5~cm^{-1}$.
Here $\epsilon_0=1.84$ is adopted as given in \cite{Zumofen}.  Unpolarised light will produce a single peak at $\rm 2138.5~cm^{-1}$ very similar to
that seen for amorphous pure CO ice. It is assumed that the dielectric functions for $\alpha$-CO and amorphous CO are identical when subjected
to unpolarised light.

The dielectric function of $\alpha$-CO for p-polarised light is therefore modeled by two Lorentz oscillators \citep{Chang}. 
The plasma frequencies of the two modes are assumed
to be identical. A value
of $\rm \omega_p=120~cm^{-1}$ is found to reproduce the strength of both modes in the laboratory spectrum of \cite{Chang}. This value
is allowed to vary slightly to account for uncertainties in the determination of the sample thickness. The LO-TO splitting model is corrected
for grain-shape effects using the same CDE grains as for the middle component. After correcting for grain shape, $\rm \omega_p=140~cm^{-1}$
gives a significantly better fit to all the astronomical spectra. The difference compared to the
laboratory data is considerable, also since a crystalline ice in space is not expected to have a different oscillator density than the laboratory sample.
Further laboratory experiments are needed to confirm if a discrepancy indeed is present.

\begin{table}
\centering
\begin{flushleft}
\caption{Polarisation fractions assuming LO-TO splitting}
\begin{tabular}{lll}
\hline
\hline
Source & $P_{4.7}$ & $P_{2.2,\rm max}^d$\\
& [\%] & [\%]\\
\hline
L 1489 &41& 70$\rm^a$\\
Reipurth 50 &57& 60$\rm^b$  \\
HH 100 IR &9& --\\
IRS 63 & 18& --\\
Elias 32 & 7& --\\
SVS 4-9 & 23 & --\\
CK 1A & $<10$ & 5$\rm^c$\\
CK 1B & $<10$ & 5$\rm^c$\\
\hline
\end{tabular}
\label{PolFracs}

\begin{list}{}{}
\item $\rm^a$ \cite{Whitney97}
\item $\rm^b$ \cite{Casali91}
\item $\rm^c$ \cite{Sogawa97}
\item $\rm^d$ Observed value

\end{list}

\end{flushleft}
\end{table}

The necessary linear polarisation fraction at $\rm 4.7~\mu m$ to explain the blue component with the LO mode is found through:

\begin{equation}
C_{\rm total}=C_{\rm unpolarised}+\frac{P_{4.7}}{1-P_{4.7}}[\beta C_{\rm LT,p}+(1-\beta)C_{\rm LT,s}],
\label{Poleq}
\end{equation}
where $P_{4.7}$ is the polarisation fraction at $\rm 4.7~\mu m$ while $C_{\rm unpolarised}$, $C_{\rm LT,p}$ and $C_{\rm LT,s}$ are the absorption cross sections for unpolarised, p-polarised and s-polarised light, respectively. $\beta$ is a factor accounting for the fraction of the light
seen as being ``p-polarised'' by the CO molecules on a grain surface. In general, $\beta=1/2$
for randomly oriented particles, which can be realised by volume-integrating the projection on the plane of the incoming
polarised light of the surface normal vector. If the grains are aligned along magnetic field lines, $\beta$ can deviate considerably from $1/2$,
but will never exceed unity. The CO ice is assumed to be entirely crystalline and the absorbing grains are assumed to be randomly oriented. It is important not
to confuse the absorbing grains with the background grains producing the scattered and polarised light, since their properties may be different.

\begin{figure*}
\centering
\includegraphics[width=8.5cm]{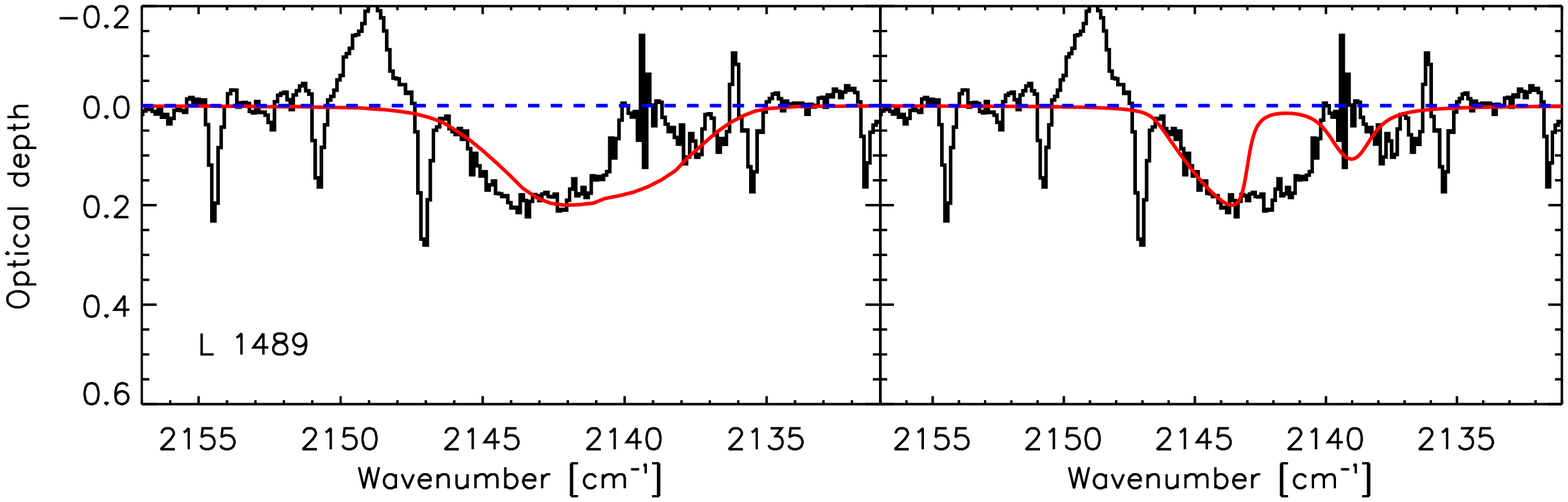}
\includegraphics[width=8.5cm]{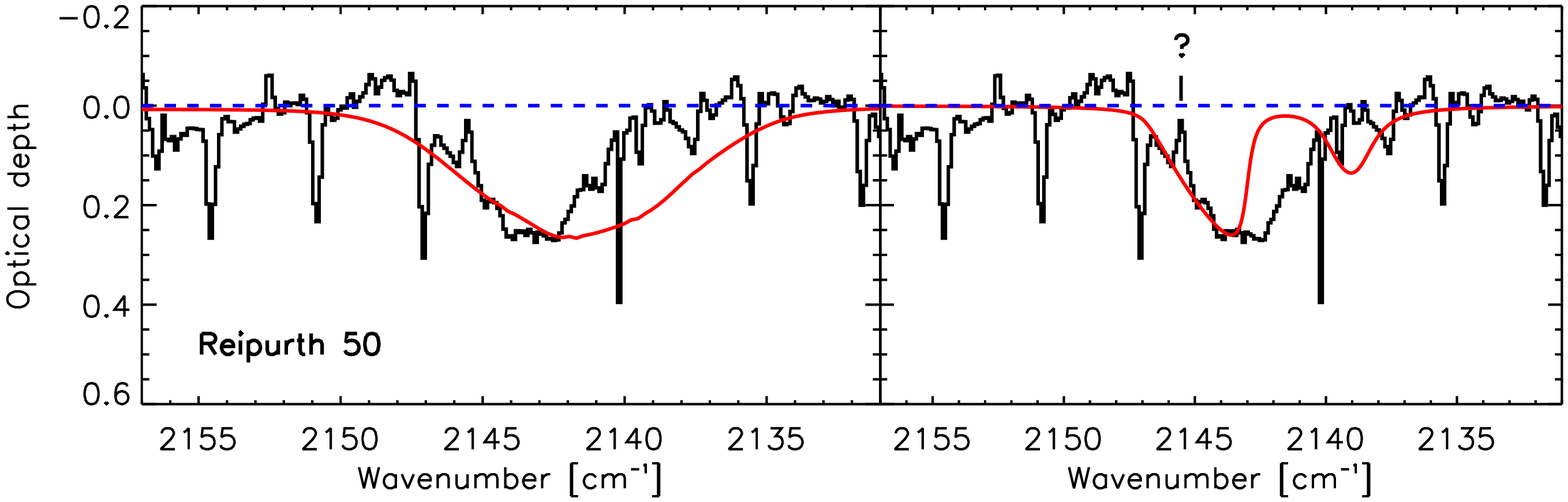}
\includegraphics[width=8.5cm]{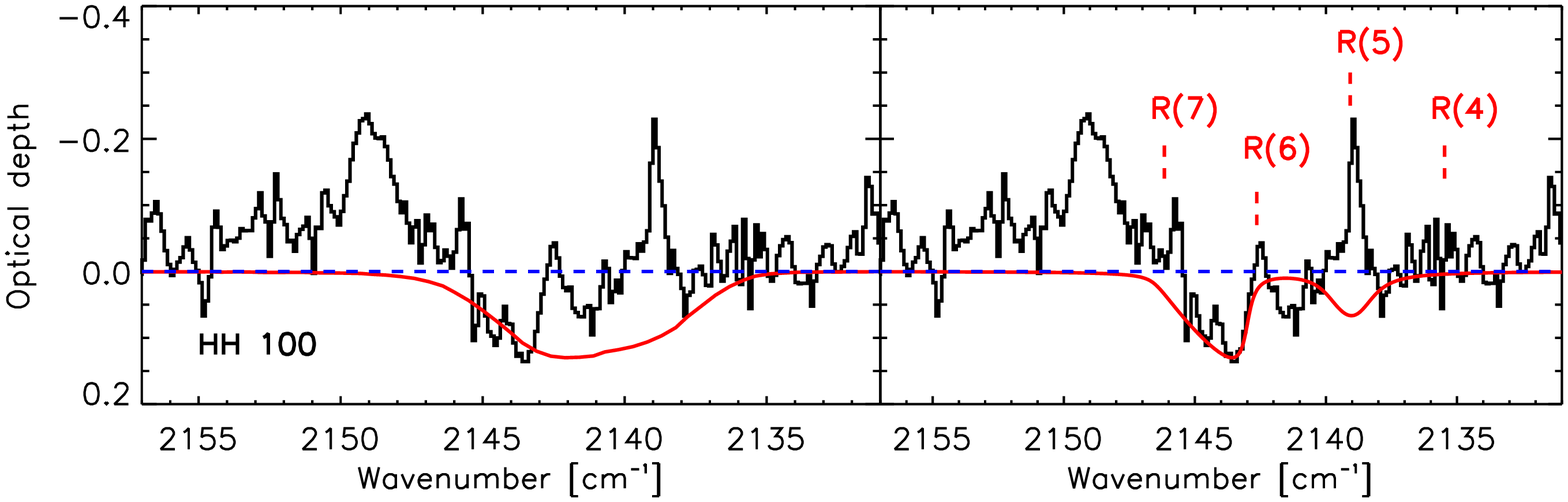}
\includegraphics[width=8.5cm]{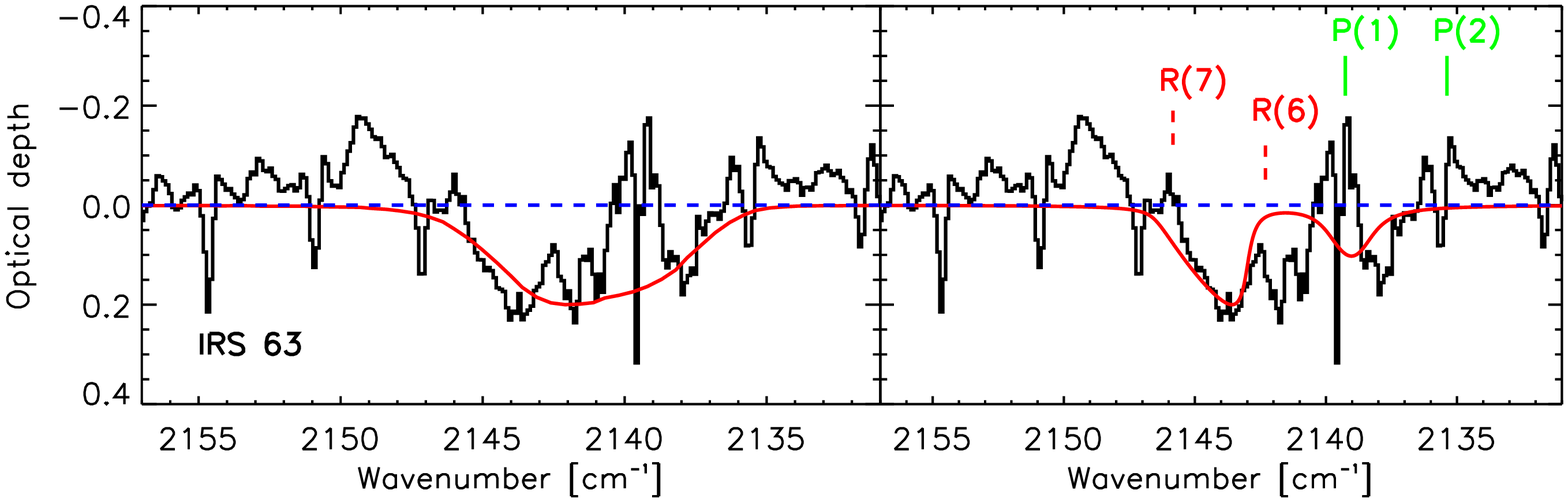}
\includegraphics[width=8.5cm]{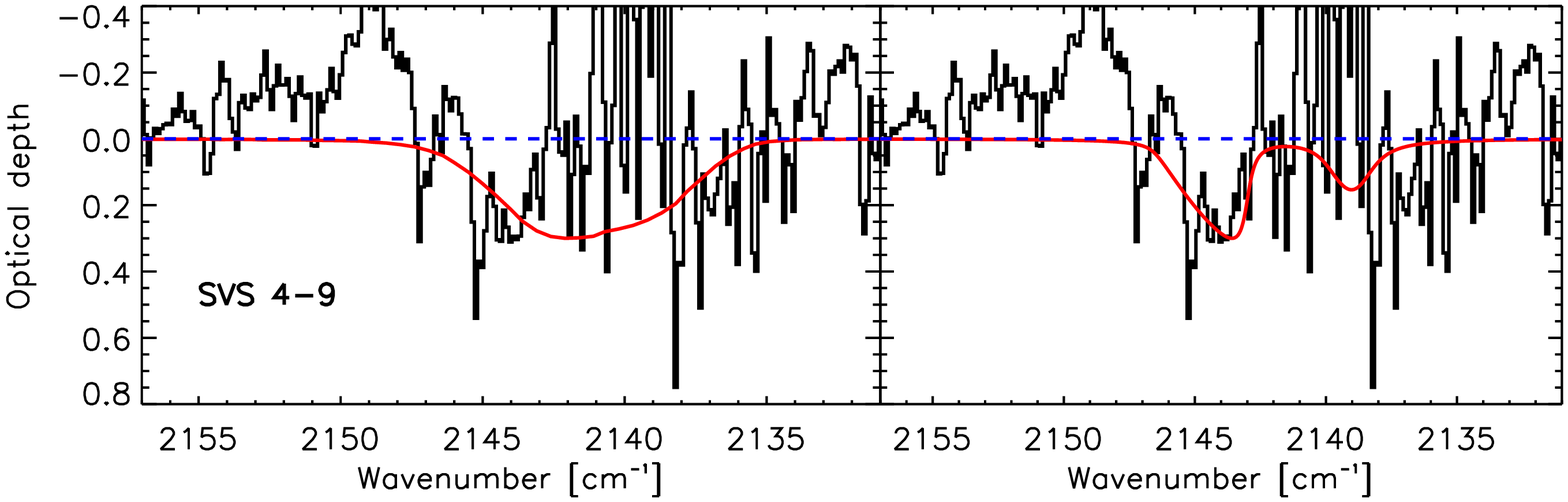}
\includegraphics[width=8.5cm]{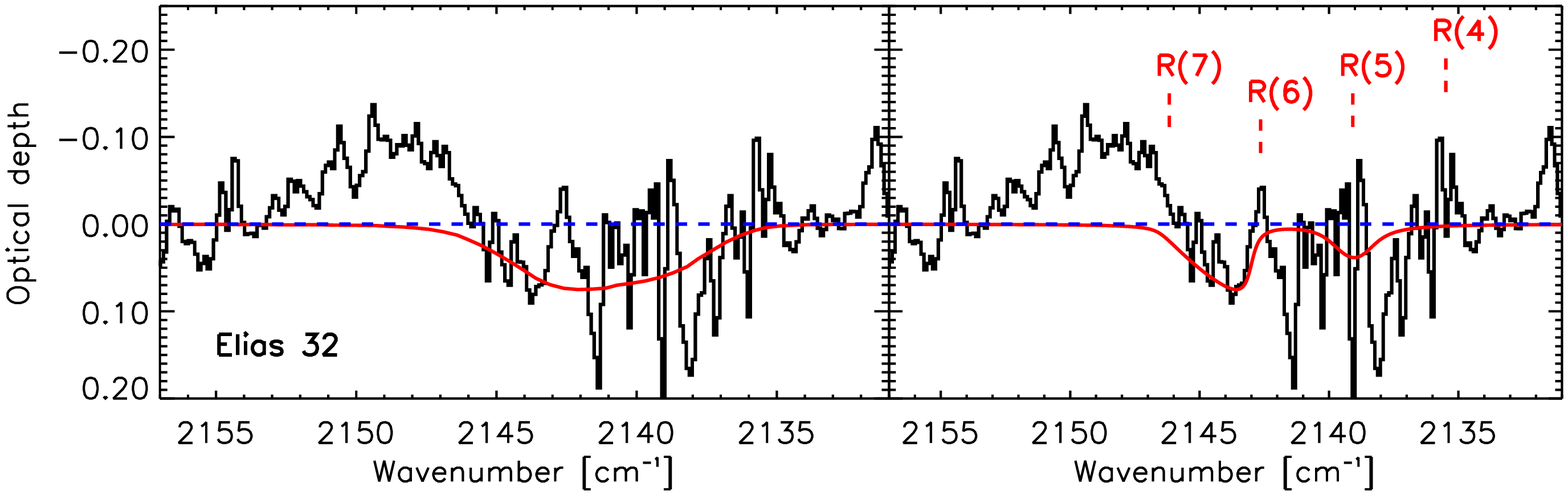}
\caption{Spectra of sources with a clearly defined blue component. The Lorentz oscillator model for the
middle component and the phenomenological red component have been subtracted. The plots on the left
show the spectra compared to the grain-shape corrected laboratory spectrum of $\rm N_2$:$\rm O_2$:$\rm CO_2$:CO=1:5:0.5:1 at 12 K by EAS. The
plots on the right show the grain shape corrected LO-TO splitting model as described in the main text. All grain shape corrections use CDE grains. 
Positions of the gas-phase $\rm ^{12}CO$ v=1-0 (solid lines) and v=2-1 (dashed lines) ro-vibrational transitions are also indicated, when present.
}
\label{LOTOvsCO2}
\end{figure*}

The model fits are shown in Fig. \ref{LOTOvsCO2} where they are compared to the $\rm CO_2$-rich laboratory spectrum proposed by \cite{AdwinL1489}
and EAS to fit the blue component, in particular in the case of L 1489. The derived polarisation fractions
are given in Table \ref{PolFracs}.
For L 1489 and Reipurth 50 the $\rm CO_2$ laboratory profile indeed gives a good fit to the blue wing of the residual. For the other
sources the fit is worse since the residual feature seems to be a factor of 2-3 narrower than the laboratory profile, although this may be an effect of
line emission from hot CO gas as indicated in Fig. \ref{LOTOvsCO2}. In no case does the laboratory mixture give a good fit to the red side.  It may be possible to construct a more complex ice mixture, which fits better. Also a better fit to the red side can be obtained by subtracting a middle component of slightly smaller optical depth. However, since the
slope of the blue wing is similar in all the sources, the exact same complicated mixture is required along all lines of sight, which seems very unlikely.

On the other hand, the extremely narrow
LO mode profile accurately reproduces the blue wing in the residual for all the
sources. Taking into account the line emission from CO gas which clearly fills in the red side of the band in all sources but L 1489 and Reipurth 50, it is seen that also the LO-TO splitting scenario has problems reproducing the red wing.
In L 1489 and Reipurth 50, excess absorption is clearly seen at 2141 and $\rm 2142~cm^{-1}$, as previously mentioned. Nevertheless, the polarisation scenario has several advantages over the $\rm CO_2$-rich mixture scenario. First, since the polarisation fraction at
$\rm 4.7~\mu m$ is predicted, an independent test is available. L 1489 and Reipurth 50 need very high polarisation fractions ($>50\%$)
to explain the blue component. However, these two sources also have some of the highest $K$-band polarisation fractions
known in YSOs, as is indicated in Table \ref{PolFracs}. It is expected that the linear polarisation fraction decreases with
increasing wavelength if more of the light received originates directly from the emitting source rather than being scattered. In these sources,
the polarisation fraction is seen to decrease from the $H$ to the $K$-band and the maximum polarisation fraction occurs offset from
the mid-infrared position \citep{Casali91,Whitney97}. It is not known if this behavior in general continues to $\rm 4.7~\mu m$. 
If the grains are small and the emitting source is obscured at $\rm 4.7~\mu m$ e.g. by a circumstellar disk, the
polarisation fraction can be largely independent of wavelength up to $\rm 5~\mu m$. Although few observations
of polarisation fractions have been made at longer wavelengths than $\rm 2.2~\mu m$, some protoplanetary nebulae have been
found to show this behaviour, such as the Egg nebula, which has a constant linear polarisation fraction of
50\% from 2.2 to $\rm 4.5~\mu m$, thus requiring grain sizes of less than $\rm \sim 1~\mu m$ \citep{Kastner}.
Also, the polarisation fraction at $\rm 4.7~\mu m$ for W 33A is known \citep{Chrys96}. The measured value of 10-13\% is consistent with
an upper limit of 15\% from the decomposition of the ISO-SWS spectrum.
Curves of constant polarisation fraction have been drawn in Fig. \ref{BlueMiddle}, showing that the majority of the sources in the sample
will require polarisation fractions of less than 30\%. 

The main problem in the determination of the polarisation fraction from the CO profile are the systematic
uncertainties introduced by the assumptions on the fits of the middle and red components as well as the assumptions
regarding the dielectric function of LO-TO split $\alpha$-CO, especially concerning the width and strength of the LO component.

All the sources are thus consistent with a LO-TO splitting scenario, but the result must be confirmed by imaging polarimetry at
$\rm 4.7~\mu m$ as well as further laboratory experiments using polarised light and high resolution spectrometers.
It will have important consequences for the understanding of the processes governing the
freeze-out of CO in the circumstellar medium if the blue component, and indeed also the middle component, is carried by crystalline CO.
Note that the formation of crystalline CO can occur at low temperatures, depending on deposition rate and the
adsorbing surface and is not necessarily a sign of processing of the ice \citep{Collings}. 

\section{Discussion}
\label{IceDis}
\subsection{Implications for the evolution and processing of CO-rich ices}
The most intriguing lesson learned from the presented data set is that the fundamental structure of the CO ice on grain mantles
seems to be practically identical along all observed lines of sight. It is
well-established that the absorption profile of solid CO is sensitive to the conditions under which it is formed. This is an observation which  suggests that the CO ices are formed under similar conditions regardless of the macroscopic context of the circumstellar medium or that the CO ice profile is invariant to differences which may exist in composition and grain shape distributions.  

It is well known that CO can efficiently freeze out on a given grain when it passes through a region of a dense cloud
with the temperature and density most suited for the formation of solid CO. In addition, an efficient method of
desorption at low temperature must exist, since CO is found in the gas phase in abundances even in the densest and coldest clouds \cite[e.g.][]{Bergin01,Lada94}.
An important desorption mechanism seems to be via the release of chemical energy through reactions with radicals produced by a cosmic ray induced UV field \citep{Shen03}, although other mechanisms connected with cosmic ray particles may contribute. At the same time ambient UV photons may only play a role on the surface of the cloud due to extinction. A given grain may also experience that all or parts of the ice mantle are evaporated and refrozen several times on the evolutionary pathway from dense cloud to protostellar envelope to
circumstellar disk. 

In the light of this active processing which is expected to take place for any grain mantle, it is highly surprising that we see almost no change in the fundamental three profiles of solid CO for different lines of sight. The observational constraints presented here are thus that any processing, be it chemical or physical, must leave the profiles of the three components invariant but may change the relative intensities. 
Typical lines of sight have 60-90\% of the solid CO in a pure or nearly pure form. Models must be able to explain why this component of the
CO has not experienced any kind of mixing with other species neither during formation nor as a result of subsequent processing. 
The results presented here thus seem to be
consistent with a layered structure of the ices, which at least separate the CO ice from other species.
Furthermore, the association of the red component with a simple $\rm H_2O$-rich ice mixture is problematic. If the red component is indeed associated with
$\rm H_2O$, it must be explained why it has the same profile for the quiescent medium (e.g. CK 2), low mass YSOs (e.g. Elias 32), low mass YSOs in
high mass star forming regions (e.g. TPSC 78), circumstellar disks (CRBR 2422.8) and high mass stars with abundant methanol (e.g. W 33A).
Similar to the case for the middle component, this is unexpected since the profile of the red component should vary with the different temperatures and the different abundances of secondary species known to be associated with the water ice, such as methanol and ammonia, in the varity of circumstellar and interstellar regions surveyed. These differences in temperatures, ice structure and composition
are observed through clear variations of the $\rm 3.08~\mu m$ $\rm H_2O$ band and should be reflected in the shape of the red component as well.

The implication is that solid CO seems to take part in little observable
chemical processing around young stellar objects. 
Only the two sources L 1489 IRS and Reipurth 50 show some evidence for additional environments to the CO molecules
through the excess absorption at 2141-$\rm 2142~cm^{-1}$ and even here the environments is not likely to
have a very high column density compared to the pure CO ice, since the features are so shallow ($\tau\sim 0.2$). Chemical processing
may still take place, but the products of any chemical reactions involving CO must be efficiently removed from the environment of the
remaining CO. 

At the same time the variations in relative intensities of the three different components are consistent with physical processing affecting
the total CO band profile, such as the evaporation of the volatile pure middle CO component prior to a hydrogen-bonding red component as a result of thermal processing. This is
supported by the observation that the red component is detected along every line of sight with solid CO. The presence of an exclusion region in 
Fig. \ref{RedMiddleRat} may also support the mixing of pure CO into the water ice component as discussed in Sec. \ref{Interpretations}.

\subsection{Is the grain shape constrained?}

Even though the astronomical spectra can be so well fitted with the simple physical model presented here, it is
a genuine worry that this is simply due to a degeneracy  similar to that which haunts the mix-and-match approach.
It is not possible to exclude a different combination of dielectric function and grain-shape model. It is, however, far easier to
theoretically explain the common presence of pure CO rather than some specific and complicated mixture. 
Perhaps the most significant evidence for the applicability of a model using irregular grains simulated by a CDE shape distribution, is the evidence given when applying the
same combination of pure CO and grain shape correction to other features. In particular it is shown by \cite{AdwinL1489} and in this
work that the stretching vibration profile $\rm ^{13}CO$ along the line of sight towards two YSOs (one high mass and one low mass)
is also consistent with pure CO independently of the grain shape. Furthermore, we have shown that the blue component, which is possibly the
LO component of crystalline CO, has a shape which is consistent with CDE grains as well. 

Thus, under the assumption that the optical constants of the middle component are known, a strong constraint on
the grain shape is given.
The used CDE grains is not a realistic model for interstellar grains. The CDE model assumes grains of solid ice and includes grain shapes like thin needles and disks, which are not likely to be formed. However, they are known to simulate irregular grains well \citep{BH} and thus provide a simple mathematical expression for this purpose. One possible scenario is that the pure CO exists as small irregular clumps on top or inside the water ice mantles. This may unify the observational constraints given by the CO band with the constraints on the icy grain size distribution given by the $\rm 3.08~\mu m$ water band \citep{Manu02}.   However, detailed calculations of the grain shape effects of
simulated interstellar grains using for instance fractal models \citep{Fogel} are necessary to confirm the conjecture
that CDE grains are appropriate for an irregularly shaped CO ice mantle. To further refine the adopted grain shape correction, additional
solid state features from other species must be included. Ideally, it must be demanded that all observed solid state features are consistent
with the same grain shape correction. However, since solid CO supplies one of the simplest and most easily observable bands, it likely provides the
best single constraint. Other features which can constrain the grain shape correction further include 
the $\rm CO_2$ stretching vibration band at $\rm 4.27~\mu m$ and the $\rm CO_2$ bending mode at $\rm 15.2~\mu m$.
Also, having shown that a simple formula for the absorption cross section can
reproduce the observed middle components, a convenient template is provided against which more detailed models of grain shapes can be tested.

\subsection{Strategies for comparison with solid CO laboratory data}
\label{LabStrategy}

The presented data provide new challenges for laboratory studies of CO-rich ice mantles and some technical requirements become
evident. The spectral features distinguishing the solid CO profiles of different sources show structures with widths less than $\rm 1~cm^{-1}$
and much of the modeling requires high quality laboratory spectra to investigate e.g. the structure of LO-TO split crystalline CO, which can
have widths as small as $\rm 0.2~cm^{-1}$. Accurate determinations of optical constants require laboratory spectra which are fully resolved.
Thus, to match the present resolution of astronomical spectra, it is essential for the further study of interstellar solid CO to have laboratory data with a spectral resolving power of at least $\rm 0.1~cm^{-1}$.

The accurate astronomical CO ice profiles encourage a change of strategy when comparing observed solid state profiles to laboratory data. As mentioned in Sec. \ref{pheno} a mix-and-match approach is often employed to analyse the observed CO ice profiles. However, due to the many parameters governing a given laboratory simulation 
(abundances of mixture constituents, deposition and annealing temperatures, irradiation parameters and the details of the laboratory setup)
and given the low spectral resolution of the available laboratory spectra, degeneracies are introduced when trying to constrain the compositions of CO-rich ices in space using a mix-and-match approach. As has been shown in \cite[e.g.][]{AdwinL1489,Manu02} and this work, both the quality and quantity of astronomical near to mid-infrared spectroscopy is now sufficient that more advanced physical models can be applied to
the shape of the observed solid state features. In particular, this requires on the laboratory side high resolution optical constants, but also detailed
experimental and theoretical studies of the microscopic structure of simple ices under interstellar conditions. On the astrophysical side, it is evident that solid state features in general also probe non-chemical phenomena in addition to the chemistry such as grain shape, size distributions
and the overall temperature and density structure of the interstellar, circumstellar and disk material, which contain dust. The macroscopic properties of the objects observed are thus inseparable from the chemical properties of the ice mantles and must be modeled concurrently. 
Ideally, a single dust model including grain shapes and sizes, physical and chemical structure of the grain mantles and a macroscopic structure of the
material along a given line of sight should be used to ensure consistency between the interpretations of different solid state features. 

\section{Conclusions and future work}
\label{Concl}
Medium resolution $M$-band spectroscopy of a large sample of low mass young stellar objects has been used to explore the line profiles of the $\rm 4.67~\mu m$ stretching vibration mode of solid CO. A simple phenomenological three-component decomposition reveals
a remarkable similarity of the structure of the CO stretching vibration mode at $\rm 2140~cm^{-1}$ between all observed lines of sight. 
A simple physical model is presented, which can reproduce the shape of all profiles to a high degree of accuracy.
\begin{itemize}
\item The phenomenological decomposition of the profile is found to provide good fits for all observed sources. The decomposition
uses three different components: a red Lorentzian profile centered on $\rm 2136.5~cm^{-1}$, a middle Gaussian profile centered on $\rm 2139.9~cm^{-1}$ and a blue Gaussian profile centered on $\rm 2143.7~cm^{-1}$. Also the widths of the three components are kept constant
such that only the relative intensities are varied between the sources. This approach reduces the information contained in
a single line of sight to three linear parameters.
\item In 30 of the 39 sources a broad and shallow absorption band is detected centered in the spectral range 2165--$\rm 2180~cm^{-1}$. The central optical depth of the band is found to correlate with the optical depth of the red component of the CO ice band for low mass YSOs only. Higher mass sources
tend to show excess absorption at $\rm 2165~cm^{-1}$.  We propose that in addition to the traditional XCN band a second
weak absorber centered at $\rm 2175~cm^{-1}$ is present. Furthermore, due to the correlation with the red CO component, 
the  $\rm 2175~cm^{-1}$ band may be due to CO in a new, unidentified binding site.
\item A simple physical model of the dielectric functions of the CO ice using Lorentz oscillators is described. It is shown that the middle component of the CO ice can be entirely explained along all lines of sight by modeling pure CO with a single Lorentz oscillator and grain-shape correcting the profile
with a continuous distribution of ellipsoids (CDE). We suggest, that the CDE particles may work so well, because they simulate irregular grain surfaces or small irregularly shaped CO clumps on top of the water ice mantle. The same model suggests that the different optical constants of pure CO obtained from laboratory experiments vary only due to a varying oscillator density (or porosity) of the ice. A variation of ice porosity in the
lab can be caused by slightly different experimental setups.  The observed middle components show no direct evidence for
the presence of contaminating species such as $\rm N_2$, $\rm O_2$ and $\rm CO_2$ mixed with the carrier of the van-der-Waals interacting component and their concentration is likely less than 10\%.
\item The blue component can be explained by the LO mode of pure $\alpha$-CO, which appears when the background source is
linearly polarised, although it cannot be ruled out that the component is carried by a $\rm CO_2$-rich ice. If the blue component is due to the LO mode of crystalline CO, the polarisation fraction at $\rm 4.7~\mu m$ is predicted in a number of sources. The derived polarisation
fractions seem to correlate well with measured $K$-band polarisation fractions from the literature. The distribution of the relative contributions of the middle and blue components support the polarisation scenario. An important implication of LO-TO split CO is that a large fraction of the pure interstellar CO ice must be crystalline. This would put strong constraints on the processes governing the formation and structure of solid CO in space. 
\item The red component does not fit well with any simple $\rm H_2O$-rich laboratory mixture due to the total absence of the
$\rm 2152~cm^{-1}$  feature, which is due to CO-OH bonds. Although thermal and energetic processing tend to destroy
the CO-OH bonds, it also irreversibly changes the shape of the red component. Since CO adsorbed on many types of surfaces can produce a profile similar to the red component, it is suggested that alternative candidates which can explain the correlation with the
$\rm 2175~cm^{-1}$ are tested. Conversely, statistics of the relative
contributions of the middle and red components support a scenario where pure CO migrates into a porous water ice upon warm-up. This scenario
also predicts a strong $\rm 2152~cm^{-1}$ feature. There is thus a significant discrepancy between the data and all interpretations of the red component, which may require additional laboratory studies to solve.
\end{itemize}

A number of observations and laboratory experiments are necessary to both confirm and to elaborate on the results presented here.
Further modeling will also significantly improve the understanding of solid state features of simple interstellar ices.
\begin{itemize}
\item Modeling of the expected ice mantle structure using existing dynamical models of collapsing protostars are necessary
to theoretically understand the degree of mixing of the principal ice species which can be expected to exist at different stages of the
star formation process. Useful output from such a model would include the detailed composition of the ice as a function of mantle depth
for a given grain. Clearly, such a model depends on the physical and chemical processes governing the adsorption of ice on surfaces, the
surface chemistry of the grain, but also on the time the grain spends in different physical environments (density and temperature)
during its lifetime.
\item Calculations of 'realistic' grain shape effects are required to explore which types of irregular grain surfaces can produce a grain shape correction similar to CDE particles. 
\item UHV laboratory spectroscopy with resolving powers of at least $\rm 0.1~cm^{-1}$ is needed to confirm the adopted dielectric functions for LO-TO split $\alpha$-CO, to derive optical constants for other relevant CO-rich ices with narrow absorption profiles and to
further explore the effect of polarised light on astrophysical ices.
\item Future work should also include comparisons of other solid state bands of sources in the same sample such as $\rm H_2O$ and $\rm CO_2$ with the results presented in this work. A physical model of the protostellar envelopes surrounding the sources using gas phase studies of rotational lines is required to fully understand the context of the observed ices and to separate the possible contribution from foreground material.

\end{itemize}

\begin{acknowledgements}
The authors wish to thank the ISAAC staff including Chris Lidman, Gianni Marconi, Olivier Marco, Rachel Johnson, Andreas Jaunsen and Vanessa Doublier for their help and assistance over several years. We are grateful to Fred Lahuis for providing us with his reduced ISO-SWS spectra. The referee, T. Nagata, is thanked for comments which helped to improve the quality of the manuscript.  This research was supported by the Netherlands Organization for Scientific Research (NWO) grant 614.041.004, the Netherlands Research School for Astronomy (NOVA) and a NWO Spinoza grant.
\end{acknowledgements}

\appendix

\section{Solid state line shapes}

\label{lineshapes}
We review here the physical argument for using Lorentz oscillators as a first order approximation to the 
dielectric functions of simple solid state species.
The formalism presented is mostly following \cite{Gadzuk}.

Common for both gas-phase and solid state lines is that any line shape from a spectroscopic transition is fundamentally the Fourier transform
of the auto-correlation function of the relevant time dependent physical variable, $q(t)$.

\begin{equation}
I(\omega) = \int_{\infty}^{\infty}\exp(-i\omega (t-t_0)) \langle q(t_0)q^*(t-t_0)\rangle dt.
\end{equation}

This can be conveniently expressed with the Fourier transform $Q(\omega) = F(q(t),\omega)$, via the convolution theorem by
using that the auto-correlation function is the convolution of $q(t)$ with itself:

\begin{eqnarray}
\lefteqn{I(\omega) = F(F^{-1}(F(q,\omega) \times F^*(q,\omega))) {} }
\nonumber \\
 & & {} = Q(\omega) \times Q(\omega)^*.
\label{autocorr}
\end{eqnarray}

The relevant solution for the pure harmonic oscillator is:

\begin{equation}
q(t) = q_0 \exp (i\omega_0 t - \eta t),
\end{equation}
where $\omega_0$ is the frequency of the oscillator and $\eta$ is the dampening factor determined by the
finite decay time of the transition modeled by the oscillator.

It is well known that this solution, when inserted into eq. \ref{autocorr} yields a Lorentzian line
profile, which collapses into a delta function when the oscillator is not damped, i.e. $\eta=0$.  Any model
specific for solid state applications must take the interaction of the oscillator with a
background surface into account. A picture which is often used, is that the oscillator interacts via elastic collisions with
a thermally fluctuating background of atoms or molecules. Every elastic collision will change the phase of the oscillator with
a random phase shift. Such a phase shift destroys the correlation between the oscillator coordinate before and
after the collision. If the collisions are random, but occur with a certain probability, an exponential correlation decay
is introduced and the correlation function in eq. \ref{autocorr} is modulated with a factor $\exp(-t/\tau_c)$, where
$\tau_c$ is the average lifetime of the oscillator before a collision. It is straightforward to show that
such a scenario will again result in a Lorentzian line shape broadened by $1/\tau_c$.

The effect is called de-phasing and is an often used concept in vibrational spectroscopy. Naturally
higher order effects may produce more complex line shapes, of which the best known examples include {\it non-homogenous}
broadening caused by non-random intermolecular distances. E.g. low surface coverage is known to have
profound effects on the profile of CO adsorbed on metals \citep{Somorjai94}. However, more advanced models may still
be result in basic Lorentzian profiles modified with frequency-dependent width functions as in \cite{Fano} and \cite{Kubo}.
We conclude that the use of harmonic oscillators and Lorentzian profiles provide a physically sound
{\it starting point} for the theoretical study of the structure of interstellar ices using simple molecules
such as CO as environmental probes.

\section{Notes on individual sources}
\label{Comments}
\subsection{Ophiuchus}
\paragraph{IRS 43} is the closest source to the edge-on disk CRBR2422.8-3423 \citep{ThiCRBR}. It has a reasonably deep ice band
and shows broad ro-vibrational lines in emission from hot CO gas.
\paragraph{GSS 30 IRS1} is associated with a large well-studied reflection nebulosity. Almost no CO ice is detected along the line of sight towards this source although it is located in the most embedded region of the $\rho$ Ophiuchi clouds according to the extinction map
by \cite{Cambresy99} ($A_V>10$). The $M$-band spectrum of this source is distinguished from other sources by showing very strong
ro-vibrational lines in emission from gaseous CO \citep{Pontoppidan}.

\subsection{Serpens}
\begin{figure}
\centering
\includegraphics[width=8.5cm]{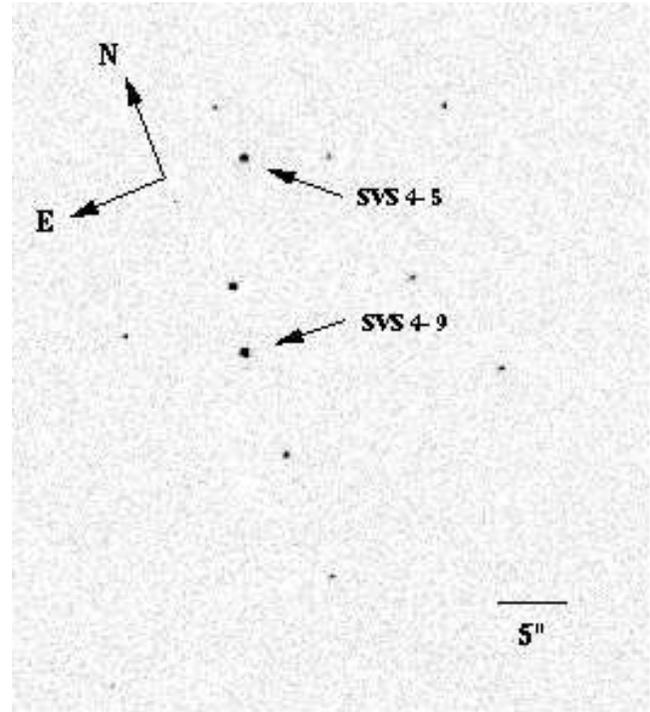}
\caption{$\rm 4~\mu m$ acquisition imaging of the SVS 4 cluster of young stellar objects in Serpens indicating the location of
the two observed sources. The seeing is measured to $\rm 1.7~pixels = 0\arcsec25$.  }
\label{SVS4}
\end{figure}

\paragraph{EC 90} also known as CK 1 or SVS 20 is a $1.6\arcsec$ binary young star, known to show signs of outflow activity \citep{Huard}. The CO
ice band has previously been observed at low resolution by \cite{Chiar94}, but without resolving the binary.
Solid $\rm CO_2$ ($\tau=2.6$) at $\rm 4.27~\mu m$ has been detected by \cite{Guertler}. We
obtained well separated spectra of both components showing deep CO ice in both sources as well as cold CO gas. EC 90B additionally shows
hot CO gas phase lines blueshifted by $\rm \sim 100~km~s^{-1}$ showing that the outflow activity is probably associated with this component.
The depth of the CO bands are $\tau=2.1$ and $\tau=1.2$ for EC 90A and EC 90B north, respectively. This shows that the distribution of CO ice
around young stars can vary significantly on physical scales of a few hundred AU.
\paragraph{SVS 4-5 and 9:}
SVS 4 is a dense cluster of YSOs near the Serpens cloud core. The cluster is so dense, that it has often been confused with a single
source. It was first resolved by \cite{EC89}, who counted 11 bright sources in the cluster.
Our acquisition image obtained under exceptional seeing conditions (see Fig. \ref{SVS4}) confirms the number of sources and shows
that no fainter sources are present to a limit of $M_{\rm 4~\mu m}=11$.
Low resolution $M$-band spectra of the two brightest $\rm 2~\mu m$ sources have been previously obtained by \cite{Chiar94} who observed SVS 4-9
and SVS 4-10, 5\arcsec to the north. We
obtained simultaneous spectra of the two brightest $\rm 4~\mu m$ sources as indicated in Fig. \ref{SVS4}, the southern source, SVS 4-9, being the same as observed by \cite{Chiar94} while the northern source, SVS 4-5, is new.

Deep CO ice is detected in both sources. The northern source has a typical narrow CO band, while the ice in the southern source shows a
much broader feature consistent with the spectrum from the literature.

\subsection{Chameleon}
These are the first ground-based observations of ices in the molecular cloud complex in Chameleon. The observations are difficult due to
the extremely southern location of the cloud ($\delta \sim -77\degr$). Consequently the sources never rise above an airmass of 1.7 as
seen from Paranal, which makes proper correction for telluric features difficult. Additionally, gas phase observations are
near impossible since the velocity shift compared to the telluric lines will be very small.
\paragraph{Cha INa 2} (source 1 in Table 2 of \cite{Persi99}) is the only source with near-infrared excess detected by ISOCAM in the Chameleon I North a cloud. It has a class I SED and is surrounded by a small reflection nebula. It is estimated that the source is only extinct by 17 magnitudes \citep{Persi99}. The source may be creating the large CO outflow observed in the area. This is one of the few sources which has a CO ice profile showing only a contribution from the red component. 

\subsection{Corona Australis}
\paragraph{RCrA IRS 7 A and B}

\begin{figure}[ht]
\centering
\includegraphics[width=8.5cm]{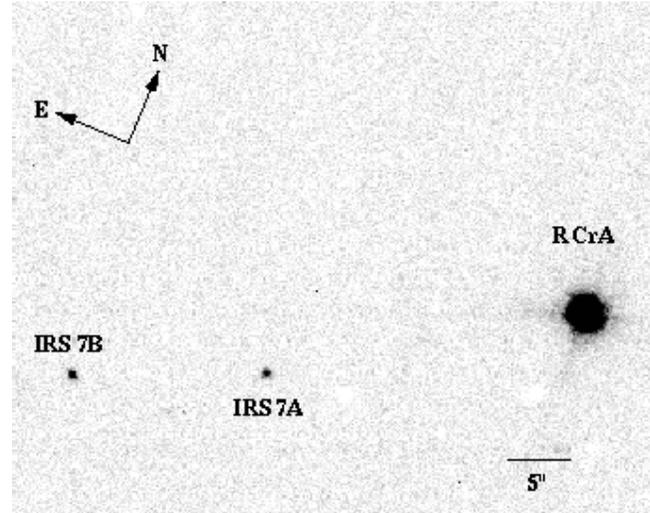}
\caption{$Mnb$ image of the region around RCrA IRS 7 with the sources A and B indicated. Neither IRS 7A nor B is detected at $\rm 3.21~\mu m$ due to the presence of a deep water band. IRS 7B is detected in the $K$-band as a compact, but resolved reflection nebula.The positions of the two sources are IRS 7A: $\alpha = 19~01~55.16$, $\delta = -36~57~20.9$ and IRS 7B: $\alpha = 19~01~56.25$, $\delta = -36~57~26.9$, J2000. }
\label{RCRA7}
\end{figure}

There is some confusion in the literature regarding the identification of IRS 7 in the CrA cloud. At 2 and 6 cm the cloud is dominated by a strong
binary radio source with a separation of $14\farcs2$ \citep{Brown87}. It has been suggested that the two radio sources represents a strong bipolar outflow, since no obvious near-infrared counterparts to the radio sources were known. We detect two bright point sources at $\rm 4.8~\mu m$ located
at the positions of the radio sources, and can therefore confirm that it is unlikely that the radio emission is from a bipolar outflow from a single
source. This was also found by \cite{Wilking97} who imaged the source at $\rm 10~\mu m$ and detected IRS 7A, but not IRS 7B, perhaps indicating that IRS 7B has a very deep $\rm 9.7~\mu m$  silicate absorption feature. $\rm 2-5~\mu m$ spectra of IRS 7 was obtained by \cite{Tanaka94}, but it is unclear which source was observed due to the confusing $K$-band field. A narrow band $M$  VLT-ISAAC
image obtained at a seeing of $0\farcs35$ is shown in Fig. \ref{RCRA7}.

\subsection{Orion}
\paragraph{TPSC 1 and TPSC 78} are two extremely red sources located in the trapezium cluster within 2\arcmin of the
BN/KL region, with $K-L > 6.1$ and 4.85, respectively \cite{TPSC}. Their $M$-band spectra show very broad CO ice bands with the narrow
component almost entirely missing as well as strong gas phase absorption.
Also evident in both sources is strong absorption in the CN stretch region around $\rm 4.62~\mu m$. The ices are likely to have been
affected by the close proximity to young high mass stars.
\paragraph{Reipurth 50 IRS} is an intermediate-mass source with a large associated reflection nebula similar to that of Cha IRN, GSS 30 IRS 1 or L 1489. The source and the VLT-ISAAC spectroscopy is described in detail in \cite{Manu03}.

\bibliographystyle{aa}
\bibliography{ms3823}

\begin{thebibliography}{77}
\expandafter\ifx\csname natexlab\endcsname\relax\def\natexlab#1{#1}\fi

\bibitem[{Baratta \& Palumbo(1998)}]{BP}
Baratta, G.~A. \& Palumbo, M.~E. 1998, J.Opt.Soc.Am, 15, 3076

\bibitem[{Bergin {et~al.}(2001)Bergin, Ciardi, Lada, Alves, \& Lada}]{Bergin01}
Bergin, E.~A., Ciardi, D.~R., Lada, C.~J., Alves, J., \& Lada, E.~A. 2001,
  \apj, 557, 209

\bibitem[{Bohren \& Huffman(1983)}]{BH}
Bohren, C.~F. \& Huffman, D.~R. 1983, Absorption and scattering of light by
  small particles (Wiley-Interscience)

\bibitem[{Bontemps {et~al.}(2001)Bontemps, Andr{\'e}, Kaas, Nordh, Olofsson,
  Huldtgren, Abergel, Blommaert, Boulanger, Burgdorf, Cesarsky, Cesarsky,
  Davies, Falgarone, Lagache, Montmerle, P{\'e}rault, Persi, Prusti, Puget, \&
  Sibille}]{Bontemps}
Bontemps, S., Andr{\'e}, P., Kaas, A.~A., {et~al.} 2001, \aap, 372, 173

\bibitem[{Boogert {et~al.}(2002{\natexlab{a}})Boogert, Blake, \&
  Tielens}]{Adwin13CO}
Boogert, A. C.~A., Blake, G.~A., \& Tielens, A. G. G.~M. 2002{\natexlab{a}},
  \apj, 577, 271

\bibitem[{Boogert {et~al.}(2000{\natexlab{a}})Boogert, Ehrenfreund, Gerakines,
  Tielens, Whittet, Schutte, van Dishoeck, de~Graauw, Decin, \&
  Prusti}]{Adwin13CO2}
Boogert, A. C.~A., Ehrenfreund, P., Gerakines, P.~A., {et~al.}
  2000{\natexlab{a}}, \aap, 353, 349

\bibitem[{Boogert {et~al.}(2002{\natexlab{b}})Boogert, Hogerheijde, \&
  Blake}]{AdwinL1489}
Boogert, A. C.~A., Hogerheijde, M.~R., \& Blake, G.~A. 2002{\natexlab{b}},
  \apj, 568, 761

\bibitem[{Boogert {et~al.}(2000{\natexlab{b}})Boogert, Tielens, Ceccarelli,
  Boonman, van Dishoeck, Keane, Whittet, \& de~Graauw}]{AdwinElias29}
Boogert, A. C.~A., Tielens, A., Ceccarelli, C., {et~al.} 2000{\natexlab{b}},
  \aap, 360, 683

\bibitem[{Brown(1987)}]{Brown87}
Brown, A. 1987, \apj, 322, L31

\bibitem[{Cambr{\'e}sy(1999)}]{Cambresy99}
Cambr{\'e}sy, L. 1999, \aap, 345, 965

\bibitem[{Casali(1991)}]{Casali91}
Casali, M.~M. 1991, \mnras, 248, 229

\bibitem[{Casali \& Eiroa(1996)}]{CE96}
Casali, M.~M. \& Eiroa, C. 1996, \aap, 306, 427

\bibitem[{Chang {et~al.}(1988)Chang, Richardson, \& Ewing}]{Chang}
Chang, H.-C., Richardson, H.~H., \& Ewing, G.~E. 1988, J. Chem. Phys., 89, 7561

\bibitem[{Chen {et~al.}(1997)Chen, Grenfell, Myers, \& Hughes}]{Chen97}
Chen, H., Grenfell, T.~G., Myers, P.~C., \& Hughes, J.~D. 1997, \apj, 478, 295

\bibitem[{Chen {et~al.}(1995)Chen, Myers, Ladd, \& Wood}]{Chen95}
Chen, H., Myers, P.~C., Ladd, E.~F., \& Wood, D. O.~S. 1995, \apj, 445, 377

\bibitem[{Chiar {et~al.}(1994)Chiar, Adamsom, Kerr, \& Whittet}]{Chiar94}
Chiar, J.~E., Adamsom, A.~J., Kerr, T.~H., \& Whittet, D. C.~B. 1994, \apj,
  426, 240

\bibitem[{Chiar {et~al.}(1995)Chiar, Adamson, Kerr, \& Whittet}]{Chiar95}
Chiar, J.~E., Adamson, A.~J., Kerr, J., \& Whittet, D. C.~B. 1995, \apj, 455,
  234

\bibitem[{Chiar {et~al.}(1998)Chiar, Gerakines, Whittet, Pendleton, Tielens,
  Adamson, \& Boogert}]{Chiar98}
Chiar, J.~E., Gerakines, P.~A., Whittet, D. C.~B., {et~al.} 1998, \apj, 498,
  716

\bibitem[{Chrysostomou {et~al.}(1996)Chrysostomou, Hough, Whittet, Aitken,
  Roche, \& Lazarian}]{Chrys96}
Chrysostomou, A., Hough, J.~H., Whittet, D. C.~B., {et~al.} 1996, \apj, 465,
  L61

\bibitem[{Collings {et~al.}(2003)Collings, Dever, Fraser, McCoustra, \&
  Williams}]{Collings}
Collings, M.~P., Dever, J.~W., Fraser, H.~J., McCoustra, M. R.~S., \& Williams,
  D.~A. 2003, \apj, 583, 1058

\bibitem[{Dartois {et~al.}(2002)Dartois, d'Hendecourt, Thi, Pontoppidan, \& van
  Dishoeck}]{Manu02}
Dartois, E., d'Hendecourt, L., Thi, W., Pontoppidan, K.~M., \& van Dishoeck,
  E.~F. 2002, \aap, 394, 1057

\bibitem[{Dartois {et~al.}(2003)Dartois, Thi, Pontoppidan, d'Hendecourt,
  Schutte, \& van Dishoeck}]{Manu03}
Dartois, E., Thi, W.-F., Pontoppidan, K.~M., {et~al.} 2003, \aap, submitted

\bibitem[{Devlin(1992)}]{Devlin}
Devlin, J.~P. 1992, J. Phys. Chem., 96, 6185

\bibitem[{Ehrenfreund {et~al.}(1997)Ehrenfreund, Boogert, Gerakines, Tielens,
  \& van Dishoeck}]{Pascale}
Ehrenfreund, P., Boogert, A. C.~A., Gerakines, P.~A., Tielens, A. G. G.~M., \&
  van Dishoeck, E.~F. 1997, \aap, 328, 649

\bibitem[{Eiroa \& Casali(1989)}]{EC89}
Eiroa, C. \& Casali, M.~M. 1989, \aap, 223, 17

\bibitem[{Elsila {et~al.}(1997)Elsila, Allamandola, \& Sandford}]{EAS}
Elsila, J., Allamandola, L.~J., \& Sandford, S.~A. 1997, \apj, 479, 818

\bibitem[{Evans(2003)}]{Evans03}
Evans, N.~J. 2003, PASP, in press

\bibitem[{Fano(1961)}]{Fano}
Fano, U. 1961, Phys. Rev., 124, 1866

\bibitem[{Fogel \& Leung(1997)}]{Fogel}
Fogel, M.~E. \& Leung, C.~M. 1997, \apj, 501, 175

\bibitem[{Gadzuk(1987)}]{Gadzuk}
Gadzuk, J.~W. 1987, in Vibrational spectroscopy of molecules on surfaces
  (Plenum Press), 49

\bibitem[{Gerakines {et~al.}(1995)Gerakines, Schutte, Greenberg, \& van
  Dishoeck}]{Gerakines95}
Gerakines, P.~A., Schutte, W.~A., Greenberg, J.~M., \& van Dishoeck, E.~F.
  1995, \aap, 296, 810

\bibitem[{Gillett \& Forrest(1973)}]{Gillett}
Gillett, F.~C. \& Forrest, W.~J. 1973, \apj, 179, 483

\bibitem[{G{\"u}rtler {et~al.}(1996)G{\"u}rtler, Henning, K{\"o}mpe, Pfau,
  Kr{\"a}tschmer, \& Lemke}]{Guertler}
G{\"u}rtler, J., Henning, T., K{\"o}mpe, C., {et~al.} 1996, \aap, 315, L189

\bibitem[{Hagen {et~al.}(1979)Hagen, Allamandola, \& Greenberg}]{HAG}
Hagen, W., Allamandola, L.~J., \& Greenberg, J.~M. 1979, \apss, 65, 215

\bibitem[{Huard {et~al.}(1997)Huard, Weintraub, \& Kastner}]{Huard}
Huard, T.~L., Weintraub, D.~A., \& Kastner, J.~H. 1997, \mnras, 290, 598

\bibitem[{Hudson {et~al.}(2001)Hudson, Moore, \& Gerakines}]{Hudson}
Hudson, R.~L., Moore, M.~H., \& Gerakines, P.~A. 2001, \apj, 550, 1140

\bibitem[{Kastner {et~al.}(2002)Kastner, Jingquiang, Siebenmorgen, \&
  Weintraub}]{Kastner}
Kastner, J.~H., Jingquiang, L., Siebenmorgen, R., \& Weintraub, D.~A. 2002,
  \aj, 123, 2658

\bibitem[{Keane {et~al.}(2001)Keane, Tielens, Boogert, Schutte, \&
  Whittet}]{Keane01}
Keane, J.~V., Tielens, A. G. G.~M., Boogert, A. C.~A., Schutte, W.~A., \&
  Whittet, D. C.~B. 2001, \aap, 376, 254

\bibitem[{Kenyon {et~al.}(1993)Kenyon, Calvet, \& Hartmann}]{KCH}
Kenyon, S.~J., Calvet, N., \& Hartmann, L. 1993, \apj, 414, 676

\bibitem[{Kerr {et~al.}(1993)Kerr, Adamson, \& Whittet}]{Kerr93}
Kerr, T.~H., Adamson, A.~J., \& Whittet, D. C.~B. 1993, \mnras, 262, 1047

\bibitem[{Kubo(1969)}]{Kubo}
Kubo, R. 1969, Adv. Chem. Phys., 15, 101

\bibitem[{Lacy {et~al.}(1984)Lacy, Baas, Allamandola, Persson, McGregor,
  Lonsdale, Geballe, \& van~de Bult}]{Lacy}
Lacy, J.~H., Baas, F., Allamandola, L.~J., {et~al.} 1984, \apj, 276, 533

\bibitem[{Lada {et~al.}(1994)Lada, Lada, Clemens, \& Bally}]{Lada94}
Lada, C.~J., Lada, E.~A., Clemens, D.~P., \& Bally, J. 1994, \apj, 429, 694

\bibitem[{Lada {et~al.}(2000)Lada, Muench, Haisch~Jr., Lada, Alves, Tollestrup,
  \& Willner}]{TPSC}
Lada, C.~J., Muench, A.~A., Haisch~Jr., K.~E., {et~al.} 2000, \aj, 120, 3162

\bibitem[{Langer \& Penzias(1993)}]{Langer93}
Langer, W.~D. \& Penzias, A.~A. 1993, \apj, 408

\bibitem[{Liseau {et~al.}(1992)Liseau, Lorenzetti, Nisini, Spinoglio, \&
  Moneti}]{LLN}
Liseau, R., Lorenzetti, D., Nisini, B., Spinoglio, L., \& Moneti, A. 1992,
  \aap, 265, 577

\bibitem[{Manca {et~al.}(2001)Manca, Martin, Allouche, \& Roubin}]{Manca01}
Manca, C., Martin, C., Allouche, A., \& Roubin, P. 2001, J. Phys. Chem. B, 105,
  12861

\bibitem[{Mathis {et~al.}(1977)Mathis, Rumpl, \& Nordsieck}]{MRN}
Mathis, J.~S., Rumpl, W., \& Nordsieck, K.~H. 1977, \apj, 217, 425

\bibitem[{Novozamsky {et~al.}(2001)Novozamsky, Schutte, \& Keane}]{Novozamsky}
Novozamsky, J.~H., Schutte, W.~A., \& Keane, J.~V. 2001, \aap, 379, 588

\bibitem[{Palumbo(1997)}]{Palumbo}
Palumbo, M.~E. 1997, J. Phys. Chem., 101, 4298

\bibitem[{Pendleton {et~al.}(1999)Pendleton, Tielens, Tokunaga, \&
  Bernstein}]{Pendleton}
Pendleton, Y.~J., Tielens, A. G. G.~M., Tokunaga, A.~T., \& Bernstein, M.~P.
  1999, \apj, 513, 294

\bibitem[{Persi {et~al.}(2001)Persi, Marenzi, G{\'o}mez, \& Olofsson}]{Persi01}
Persi, P., Marenzi, A.~R., G{\'o}mez, M., \& Olofsson, G. 2001, \aap, 907

\bibitem[{Persi {et~al.}(1999)Persi, Marenzi, Kaas, Olofsson, L., \&
  Roth}]{Persi99}
Persi, P., Marenzi, A.~R., Kaas, A.~A., {et~al.} 1999, \aj, 117, 439

\bibitem[{Pontoppidan {et~al.}(2002)Pontoppidan, Sch{\"o}ier, van Dishoeck, \&
  Dartois}]{Pontoppidan}
Pontoppidan, K.~M., Sch{\"o}ier, F.~L., van Dishoeck, E.~F., \& Dartois, E.
  2002, \aap, 393, 585

\bibitem[{Sandford \& Allamandola(1988)}]{SA88}
Sandford, S.~A. \& Allamandola, L.~J. 1988, Icarus, 76, 201

\bibitem[{Sandford {et~al.}(1988)Sandford, Allamandola, Tielens, \&
  Valero}]{Sandford}
Sandford, S.~A., Allamandola, L.~J., Tielens, A. G. G.~M., \& Valero, G.~J.
  1988, \apj, 329, 498

\bibitem[{Schmitt {et~al.}(1989)Schmitt, Greenberg, \& Grim}]{Schmitt89}
Schmitt, B., Greenberg, J.~M., \& Grim, R. J.~A. 1989, \apjl, 340, L33

\bibitem[{Schutte \& Greenberg(1997)}]{SG}
Schutte, W.~A. \& Greenberg, J.~M. 1997, \aap, 317, L43

\bibitem[{Shen {et~al.}(2003)Shen, Greenberg, Schutte, \& van
  Dishoeck}]{Shen03}
Shen, C.~J., Greenberg, J.~M., Schutte, W.~A., \& van Dishoeck, E.~F. 2003,
  \aap, submitted

\bibitem[{Sogawa {et~al.}(1997)Sogawa, Tamura, Gatley, \& Merril}]{Sogawa97}
Sogawa, H., Tamura, M., Gatley, I., \& Merril, K.~M. 1997, \aj, 113, 1057

\bibitem[{Soifer {et~al.}(1979)Soifer, Puetter, Russel, Willner, Harvey, \&
  Gillet}]{Soifer}
Soifer, B.~T., Puetter, R.~C., Russel, R.~W., {et~al.} 1979, \apj, 232, L53

\bibitem[{Somorjai(1994)}]{Somorjai94}
Somorjai, G.~A. 1994, Introduction to surface chemistry and analysis (Wiley and
  sons)

\bibitem[{Tanaka {et~al.}(1994)Tanaka, Nagata, Sato, \& Yamamoto}]{Tanaka94}
Tanaka, M., Nagata, T., Sato, S., \& Yamamoto, T. 1994, \apj, 430, 779

\bibitem[{Teixeira {et~al.}(1998)Teixeira, Emerson, \& Palumbo}]{Teixeira}
Teixeira, T.~C., Emerson, J.~P., \& Palumbo, M.~E. 1998, \aap, 330, 711

\bibitem[{Thi(2002)}]{ThiThesis}
Thi, W.-F. 2002, PhD thesis, Leiden Observatory

\bibitem[{Thi {et~al.}(2002)Thi, Pontoppidan, van Dishoeck, Dartois, \&
  d'Hendecourt}]{ThiCRBR}
Thi, W.-F., Pontoppidan, K.~M., van Dishoeck, E.~F., Dartois, E., \&
  d'Hendecourt, L. 2002, \aap, 394, 27

\bibitem[{Tielens \& Hagen(1982)}]{TielensHagen}
Tielens, A. G. G.~M. \& Hagen, W. 1982, \aap, 114, 245

\bibitem[{Tielens {et~al.}(1991)Tielens, Tokunaga, Geballe, \& Baas}]{tielens}
Tielens, A. G. G.~M., Tokunaga, A.~T., Geballe, T., \& Baas, F. 1991, \apj,
  382, 523

\bibitem[{van Broekhuizen {et~al.}(2003)van Broekhuizen, Schutte, \&
  Fraser}]{Fleur}
van Broekhuizen, F., Schutte, W.~A., \& Fraser, H. 2003, \aap, in prep

\bibitem[{Vandenbussche {et~al.}(1999)Vandenbussche, Ehrenfreund, Boogert, van
  Dishoeck, Schutte, Gerakines, Chiar, Tielens, Keane, Whittet, Breitfellner,
  \& Burgdorf}]{Bart}
Vandenbussche, B., Ehrenfreund, P., Boogert, A. C.~A., {et~al.} 1999, \aap,
  346, L57

\bibitem[{Whitney {et~al.}(1997)Whitney, Kenyon, \& Gomez}]{Whitney97}
Whitney, B.~A., Kenyon, S.~J., \& Gomez, M. 1997, \apj, 485, 703

\bibitem[{Wilking {et~al.}(1989)Wilking, Lada, \& Young}]{WLY}
Wilking, B.~A., Lada, C.~J., \& Young, E.~T. 1989, \apj, 340, 823

\bibitem[{Wilking {et~al.}(1997)Wilking, McCoughrean, Burton, Giblin, Rayner,
  \& Zinnecker}]{Wilking97}
Wilking, B.~A., McCoughrean, M.~J., Burton, M.~G., {et~al.} 1997, \aj, 114,
  2029

\bibitem[{Wilking {et~al.}(1986)Wilking, Taylor, \& Storey}]{Wilking86}
Wilking, B.~A., Taylor, K. N.~R., \& Storey, J. W.~V. 1986, \aj, 92, 103

\bibitem[{Willner {et~al.}(1982)Willner, Gillet, Herter, Jones, Krassner,
  Merril, Pipher, Puetter, Rudy, Russel, \& Soifer}]{Willner}
Willner, S.~P., Gillet, F.~C., Herter, T.~L., {et~al.} 1982, \apj, 253, 174

\bibitem[{Ziman(1979)}]{Ziman}
Ziman, J.~M. 1979, Principles of the Theory of Solids, 2nd edn. (Cambridge
  University Press)

\bibitem[{Zumofen(1978)}]{Zumofen}
Zumofen, G. 1978, J. Chem. Phys., 68, 3747

\end{thebibliography}

\end{document}